%% file: main.tex
\documentclass[11pt]{article}
\usepackage[utf8]{inputenc}%
\usepackage[tracking=true, letterspace=100, expansion=false]{microtype}
\usepackage[T1]{fontenc}
\usepackage{amsmath}%
\usepackage{amsfonts}%
\usepackage{amsthm}%
\usepackage{comment}
\usepackage{csquotes}
\usepackage{booktabs}
\usepackage{tabularx}
\usepackage{soul}
\usepackage[font=small,labelfont=bf]{caption}
\usepackage[labelformat=simple]{subcaption}
\usepackage{multirow}
\usepackage{graphicx}
\usepackage{varioref}

\theoremstyle{definition}

\theoremstyle{remark}

\usepackage[margin=1in]{geometry}
\usepackage[onehalfspacing]{setspace}
\usepackage{xurl}
\usepackage{mathtools}

\usepackage{bbm} % blackboard math for computer modern

\usepackage[title]{appendix}
\usepackage{etoolbox} % provides AtBeginEnvironment
\AtBeginEnvironment{appendices}{\crefalias{section}{appendix}}
\usepackage{pdflscape}
\usepackage{afterpage}
\usepackage{longtable}
\usepackage{changepage}
\usepackage{natbib}
\usepackage{xcolor}%
\definecolor{webbrown}{rgb}{.6,0,0}%
\usepackage{hyperref} %
\hypersetup{%
  breaklinks = true,%
  colorlinks = true,%
  anchorcolor = webbrown,%
  citecolor = webbrown,%
  filecolor = webbrown,%
  linkcolor = webbrown,%
  menucolor = webbrown,%
  urlcolor= webbrown,%
  citebordercolor= 1 0 0,%
  menubordercolor=1 0 0,%
  urlbordercolor=1 0 0,%
  runbordercolor=1 0 0,%
  pdftitle={Anonymity and Identity Online},
  pdfauthor={Ederer, Goldsmith-Pinkham, Jensen}
}

\usepackage{cleveref}
\crefname{appsec}{appendix}{appendices}
\crefname{appsubsec}{appendix}{appendices}
\crefname{assumption}{assumption}{assumptions}

\usepackage[nolist]{acronym}
\begin{acronym}
  \acro{CI}{confidence interval}%
  \acro{OLS}{ordinary least squares}%
  \acro{CLT}{central limit theorem}%
  \acro{IV}{instrumental variables}%
  \acro{ATE}{average treatment effect}%
  \acro{RCT}{randomized control trial}%
  \acro{VAM}{value-added model}%
  \acro{LAN}{locally asymptotically normal}%
  \acro{DiD}{difference-in-differences}%
  \acro{OVB}{omitted variables bias}
  \acro{FWL}{Frisch-Waugh-Lovell}
\end{acronym}

\usepackage{accents}

\newcolumntype{Y}{>{\centering\arraybackslash}X}

\input{constants}

\usepackage{float}

\makeatletter
\newenvironment{chapquote}[2][2em]
  {\setlength{\@tempdima}{#1}%
   \def\chapquote@author{#2}%
   \parshape 1 \@tempdima \dimexpr\textwidth-2\@tempdima\relax%
   \itshape}
  {\par\normalfont\hfill--\ \chapquote@author\hspace*{\@tempdima}\par\bigskip}
\makeatother

\begin{document}

\title{\textbf{Anonymity and Identity Online}\thanks{We are grateful to Jason Abaluck, Andre Alcantara, Tim Bresnahan, Nicholas Christakis, Damon Clark, Forrest Crawford, Diego Jiménez Hernández, Tanjim Hossain, Rafael Jiménez-Durán, Camilo Machado Goncalves, John Mamer, Rachael Meager, Barry Nalebuff, Matt Notowidigdo, Michael Ostrovsky, Julia Simon-Kerr, Steve Tadelis, Catherine Tucker, Mahboud Zabetian, seminar audiences at the AEA Meetings, NBER Summer Institute, Boston University, Ofcom, San Francisco Federal Reserve, University of Toronto, and Yale as well as several economists who wish to remain anonymous, for their helpful comments and suggestions. The analysis described in this manuscript was allowed to proceed by the Yale HRPP, IRB protocol ID 2000034072. Our study uses only publicly available pages on EJMR, the same pages crawled and indexed by major search engines and does not access any non-public pages or hidden URLs. Due to the presence of misinformation about our paper, we compiled an FAQ in Appendix~\ref{sec:faq}.}}
\author{Florian Ederer\thanks{Boston University Questrom School of Business, CEPR, ECGI, and NBER}
\and Paul Goldsmith-Pinkham\thanks{Yale School of Management and NBER}
\and Kyle Jensen\thanks{Yale School of Management}}
\date{\today}

\pagenumbering{Alph} 
\begin{titlepage}
  \maketitle
  \thispagestyle{empty}

\begin{adjustwidth*}{0.1cm}{0.1cm}
\begin{abstract} % 100 word version:
\noindent Economics Job Market Rumors (EJMR) is an online forum and clearinghouse for information on the academic job market for economists. It also includes content that is abusive, defamatory, racist, misogynistic, or otherwise ``toxic.'' Almost all of this content is created anonymously by contributors who receive a four-character username when posting on EJMR\@. \emph{Using only publicly available data} we show that the statistical properties of the scheme by which these usernames were generated allows the IP addresses from which most posts were made to be determined with high probability.\footnote{The scheme changed on May 17, 2023 after the first release of the abstract of this paper on May 16, 2023. The scheme changed again on July 21, 2023 after the first public presentation of this paper on July 20, 2023.} 
We recover \niceNumDistinctIPsAssigned~distinct IP addresses of EJMR posters and attribute them to \percentageOfEJMRPostsAssigned~of the roughly 7 million posts made over the past 12 years. We geolocate posts and describe \emph{aggregated} cross-sectional variation---particularly regarding toxic, misogynistic, and hate speech---across sub-forums, geographies, institutions, and IP addresses. Our analysis suggests that content on EJMR comes from all echelons of the economics profession, including, but not limited to, its elite institutions.

\end{abstract}
\end{adjustwidth*}

\bigskip
\noindent \textbf{JEL Codes}: C55, D83, D91, L86, Z13 \\
% D12 Consumer Economics: Empirical Analysis
% D83 Search; Learning; Information and Knowledge; Communication; Belief
% D90 Micro-Based Behavioral Economics: General
% D91 Micro-Based Behavioral Economics: Role and Effects of Psychological, Emotional, Social, and Cognitive Factors on Decision Making
% C38 Multiple or Simultaneous Equation Models: Classification Methods; Cluster Analysis; Principal Components; Factor Models
% C55 Large Data Sets: Modeling and Analysis
% Z13 Economic Sociology; Economic Anthropology; Language; Social and Economic Stratification
% L86	Information and Internet Services; Computer Software
% C63	Computational Techniques
% C8	Data Collection and Data Estimation Methodology • Computer Programs
\noindent \textbf{Keywords}: cryptography, internet privacy, large language models, toxic speech

\end{titlepage}
\pagenumbering{arabic}
\setstretch{1.2}

\begin{chapquote}{Jacob Viner, quoted in \citet[p.~814]{spiegel1987viner}}
``Economics is what economists do.''
\end{chapquote}

% \section{Introduction}
% \label{sec:introduction}

Economics Job Market Rumors (\url{http://www.econjobrumors.com}), henceforth EJMR, is an anonymous internet message board featuring discussion about economics, the economics profession, and, in particular, the annual job market for PhD economists. However, the discussion board is active year-round and EJMR users also post content unrelated to economics. The site is popular: in early 2023, SimilarWeb estimated that  EJMR received 2.5 million visits per month with an average of 6.45 pages viewed per visit. In comparison, the same figures for the websites of the National Bureau of Economic Research (NBER) and the American Economic Association (AEA) were 1.1 million and 991,000 visits and 2.09 and 2.76 pages per visit, respectively.

EJMR is controversial in the economics profession. It has been called ``a breeding ground for personal attacks of an abusive kind'' \citep{blanchard2017ejmr}, a ``cesspool of misogyny'' \citep{romer2017evidence}, ``4chan'' and ``4chan for economists'' \citep{lowrey2022sexism,taylor20234racist}, and ``evidence of a toxic environment for women in economics'' \citep{wolfers2017evidence}.\footnote{The Committee on the Status of Women in the Economics Profession (CSWEP) of the AEA (\url{https://www.aeaweb.org/about-aea/committees/cswep/statement}) also condemned ``the sexist, racist, homophobic and anti-Semitic statements that have appeared on the Economics Job Market Rumors (EJMR) site, and particularly the harassment and abuse targeted at particular scholars.'' More than 1,000 signers urged the AEA in 2017 to create a moderated, well-functioning site to provide up-to-the-minute job market information (\url{https://www.iaffe.org/petition-aea-ejmr/}). However, the forum EconSpark and the information website EconTrack, both created by the AEA, were languish without much use.} A substantial fraction of EJMR consists of content that is abusive, defamatory, racist, misogynistic, or otherwise ``toxic.'' Such content exists in spite of both automated and manual moderation. Examples include ``the whole point of women is to get railed and make babies'' and ``The biggest enemies of America are: Blks'' and ``America lost its war against blks \textelp{} At least until we resolve to final solution'' and ``University of Stupid Chinese'' and ``The average woman has a 15\% smaller brain than the average man'' and ``the fastest route to a qje is to grift and be black.'' Comments also target particular individuals, including examples such as ``Should Jennifer Doleac be executed for her anti-Chinese hatred?'' and ``Anya Samek \textelp{} took advantage of her initial postdoc position organizing conferences \textelp{} and handling requests for grant proposals to steal ideas'' and ``Are Vrinda and Hampole in a secret same-sekhs love-hayte relationship?''.

In this paper, we show that EJMR's contributors post from locations, institutions, and universities that are intrinsically linked to the academic economics community, including the upper echelons of academia, government, and the private sector. This result contradicts the common claim that EJMR is not representative of the economics profession and that its most frequent users are not actually economists. We confirm that EJMR contains a substantial amount of sexist content, as first documented by \cite{wu2018gendered} and \cite{wu2020gender}. We also provide new evidence that toxic speech is pervasive across contributors and institutions, including elite universities, suggesting that EJMR perpetuates and amplifies existing inequalities in the economics profession  \citep{bayer2016diversity,antecol2018equal,lundberg2019women,dupas2021gender,hengel2022publishing}.  

EJMR used a specific algorithm to assign usernames to posts written by anonymous contributors. We show that the properties of this algorithm do not anonymize the contributors of posts, but instead allows the IP address from which each post was made to be determined with high probability. 
To recover IP addresses from the observed usernames on EJMR, we employ a multi-step procedure.\footnote{Our approach has similarities to the recovery process used in \cite{CHEN2020118} where they uniquely identify partially obfuscated IP addresses in SEC Edgar log-files. However, our setting is more complex and larger scale, relying on the avalanche property of the SHA-1 hash.} First, we develop GPU-based software to quickly compute the SHA-1 hashes used for the username allocation algorithm on EJMR\@. In total, we compute almost \englishNumHashes~hashes to fully enumerate all possible IP combinations and to check which of the resulting substrings of hashes match the observed usernames. For each post, this procedure narrows the set of possible IP addresses from $2^{32}$ to roughly $2^{16}$. Second, we measure which IP addresses occur particularly often in a short time window and use the uniformity property of the SHA-1 hash to test whether these IP addresses appear more often than would likely occur by chance. 

Our statistical test minimizes the probability of falsely assigning an IP address to a post because the p-value thresholds we employ are of the order of approximately $10^{-11}$. Our approach has several further advantages. First, even though there are \niceTotalPosts~posts on EJMR, we assign exactly \emph{zero} posts to the large set of unaddressable (bogon) IP addresses from which no traffic should legitimately appear on the public Internet. Second, our assignment method precisely pinpoints when EJMR changed its hashing algorithm and assigns exactly \emph{zero} posts if an incorrect hashing algorithm is used. Third, our methods correctly identify the location of users' IP addresses. IP addresses post during the standard work and day time hours of their geolocation and the dominant non-English language of the country of origin of the IP address is the country’s native language. Despite this conservative approach our procedure recovers \niceNumDistinctIPsAssigned~distinct IP addresses of EJMR posters and assigns \percentageOfEJMRPostsAssigned~of the roughly 7 million total posts to these IP addresses.

We then describe \emph{aggregated} features of posting behavior on EJMR\@. Despite several attempts to limit its influence, EJMR remains popular. In 2022, the platform averaged 70,000 monthly posts. This allows us to recover roughly 1,100 unique IP addresses per month. Although EJMR has been a popular website for economists since its inception, posting and engagement on the site have surged since the start of the COVID-19 pandemic, especially in the United States where posting rates tripled. This increase was primarily driven by a very large increase in off-topic forum posts that are mostly unrelated to the job market for academic economists. Off-topic posts in other countries such as Canada, the United Kingdom, Hong Kong, Australia, Germany, Italy, and France also increased, but the increases were less pronounced and more temporary.

Based on the geographic location of the IP addresses we identify, we show that the majority of posts come from large cities in the US (Chicago, New York, Philadelphia) and outside the US (Hong Kong, London, Montreal, Toronto). Smaller US cities with leading research institutions such as Cambridge and Berkeley are also towards the top of the list. Beyond the city and country, the IP addresses we recover also allow us to identify the associated internet service providers. Based on this information, we show that posting on EJMR is pervasive throughout the economics profession. Around 15\% of posts originate from university networks including, but not limited to, all top-ranked universities in the United States. A substantial number of posts also come from government agencies, companies, and non-profit organizations employing economists as well as universities around the world.

We then study the distribution of problematic content on EJMR and show that it is widespread among both residential and university IP addresses, and not concentrated among just a small set of contributors. To do so, we deobfuscate the text of EJMR posts and use state-of-the-art transformer models to show that roughly 11.8\% of posts contain text labeled as toxic, 3.3\% as misogynistic, and 3.1\% as hate speech. Posts marked as toxic, misogynistic, and hate speech are more likely to originate from residential IP addresses than from IP addresses associated with universities, but the difference is small. We also show that IP addresses are very often posting in topics that contain toxic content, even when their own content is not toxic. Moreover, the share of problematic content on EJMR is high even by the standards of anonymous online forums. We show that EJMR posts are more likely to be labeled as toxic, misogynistic, and hate speech than 69, 73, and 95 percent of the 1,000 most popular subforums (subreddits) of Reddit, a popular discussion website where users can post relatively anonymously. We also find that EJMR posts mentioning women are substantially more likely to contain hate speech, misogyny, and toxicity. Moreover, mentions of women attract even more misogyny and hate speech on EJMR than on Reddit, but the increase in toxicity is similar.

Finally, we ask what drives posting behavior. Due to EJMR's anonymous nature, we have limited explanatory variables. We first consider the set of posts associated with a university and examine how characteristics of those universities affect posting behavior. We find that universities with larger economics programs and higher ranked universities both have significantly higher numbers of posts, but do not have significantly higher average toxicity. 

We then show that intrinsic motivation for attention gives individuals reason to participate on EJMR. To do so, we employ an empirical design inspired by \cite{srinivasan2023paying}, where we use the first time an IP creates a new topic on EJMR, and examine the variation in how much attention that topic gets within the first day. We show that if this topic receives more initial attention, that IP address will post significantly more frequently \emph{on other threads} over the next three days. Hence, despite the lack of monetary or reputational incentives to accrue attention across the platform, attention on users' initial posts have meaningful effects on their use of the site. 

\section{Methods}
\label{sec:methods}

\subsection{Relationship between IP Addresses and Usernames}
\label{sec:relationship}

The vast majority of EJMR users do not log into the EJMR website using a persistent username of their own selection. Instead, the site uses a scheme it described as follows: ``EJMR allows you to post anonymously whereby your post enters the database without a record of personally identifiable information like your IP or email address. However to prevent users from voting for themselves and to help users maintain the same 4 letter indentity [sic] within a thread one way encryption is used to create a 4 letter identity. This is a combination of random strings and the user's IP address which is one way encrypted and then sliced up to create a 4 letter ID which is stored in the database.''

Shortly after introducing this scheme EJMR's pseudonymous administrator ``Kirk'' wrote ``...for example you can see this post is also from me by looking at the fddf2 on the left. But I'll give you a million US\$ if you can guess my ip.''\footnote{As this post shows, EJMR briefly experimented with five-letter usernames.} That IP address---twelve years ago---was almost certainly 188.220.40.122.\footnote{This IP address is part of a large block of consumer IP addresses which likely changes hands frequently, could be used by millions of devices, and gives little geographic detail beyond being in or proximate to London. This location is, however, consistent with an article on EJMR and ``Kirk'' in the German newspaper WirtschaftsWoche in 2011 (\url{https://amp2.wiwo.de/politik/economics-job-market-rumors-die-geruechtekueche-der-volkswirte/5971044.html}).}

How we can make such a statement? In brief, we correctly guessed the scheme by which EJMR assigned usernames for most of its history. The statistical properties of that scheme allow us to back out IP addresses from usernames for about \percentPostsAssigned\% of posts on EJMR\@. We likely identify 100\% of the IP addresses that were consistently active on the site and, in thousands of cases, IP addresses active on the site for as little as a week's time. \ul{\emph{All of this is achieved using only publicly available data.}} With the exception of the aforementioned ``million-dollar'' address, we will not disclose IP addresses associated with EJMR posts in this paper.

Before we describe how to map usernames to IP addresses, we provide a short introduction to a few concepts that may be unfamiliar to readers trained in economics. First, IPv4, or Internet Protocol version 4, is the prevailing method of addressing  devices on the internet.
An IPv4 address is a 4-byte, or 32-bit, number in the interval  $[0, 2^{32})$, or $[0, 4294967296)$. For example, 2189728028 is a valid IP address. However, this IP address would more commonly be written in the so-called ``dotted decimal'' notation as 130.132.153.28, wherein each number represents one of the four bytes, or four octets, in the binary representation of the number 2189728028. Each octet is an integer in the range $[0, 256)$. Blocks of IP addresses are assigned by the Internet Assigned Numbers Authority to ``autonomous systems:'' smaller networks of computers administered by internet service providers (ISPs), government agencies, corporations, and universities. For example, Yale University owns autonomous system number 29 (AS29) and owns two blocks of $2^{16}$ addresses (about 130,000 in total) including the above address 130.132.153.28.

Autonomous systems allocate their IP addresses to devices on their networks using a variety of mechanisms. For example, an IP address can be statically assigned to a device, in which case the device can have the same IP address for years. Or, IP addresses can be dynamically assigned to devices through methods like Dynamic Host Configuration Protocol (DHCP), as might be the case on a university wireless network.  Devices using DHCP retain IP addresses for minutes to months. Also, devices can be behind ``network address translation'' or NAT, a method that allows multiple devices to share the same ``external'' IP address while using a unique ``local'' address. ``Carrier grade'' NAT is particularly common for mobile networks~\citep{livadariu2018nat}. Multiple devices can also share the same IP address when using a proxy, a virtual private network (VPN), or an ``anycast'' domain name system (DNS) configuration. The latter of these is rare for consumer devices, however, proxies and VPNs are common.

As this description should make clear, \emph{IP addresses are not people}. Humans using devices, such as those posting to websites, change IP addresses. They use multiple devices. And, quite often, masses of people can share the same IP address. Therefore, nothing in our manuscript should be construed as identifying \emph{persons}.\footnote{The ability to link several posts coming from the same IP address reveals additional information about the persons posting content on EJMR and in some cases can allow for the identification of specific individuals. However, such an exercise is not the focus of this paper.}

Having covered IP addresses, a short description of EJMR's organization is necessary to understand how IP addresses fit into its anonymity protocol. EJMR is built on bbPress~\citep{bbpress}, which is a version of the popular WordPress blogging software that is customized to host forums. bbPress websites, such as EJMR, are organized by \emph{topics}. Each topic has a URL and a title. For example, the topic at URL \url{https://www.econjobrumors.com/topic/dream-job-imf-economist} has the title ``Dream job: IMF economist''. Topics also have \emph{posts}. When a user creates a new topic, that user simultaneously creates the first post. Subsequently, other users can add posts to the topic. Each post has some user-written content, a timestamp, and a username identifying the user who made the post.

As we described previously, most EJMR users do not log into the site using a self-selected username as do users on a typical bbPress site. Instead, the overwhelming majority of posts on the site are made by anonymous users for whom EJMR \emph{generates} usernames.  Consider the topic ``Dream job: IMF economist''. The initial poster asks about how best to gain employment at the IMF and was assigned username 270b. Shortly thereafter, EJMR assigned username dd86 to the user replying ``Your country must be a real craphole for IMF to be your dream job.''

As EJMR's notification says, the site assigns contributors a \emph{persistent} username for each topic using the contributor's IP address. That is, a contributor commenting on topic $t$ from IP address $a$ always receives the same username, regardless of date, comment content, or browser state, including user-agent and cookies.  If, at some later date, user dd86 contributed EJMR from the same IP address and offered more advice in the IMF topic, they would retain the username dd86. However, this contributor will, with probability approaching one, receive a \emph{different} username when they post on a different topic or from a different IP addresses.

How are usernames generated on EJMR? The four-letter usernames comprise solely the characters a-f and 0-9, which suggested to us that the usernames are hexadecimal encoded numbers. Hexadecimal---or base-16---encoding is compact way to write large numbers as alternative to base-10. Hexadecimal digits include the base-10 Arabic digits 0-9 and A, B, C, D, E, and F which represent 10, 11, 12, 13, 14, and 15, respectively. In base-10, the username of the IMF poster 270b is the number $2\times16^3 + 7\times16^2 + 0\times16^1 + 11\times16^0 = 9995$. Hexadecimal is common encoding by which to represent the output of ``hash'' functions. We guessed that hashing is the technique to which EJMR referred when saying ``the user's IP address ... is one way encrypted.''\footnote{Strictly speaking, hash functions are not a form of encryption, which is by definition two-way.} Hash functions---such as the SHA and MD5 family of functions---are functions that map data of arbitrary or large range to a domain of fixed size. For example, imagine a hash function $f$ that takes a sentence of text and returns an integer representing the index of the first letter of text in the English alphabet. The sentences are of arbitrary size and the domain is of finite size: $[1-26]$. This would be a bad hash function for most practical uses of hash functions. As the EJMR notice referenced above describes, hashes are ``one-way'' functions: the output is easily determined from the input, but given an output it is difficult or impossible to know the input. For example, knowing that a sentence begins with ``E'' does not tell you the sentence. Most cryptographic hash functions do not output small numbers like 26, but rather exceedingly large numbers that are more compactly written in hexadecimal than decimal format. For example, the output of the SHA-1 hash is an integer in the range $[0,2^{160})$.

The topic pages for websites built on BBPress each contain two identifiers that we thought might be inputs to the username scheme. For example, consider the EJMR topic page \url{https://www.econjobrumors.com/topic/dream-job-imf-economist}. This topic has a ``slug'' which is the string ``dream-job-imf-economist'' and it also has a numeric ID, which has the value 227259 in this case. Topic IDs on BBPress websites are auto-incrementing integer primary keys in the underlying MySQL relational database used by BBPress and WordPress. Topics can be accessed by these IDs online. For example, visiting \url{https://www.econjobrumors.com/topic/227259} redirects to \url{https://www.econjobrumors.com/topic/dream-job-imf-economist}, showing the same content to visitors.

We guessed that EJMR's usernames were generated as follows
\begin{equation}
u =  S(\mathcal{H}(M(t, a, o)))
\end{equation}
where $t$ is a topic identifier, $a$ is a visitor's IP address, and $o$ is some other data, typically a ``salt'' which is used to obfuscate the hashed data~\cite[p.~304]{ferguson2010cryptography}.  $M$ is a function that mixes the inputs, such as a ``stretch'', function \cite[p.~304]{ferguson2010cryptography}. $S$ is the function EJMR uses by which the output would be ``sliced up to create a 4 letter ID''. $\mathcal{H}$ is a hash function \cite[p.~77]{ferguson2010cryptography}. Because the EJMR usernames are in hexadecimal, we suspected $S$ to be a simple function of $\mathcal{H}$'s output.

We had in our possession three different EJMR usernames for which we knew both the topic ID and the IP address from which the post was made. This gave us three concordant sets of $u$, $t$, and $a$.  We suspected $a$ was restricted to IPv4 addresses, which are more commonly used than IPv6 addresses. Later, we verified that EJMR's webserver does not respond to IPv6 internet traffic, only IPv4 traffic. Therefore, each EJMR user has an IPv4 address. We observed $u$ and $t$ on EJMR\@. We began a search for $o$, $M$, $\mathcal{H}$ and $S$ with simple guesses by which we attempted to recreate our three observed values of $u$ for our three sets of $u$, $t$, and $a$.\footnote{In total, we posted five times on EJMR: three times to verify the hashing scheme and two times to produce a brief set of videos to document how the hashing scheme worked. These videos can be viewed at \url{https://www.youtube.com/watch?v=on5YCsEhGrY&list=PLWWcL1M3lLloToQOE_j1Ys8dQZlckMIIp}.} Our search was short. We presumed that there was no salt and thus set $o$ to null. We guessed  $M$ to be either the concatenation of $t$ and $a$ as strings, or $a$ and $t$. We further guessed  $S$ to be a function that merely returned a substring of the hash. Finally, we guessed that $\mathcal{H}$ was a common hash function such as MD5, SHA-1/224/256/384/512, or CRC32. 

We found that $o$ is indeed null. $S$ is the string concatenation of $t$ and $a$, where $a$ is in the dotted decimal notation. $\mathcal{H}$ is the SHA-1 hash, and $M$ returns characters 10-13 of the hexadecimal hash (1-based indexing). That is, if a user visits EJMR from the IPv4 address 131.111.5.175 and posts on the topic with id 227259, EJMR assigns the username c2b1. This is the four-character interval at position 10-13 in e8b5eae32c2b197a0ac4cb889a9bbb8f417f3bff which is the hexadecimal encoding of the SHA-1 hash of the string ``227259131.111.5.175'' (ASCII encoded). In other words, the EJMR username is the hexadecimal representation of the two bytes of data beginning at the 40th bit of the 20-byte big-endian SHA-1 hash. In plain English, EJMR combines a visitor's IP address with an integer topic id, hashes that with SHA-1, and uses a part of that hash as the username.

\begin{figure}[tbp]
    \centering
    \includegraphics[width=\linewidth]{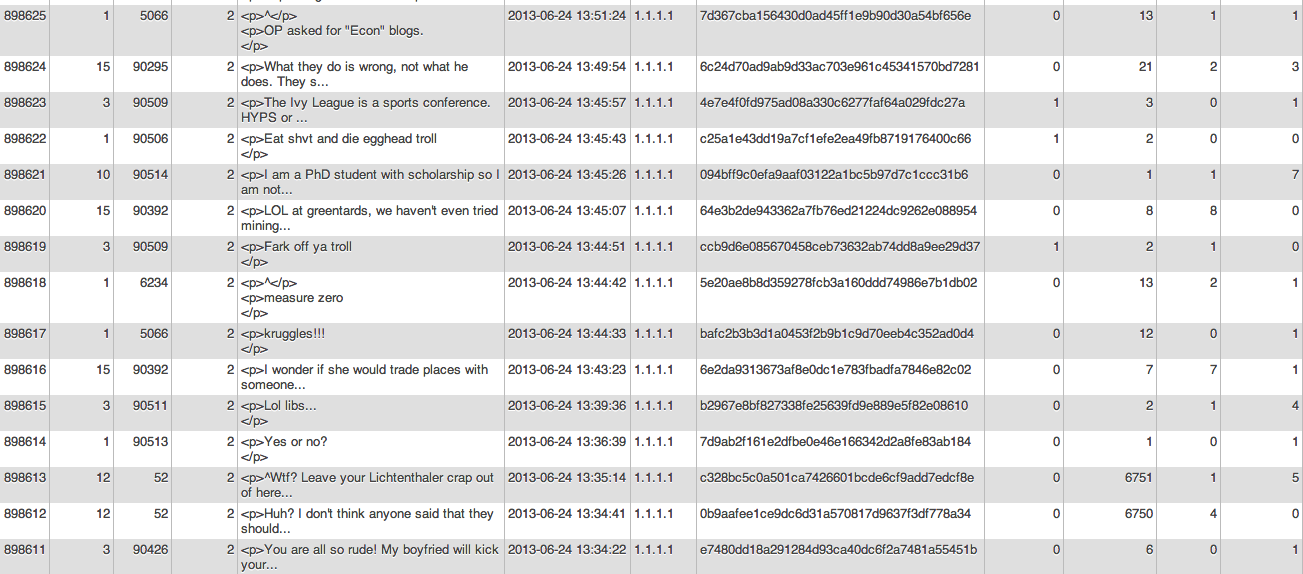}    
    \caption{Screenshot of the EJMR MySQL database posted by the administrator ``Kirk.''}
    \label{fig:kirk-imgur-database-screenshot}

\end{figure}

The above is an accurate description of the EJMR username scheme for the period from July 8, 2013 to May 17, 2023. Because this scheme is no longer in use, a skeptical reader may rightly ask how they can verify this claim. We have three responses. First, the results that follow will, in numerous ways, show statistical patterns that would be nigh impossible if we were incorrect about the username generation scheme. Second, on February 7, 2023, we recorded a brief video in which we show how an EJMR username could be computed \emph{prior to posting on the site} with a knowledge of a topic ID and one's IP address. This video can be viewed at \url{https://www.youtube.com/watch?v=on5YCsEhGrY&list=PLWWcL1M3lLloToQOE_j1Ys8dQZlckMIIp}. Third, and most importantly, our claim is supported by the public posts of EJMR's administrator. On July 3, 2013, the site administrator wrote ``Here is a direct screenshot of all of the fields for each post \url{http://i.imgur.com/1htoXw7.png}''. This post can be viewed in the WayBack Machine\footnote{\url{https://web.archive.org/web/20230531180223/https://econjobrumors.com/topic/kirk-31\#post-913648}} and the screenshot appears in Figure~\ref{fig:kirk-imgur-database-screenshot}. The screenshot shows 15 rows, one for each of 15 different posts. Each post has 12 columns. Column 3 is the topic ID and column 8 shows the SHA-1 hash of a topic ID and the IP address from which the post was made. Readers can easily verify that there is \emph{one and only one} IPv4 address that, when pre-pended with the topic ID, produces the SHA-1 hash shown in the screenshot. For example, there is only a \emph{single} IPv4 address ending in ``.42'' that, when pre-pended with 6234, produces the SHA-1 hash 5e20ae8b8d359278fcb3a160ddd74986e7b1db02. 

\begin{figure}[ht]
    \centering
    \includegraphics[width=\linewidth]{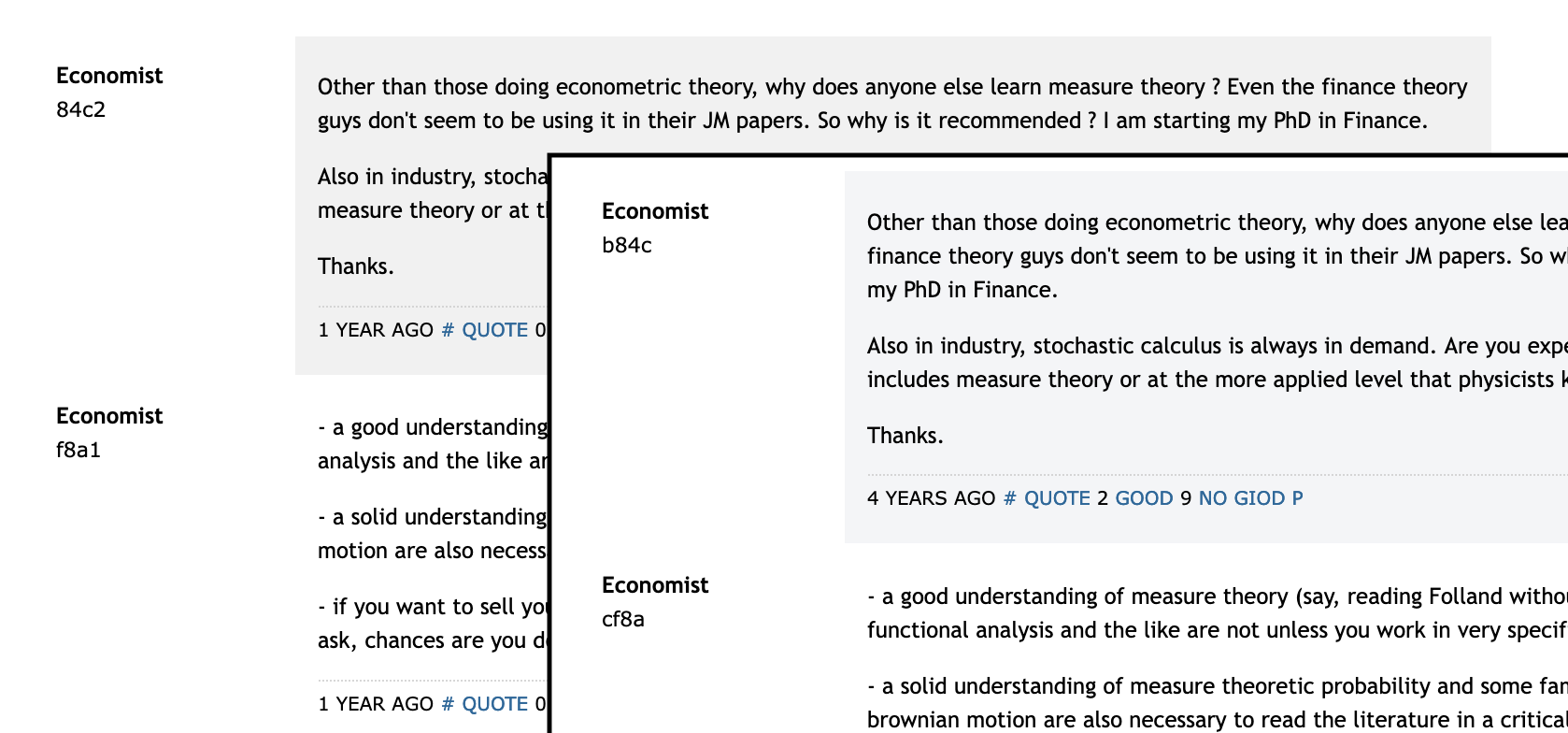}    
    \caption{Screenshot of an EJMR topic before and after July 8, 2013, from the WayBack Machine. The leftmost posts shows usernames created using positions 10, 11, 12, and 13 in the SHA-1 hash of the topic id, IP address combination. The inset on the right shows the same posts after July 8, 2013 where the usernames are constructed using positions 9, 10, 11, and 12 of the same hash. We believe this change to be a result of administrator error.
    }
    \label{fig:2013-hash-changes}

\end{figure}

At the time this screenshot was shared on EJMR, the website saved the entire SHA-1 hash for each post, but \emph{displayed} just characters 10-13 from the hash. In response to some EJMR user criticism, on July 8, 2013, the site administrator began storing just positions 10-13 in the database instead of the whole hash. The site administrator also elected to purge old hashes from the database. But, for old posts (i.e., posts before July 8, 2013) EJMR began showing positions 9-12 of the SHA-1 hash of each topic-IP pair, as shown in Figure~\ref{fig:2013-hash-changes}. This was mostly likely due to an error on the part of the administrator. BBPress is built with the PHP language, which uses zero-based indexing. So, the PHP code for the EJMR username scheme looks something like \texttt{substr(sha1(\$topic\_id . \$user\_ip), 9, 4)}, which means ``take the substring from position 9 for 4 characters'' where position 9 is the \emph{tenth} character in the hash because the first character is \emph{zero}. It seems likely that, in an effort to discard whole SHA-1 hashes, the site administrator issued a MySQL command like \texttt{update posts set the\_sha1\_hash = substring(the\_sha1\_hash from 9 for 4)}. However, MySQL uses 1-based indexing instead of 0-based indexing. Therefore, the effect of this command would be to ``shift'' the username left for posts made before July 8, 2013. Having issued this SQL command, without IP addresses or a database backup, it would be impossible to ``correct'' the usernames. Using the WayBack Machine, readers can verify that the EJMR usernames before May 17, 2023 were as we claim here. For example, the above-mentioned post %\footnote{\url{https://web.archive.org/web/20150715234225/https://www.econjobrumors.com/topic/why-is-measure-theory-considered-useful-for-theory-and-finance\#post-898618}}
with hash 5e20ae8b8d359278fcb3a160ddd74986e7b1db02 in Figure~\ref{fig:kirk-imgur-database-screenshot} had username 8d35 when captured by the WayBack Machine in 2015. This is position 9 to 12 in the hash because this post was among those ``shifted'' left.

The discussion above should make it clear that the EJMR username scheme sat in plain sight for the past decade. Further, the scheme is unsophisticated. It contains no salt ($o$), stretching function ($M$), or cipher (nontrivial $S$). The scheme accomplishes the objective of ensuring that returning visitors receive a consistent username in each topic. However, it is an \emph{exceedingly bizarre} choice if one wishes to obfuscate the IP address from which a post originates. The most common use of cryptographic hashes like SHA-1 is to \emph{identify} content. For example, SHA hashes are used to identify content in the git version control system and on the Bitcoin blockchain. The username scheme that EJMR uses nakedly advertises roughly 16 bits of information from each post's 32-bit IPv4 address origin. Due to the trivial choices for $M$, $S$, and in particular $o$ (no salt), that information is readily recoverable, as we describe in the next section.

\subsection{Mapping Usernames to Candidate IP Addresses}
\label{sec:mapping}

To answer basic questions about the distribution of toxic speech on EJMR, we
need to map EJMR posts to the IP addresses from which they were made. This
presents a few challenges. To understand these challenges and how we overcame
them, some background on the SHA-1 hash is helpful. The SHA-1 hash was created
by the United States National Institute of Standards and Technology in
1995~\citep{standard1995fips}. Like its predecessor MD5 and its successor
SHA-256, the SHA-1 hash uses the so-called Merkle–Damgård
construction~\citep{damgaard1990design}. Each of these hashes is widely used.
For example, as mentioned above, the version control system \texttt{git} uses
SHA-1 to identify source code changes~\cite{spinellis2012git} and SHA-256 is
used in the proof-of-work system of Bitcoin~\citep{nakamoto2008bitcoin}. Each of
these hashes has a desirable property called the ``avalanche'' effect whereby
small changes in the input produce large changes in the
output~\citep{motara2016sha}. Ideally, a one-bit change in the input causes each
output bit to flip with probability $0.5$. The avalanche effect leads to the
``uniformity'' property of these hashes such that the inputs map uniformly to
the output domain. In the case of SHA-1, this means that inputs, such as the
topic-IP concatenation of EJMR, are uniformly mapped to the range $[0,2^{160})$.

The uniformity property of SHA-1 implies that every hash value in the output
range is generated with roughly the same probability
\citep{cormen2022introduction}. Of course, the EJMR username is a substring of
the SHA-1 hash (four hexadecimal characters). To verify that the SHA-1 hash is
also uniform over this substring, we conducted two experiments. First, we
randomly chose a topic ID and computed the EJMR username for all IPv4 addresses.
Second, we chose a random IPv4 address and computed the EJMR username for all
extant EJMR topic IDs. In both cases, we found the SHA-1's uniformity to be
preserved over the substring used for the EJMR username. Or, more formally, we
could not reject the hypothesis that it is uniform using a chi-squared
statistic. Therefore, given a post with a username, we expect that about
$2^{32}/2^{16}=2^{16}= 65,536$ IP addresses are possible origins of the post.
Furthermore, these are uniformly distributed over the range of IPv4 addresses.

To see why this is helpful, let us return to our example in
Section~\ref{sec:relationship} in which the IP address 131.111.5.175 posting on
the topic ``Dream job: IMF economist'' with the topic ID 227259 was assigned the
username c2b1. There are exactly 65,028 IPv4 addresses that, when prepended with
topic ID 227259, create a hexadecimal SHA-1 hash with c2b1 at positions 10-13.
Due to the uniformity property of the SHA-1 hash, these matching IP addresses
are uniformly distributed over the entire IPv4 address range from 0.0.0.0 to
255.255.255.255.

Now imagine the same IP address 131.111.5.175 posts on the topic ``Are US-based
journals biased towards US data?'' with the topic ID 227279. In this case, it
would be assigned the username 91c2. There are exactly 65,635 IP addresses that
would also receive the username 91c2. But there is \emph{only one} IP address,
the true IP address 131.111.5.175, which appears in the two sets. As this
example illustrates, ``true positive'' IP addresses---those from which posts on
EJMR were actually made---\emph{stick out} because these IP addresses explain
more observed usernames on the site than \emph{false positive}, or ``noise,'' IP
addresses. 

To determine the ``true'' IP address for an EJMR post, we need to know the
roughly 65k \textit{candidate IP addresses}. But determining those is not
trivial because the hash is one-way: one cannot determine the input to a SHA-1
hash given the output, much less a fraction of that output. The only feasible
way to determine which IP addresses map to a topic-username pair is to compute
the username for \emph{each} of the $2^{32}$ IPv4 addresses. This is
conceptually easy, but slow in practice. Because there are
\niceTotalTopics~topics on EJMR and $2^{32}$ IPv4 addresses, we need to perform
about $2.98 \times 10^{15}$ (2.98 quadrillion) SHA-1 hashing operations and
check which of the computed hashes corresponds to topic-username combinations
observed on EJMR\@. This computation requires only a handful of lines in
high-level languages like Python. However, our initial tests suggested that such
an effort would take over 60 years on a single core of a typical modern CPU\@.
Fortunately, this variety of computation is made easier by graphical processing
units (GPUs) which are essentially massively parallel computers but which
require specialized programming frameworks such as
Nvidia's CUDA~\citep{nickolls2008scalable}. 

After obtaining every topic-username combination observed on EJMR (of which
there were \niceUniqueTopicIDUsernamePairs~in our data set) and developing
software that runs on Nvidia GPUs, we determined the candidate IP addresses
for each topic-username pair.  The heart of the
software is based on an open-source implementation of the SHA-1 algorithm
designed for Nvidia devices from the Mochimo Cryptocurrency project
\citep{mochimo}. Some aspects of our task allowed for optimizations that
substantially sped up this task. First, because the topic ID is \emph{prepended}
to the IP address before hashing, the SHA-1 algorithm could be ``primed'' once
for each topic and fed to each compute core of the GPU\@. Second, because the
IPv4 addresses are only 32-bit integers, they could be enumerated on the GPU
device rather than passed in as strings, thereby limiting GPU-CPU data transfer,
which can otherwise be a bottleneck. Third, we required only one pass over the
$2^{32}$ IPv4 addresses for each topic. Roughly speaking, our algorithm passes
the GPU a ``primed'' SHA-1 hash and a list of the observed usernames for a
particular topic. Each core of the GPU considers a single IP address and checks
if that IP address would produce a username that is observed. If so, the
username-IP pair is appended to a list of results that is ultimately passed back
to the CPU for output. This process repeats until all $2^{32}$ IPv4 addresses
are checked for a single topic. 

The IP enumeration task is ``embarrassingly parallel'' because the topics can be
enumerated independently. This task took approximately 240 hours (ten days) of
total computing time on Nvidia A100 devices, each of which has 6,912 cores
\citep{choquette2021nvidia} operating in parallel. We used multiple A100s so the
actual time was significantly lower.\footnote{The device we used---A100 GPUs
with 40g of memory---retail for roughly \$8,000 at this time. These devices can
also be rented hourly. For example, the P4d instances on Amazon's EC2 contain
eight A100 devices. Our study could be reproduced for roughly \$1,000 on an AWS
P4d instance at the hourly on-demand price although, as we describe later, we
repeated this hash inversion in triplicate for the purpose of our statistical
analysis. Furthermore, because our method used only approximately 0.5Gb of
memory per GPU device, it can almost certainly be reproduced with older, less
expensive CUDA devices. The intermediate output of the hash inversion required
roughly 3Tb of storage space. That said, the statistical analysis of the output
of this process also required computers with at least 100Gb of RAM.} The source
code for this enumeration task is available at
\url{https://github.com/to-be-determined} in the ``cuda-sha'' directory.
\footnote{This code will be made available when our manuscript is published.}

\subsection{Applying a Statistical Test of Noise to Candidate IP Addresses}
\label{sec:identification}

The preceding work leaves us with a set of about 65k candidate IPs for each post
(topic-username pair) that we observe on EJMR. If the posts we observed were all
in a single topic, we could make no further progress. We would have no basis to
believe that any single IP address in the candidate set is more likely than the
others to be the ``true'' address for a post. However, when we observe posts
across many topics, ``true'' IP addresses begin to stand out, as we previously
described with the hypothetical posts from 131.111.5.175. They stand out because
``noise'' IP addresses have a particular, rather simple, statistical
distribution that we describe here.

To understand the distribution of noise IPs, imagine a world in which we are
incorrect about the scheme by which EJMR assigns usernames to posts. Say, for
example, that EJMR is not using the SHA-1 hash. In this world, our candidate IP
addresses for a topic-username pair would still be a set of IP addresses
uniformly drawn from the possible 2$^{32}$ IPs, but there would be no meaningful
information in this set: it is random. Now imagine that we examine more posts
(topic-username pairs) in other topics. As we accumulate more posts, we can
count the number of times an IP address appears across all candidate IP sets.
This count will be our null distribution and, as we will see, the distribution
of ``true'' IPs will be substantially different from the null.

To begin, some formal notation is helpful.
Let $a$ denote one of the $2^{32}$ possible IPv4 addressess; $u$ denote one of
the $16^4$ possible EJMR usernames; $t$ denote one of the
$T=$~\niceTotalTopics~topics on EJMR; $k_{t}$ denote the number of usernames
observed in topic $t$;\footnote{Repeated posts with the same username in a
given topic are redundant and provide no additional information helpful to
pinpointing the source IP. We therefore ignore duplicate topic-username pairs.} and,
$A_{(t,u)}$ denote the set of candidate IP addresses associated with a $(t,u)$
pair.\footnote{Because the uniformity property holds in expectation, the exact
number of matching IP addresses for any single topic-username observation can
vary. The number of IP addresses matching for any of the
\niceUniqueTopicIDUsernamePairs~topic-username combinations observed on EJMR
varies between \niceMinIPsPerTopicUsername~and \niceMaxIPsPerTopicUsername~with
a mean of \niceMeanIPsPerTopicUsername, which is $2^{16}+1$.} 

Finally, let $n_{a}$ be the number of times we observe the IP address $a$ in the
set of candidate IP addresses $A_{(t,u)}$ for all topic-username combinations.
The distribution of $n_{a}$ will form the basis for our test.

When there is a single topic ($T=1$) with one username, the distribution of
$n_{a}$ is straightforward: every IP either maps to the observed username or
not. That is, $n_{a}$ is either one or zero, and is zero with probability
\begin{align*}
     Pr(n_{a} = 0) &={2^{16} - 1 \choose 1 }\big/ {2^{16}  \choose 1 } = \frac{65535}{65536}
\end{align*}
and hence $ Pr(n_{a} = 1) = 1/2^{16}$.

Now consider a case with a single topic and \textit{two} observed usernames.
Recall that each IP address has only a single username for a given topic and
therefore $n_{a}$ is still either one or zero. But in this case, it is zero with
probability
\begin{align*}
    Pr(n_{a} = 0) &= {2^{16} - 1 \choose 2 }\big/ {2^{16}  \choose 2 }
\end{align*}

This is a hypergeometric distribution with two draws where there are $2^{16} -
1$ ``red balls'' in one ``urn'', and $1$ ``blue ball'' in the other urn. For a
given $t$ and $a$ there is a single $u$---that is the blue ball. The probability
of observing zero counts for that IP address is the likelihood of never
selecting the blue ball after two draws. Thus, in the case of $k_{t}$ usernames in a
single topic $t$, this is
\begin{align*}
    Pr(n_{a} = i) &= (1-q(k_{t}))^{1-i}q(k_{t})^{i}\\
    q(k_{t}) &=  1- {2^{16} - 1 \choose k_{t} }\big/ {2^{16}  \choose k_{t} }.
\end{align*}

Now consider $T = 2$ topics. $n_{a}$ can now be zero, one, or two. For the
moment, assume that each topic has the same number of usernames. That is, $k_{t} =
k$ for each topic. Each topic is an independent observation for that IP address
under the null distribution. Hence, $n_{a}$ is distributed binomial with
probability $q(k)$
\begin{equation*}
    Pr(n_{a} = i) = {T \choose i } (1-q_{k})^{T-i}q_{k}^{i}.
\end{equation*}

Of course, topics generally do not have the same number of usernames.  Instead $k$
varies between topics because some topics are popular---they have more
posts---and some are less so. Thus, rather than being binomial, $n_a$ is
distributed \textit{Poisson-binomial}, which is a mixture of binomials with
different event probabilities. Hence, for $T$ topics and $k_{t}$ observed usernames per topic, the probability distribution for $n_{a}$ under the null is
\begin{align}\label{eq:null-hypothesis}
     Pr(n_{a} = i) &= \sum_{A \in F_{i}} \prod_{t \in A} q(k_{t}) \prod_{t' \in A^{c}} (1-q(k_{t'})),
\end{align}
where $F_{i}$ is the set of all subsets of $i$ integers that can be selected
from the topic indices $\{1, \ldots, T\}$. Hence, when $T=2$, $F_{1}$ is $\{1\}$
or $\{2\}$, and the probability corresponds to the statement that either we
observe the IP in topic 1 (which has probability $q(k_{1})$) and not in topic 2
(with probability $1-q(k_{2})$), \emph{or} vice versa. 

We now have a statistical test that allows us to say if an observed $n_a$ is
likely under the null, given a collection of topics $T$ and the posts (usernames) therein. If $n_a$ is large, the probability of observing $n_a$
under the null is small. We denote this probability (or p-value) as $p_a$.
If $p_a$ is less than some threshold $p^{*}$, we will say an IP address is
``active.'' Since we are interested in avoiding false positives (e.g.,
we want to control the overall size of our family of statistical tests across IPs \citep{hochberg1987multiple,benjamini1995controlling}), we want
to choose a $p^{*}$ that is sufficiently conservative. We describe our method
for doing so in the following section.

\subsection{Assignment of Posts to IP Addresses}

This test gives us the ability to say which IP addresses are active on EJMR for a collection of topics and posts, or equivalently, in
some window of time. However, it does not allow us to assign a \emph{particular}
post to a \emph{particular} IP address.
To do that, one could use a generative Bayesian model describing humans, some of whom are
EJMR contributors, probabilistically acquiring and releasing IP addresses,
viewing EJMR topics, and selecting in which topics to post according to their
individual preferences. Such a model is likely under-specified and, for the
moment, beyond our abilities. In its stead, we present a model that is
\emph{practical} in the sense of being both intuitive and tractable.

The intuition for our practical model is as follows. Consider a post on EJMR\@.
Like all other posts, this post is characterized by some $(t, u)$ pair for which 
we have a candidate set of IPs $A_{(t,u)}$.
These IP addresses can ``explain''
the post's username. But imagine that one of these IP addresses \emph{also}
explains 20 other posts made around the same time. What is the likely origin of
the post? It is, we contend, this \emph{highly explanatory} IP address. This
will be an ``active'' IP address whose observed $n_{a}$ is unlikely under the
null distribution described in the previous section.

This is effectively an optimization
problem. We adopt a simple rule by which we choose to assign posts to the
smallest set of IP addresses that can explain them in a given window of time.
However, we do not assign \emph{all} posts to IP addresses. We use the
statistical model in \Cref{sec:identification} to limit our candidate IP
addresses to those which are above some threshold of apparent activity that is
improbable by chance: They have an improbable $n_a$ under the null.
The structure of our data allows us to determine this
threshold in a manner that minimizes incorrect attributions. In other words,
rather than saying ``this post came from this IP with probability X'' we say
``this post came from this IP if you use this sparsity rule'' and we describe
some of the error properties of that rule.

Our assignment procedure is structured as follows. First, we order all posts by
time and bin them by GMT date. We consider all the posts on a single day. From
this day we extend a window of time three days into the past and three into the
future, thereby collecting a week's worth of posts. For each of the posts in
this week, we gather the candidate IPv4 addresses which we obtain from the IP
enumeration procedure described before. For each of the roughly 4.3 billion IPv4
addresses, we count $n_a$ in that window of time.\footnote{Recall that given the
nature of the username allocation algorithm, an IP address posting multiple
times in the same topic is assigned the same username.} We then compute $p_a$
for each IP address. Recall that these counts follow a Poisson-binomial
distribution under the null.

For each post on the target day, we assign the post to the IP address with the
lowest p-value, but \emph{only} if that p-value is below some threshold $p^{*}$,
which is determined in a manner we describe shortly. Having attempted to assign
all the posts on the target day, we move to the next day and repeat the
procedure.

To recover IP addresses that do not post as frequently during our relatively
short 7-day window but still post regularly over longer time periods, our
procedure considers two additional time windows of 31 and 91 days. 

How should we choose $p^{*}$? In other words, how can we know when an IP
address is overrepresented in a particular window of time and that its
explanatory power does not arise by chance in the noise component of equation
(\ref{eq:null-hypothesis})? Fortunately, we have a very accurate way to empirically
generate
the noise distribution by ``shifting'' the characters from which the EJMR
username is drawn in the SHA-1 hash. Instead of using positions 10-13 of the
hash, we use positions 11-14 and ask the following question. What if the IP
addresses that could have generated the username for this post were not those
with a SHA-1 hash containing the username at position 10, but rather at position
11? Both sets of IPs are roughly 65k in size. The position-10 set is guaranteed
to contain the true IP address and all the other IP addresses contained in it
are ``noise.'' In contrast, the position-11 set contains \emph{only} noise. 

We use this insight to select $p^{*}$. We repeat the entire IP enumeration as
if the EJMR username were drawn from positions 11-14 and then repeat the post-IP
assignment procedure described above. Then, we calculate the p-value thresholds
$p^{*}$ such that we would obtain \emph{zero} assignments of posts to an IP
address for position-11. That is, using the (incorrect) position-11 hashing set
and these thresholds, \emph{none} of the roughly 7 million posts observed on
EJMR would be assigned an IP address. The p-value thresholds we found are
$\pSevenDays$~for the 7-day window, $\pThirtyOneDays$~for the 31-day window, and
$\pNinetyOneDays$~for the 91-day window. We then use these same $p^{*}$
thresholds for the correct hash positions (position-9 before July 13, 2013 and
position-10 thereafter until May 17, 2023). 

The p-value thresholds above are quite small. This is because our method of
determining $p^{*}$ naturally adjusts for multiple hypothesis testing. For each
window (7, 31, and 91 days), we are conducting approximately 7 million $\times$
65k $\approx$ half a trillion hypothesis tests. Thus, a p-value threshold with a
low \emph{overall} error rate will be quite small (e.g.,
the overall size of our family of statistical tests across IPs \citep{hochberg1987multiple,benjamini1995controlling} is tightly controlled to avoid false positives). The number of IP addresses
that never posted to EJMR but that we mistakenly assign to a post is, in
expectation, less than one and likely close to zero. However, there are
potentially other varieties of error, which we discuss later in this section.

\begin{figure}[!ht]
    \centering
    \includegraphics[width=0.7\linewidth]{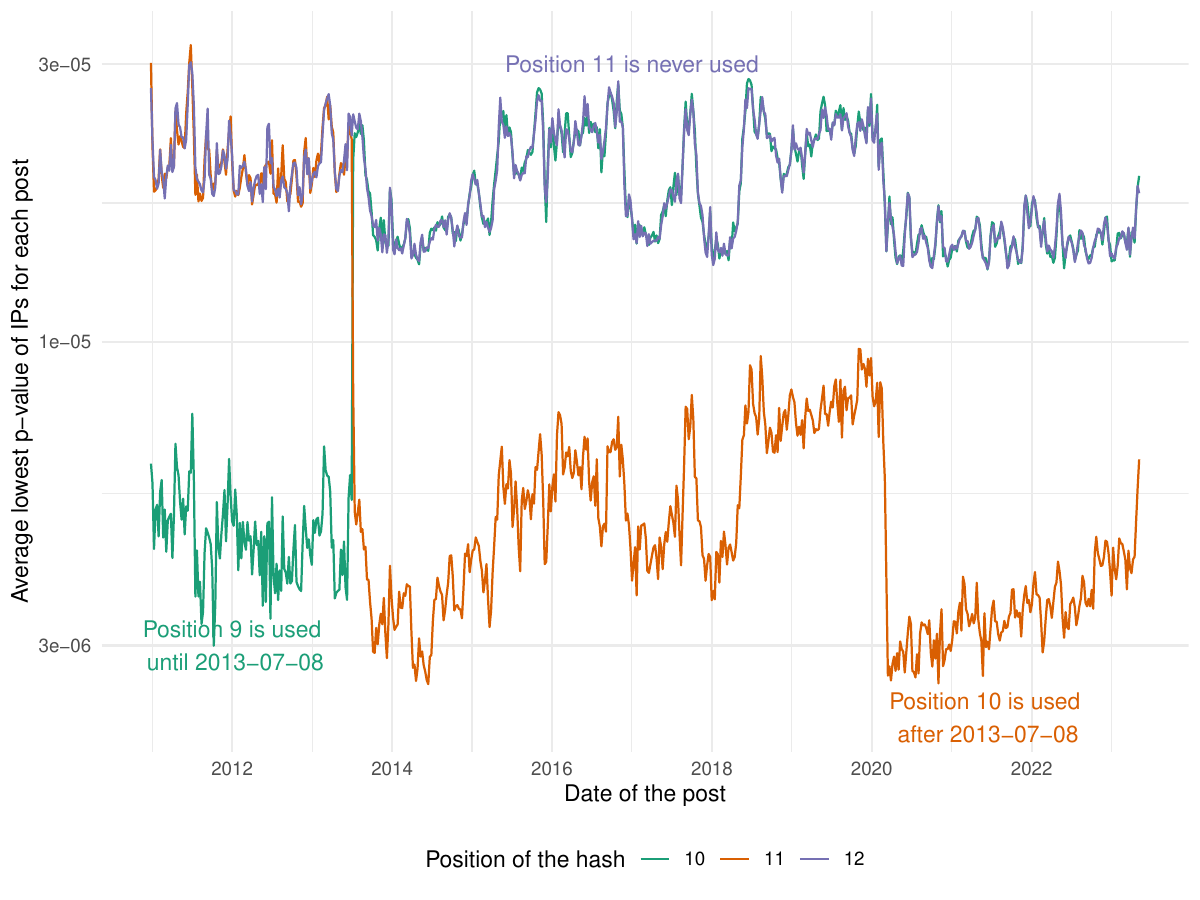}
    \caption{Average minimum p-value of posts in a given week for different hash positions over time. 
    We employ the following two-step procedure. First, we find the IP address that has the lowest p-value for a given post and refer to its p-value as the minimum p-value of a post. Second, we calculate the mean of the minimum p-value of a post across all posts in a given week. The graph clearly shows what hash position was used for the EJMR username at each date. The orange line showing the position-10 p-values is toward the bottom of the graph because EJMR usernames started at position-10 for most of the website's history. On July 8, 2013, the site administrator likely made a database error that ``shifted'' the usernames one position left. For this reason, the position-9 p-values are lower before this date. When a position is not ``in use,'' its p-values closely track the position-11 ``noise'' distribution. Note that none of the posts to which we actually assign IP addresses would be visible on this graph as our $p^{*}$ thresholds are on the order of $10^{-11}$. These low p-values are in effect what is pulling down the green and orange lines which are weekly averages.}
    \label{fig:assignment_graph}
\end{figure}

We completed the assignment procedure using three different starting positions of the hash (9, 10, and 11). \Cref{fig:assignment_graph} shows the average minimum p-value of IP addresses for all posts in a given week for each of these hash positions over time. The value at each point in this graph is most clearly described by the two-step procedure we use to calculate it. First, we find the IP address that has the lowest p-value for a given post and henceforth refer to this p-value as the minimum p-value of a post. Second, we calculate the mean of the minimum p-value of a post across all posts in a given week. The graph clearly shows what hash position was used for the EJMR username at each date. The orange line showing the position-10 p-values is toward the bottom of the graph because EJMR usernames started at position-10 for most of the website's history. On July 8, 2013, the site administrator likely made a database error that ``shifted'' the usernames one position left. For this reason, the position-9 p-values are lower before this date. When a position is not ``in use,'' its p-values closely track the position-11 ``noise'' distribution. Importantly, none of the posts to which we actually assign IP addresses would be visible on this graph as our $p^{*}$ thresholds are much smaller and on the order of $10^{-11}$. These low p-values of posts to which we assign IP addresses, are in effect what is lowering the green and orange lines which are averages across all posts in a week. 

Having confirmed the cut-off date between position-9 and position-10, we elected to use only the position-9 assignments prior to July 8, 2013 and only the position-10 assignments afterward.\footnote{On the cut-off date we allow either hash position.} In total, we assigned IP addresses to \niceNumEJMRPostsAssigned~of the \niceTotalPostsWithTopicIDAndUsername~EJMR posts for which we have both a topic ID and username, or \percentageOfEJMRPostsAssigned~of posts over the period spanning \niceFirstPostDateHex~to \niceLastPostDate. These posts originate from just \niceNumDistinctIPsAssigned~distinct IP addresses.\footnote{We also have \niceTotalUnassignablePosts~posts that have either no topic ID or no username. These cannot be assigned to IP addresses.} Most of our assignments come from the 7-day window procedure. Roughly speaking, if an IP address was the source of posts on more than about a dozen topics in a week, that IP address is identified by our method.

\begin{figure}[!ht]
    \centering
    \includegraphics[width=0.8\linewidth]{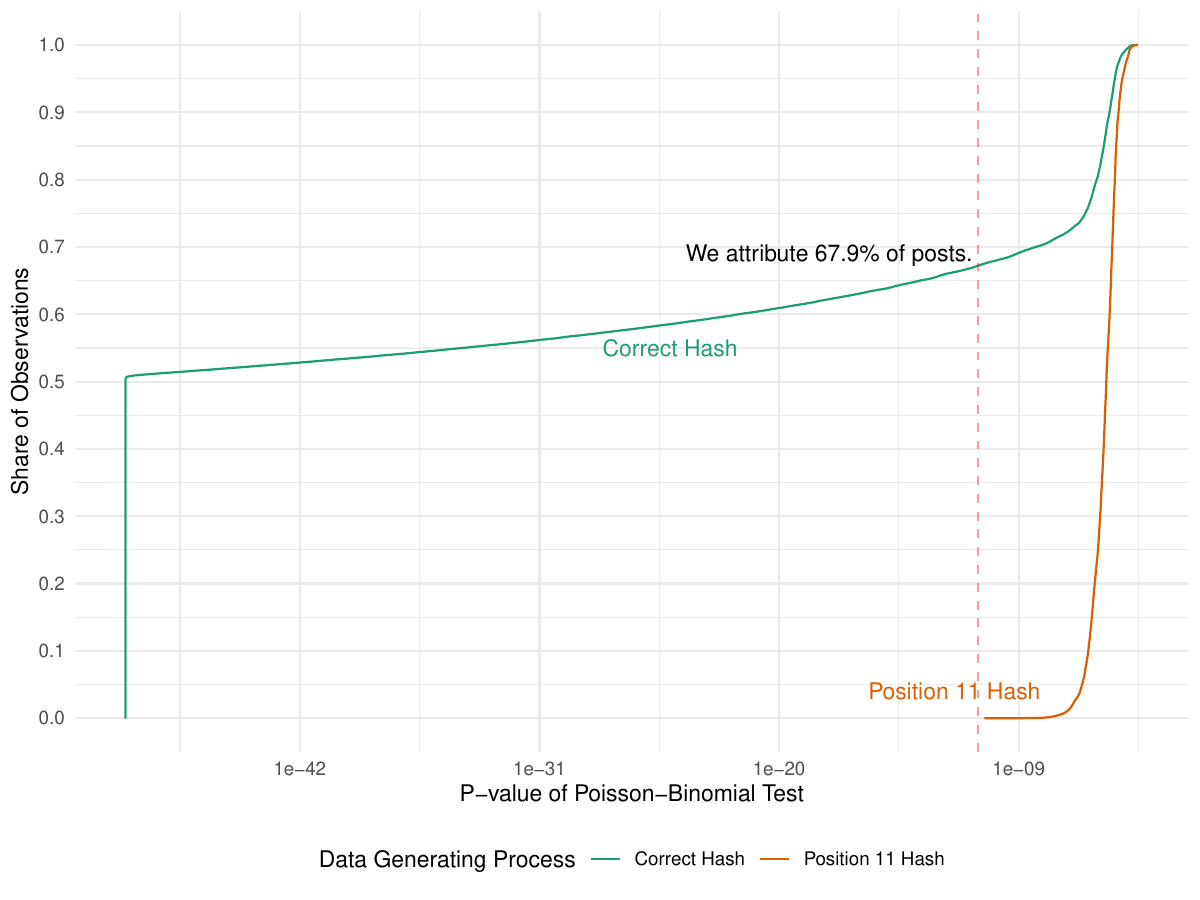}
    \caption{Cumulative distribution functions of the minimum p-value of posts for different hash positions. The orange line plots the CDF of the minimum p-values for posts calculated based on the incorrect position-11 hash. The green line shows the CDF under the correct hash (position-9 before July 8, 2013 and position-10 afterward). The approximate p-value threshold $p^{*} \approx 10^{-11}$ is represented by the dashed vertical line. For position-11, none of the \niceTotalPostsWithTopicIDAndUsername~EJMR posts for which we have both a topic ID and username would be assigned to an IP address because no IP has a sufficiently low minimum p-value. In contrast, \percentageOfEJMRPostsAssignedOfAssignable~of these posts have minimum p-values below $p^{*}$ and for more than 50\% of posts the minimum p-values are smaller than $10^{-50}$.} 
    \label{fig:assignment_graph_cdf}
\end{figure}

\Cref{fig:assignment_graph_cdf} plots the cumulative distribution functions of the minimum p-values of the posts for different hash positions. The orange line plots the CDF of the minimum p-values for posts with the incorrect position-11 hash which only contains noise. Comparing this line with the approximate p-value $p^{*} \approx 10^{-11}$ which is represented by the dashed vertical line, it is evident that not a single post of the \niceTotalPostsWithTopicIDAndUsername~EJMR posts for which we have both a topic ID and username would be assigned to an IP address because no post would have a sufficiently low minimum p-value. In contrast, the green line shows the CDF under the correct hash (position-9 before July 8, 2013 and position-10 afterward). \percentageOfEJMRPostsAssignedOfAssignable~of posts have minimum p-values below $p^{*}$ and for almost 30\% of posts the minimum p-values are smaller than $10^{-50}$.

Earlier, we claimed our method was unlikely to assign posts to IP addresses that are truly inactive on EJMR\@. We identified a convenient way of testing that claim because the IPv4 address space contains certain ``reserved use'' IP addresses from which no traffic should ``legitimately appear on the public internet'' \citep{RFC5735}. These are the so-called \emph{bogon} addresses, which are nearly 600m in number and thus occupy 13.8\% of the entire IPv4 address space. We know that any assignments we made to bogon IP addresses were surely in error. However, out of the \niceNumEJMRPostsAssigned~posts to which we assigned IP addresses, the number of posts our procedure assigned to bogons is \emph{zero}.

There is, however one type of notable---but estimable---error in our assignments
that is possible. This error arises from IP addresses that post with high
frequency ``stealing'' posts from the \emph{true} posting IP address. To gain an
intuition for this type of error, imagine the following situation. A one-time
EJMR user posts to some topic and receives username ab34. However, it just so
happens that a highly active IP address would also receive that same username if
it posted in that topic. That is, an assignment on this highly active IP address
occurs by chance in what we have been calling the ``noise'' component of the
SHA-1-based username. In our assignment scheme, we would mistakenly assign the
post to the highly active IP\@.  We do not have a
precise estimate for how often such a misassignment occurs.
However, we have a rough estimate. First, note that this situation is fairly
rare. Our event windows are small and the number of highly active IPs at any
given time is miniscule relative to the total number of IPv4 address. Second, 
this kind of error is maximized when the window size used for
assignment is large and when the IP in question is most highly active. That is,
highly active IP addresses can ``steal'' the most posts and the opportunity to
``steal'' is largest when the window size is largest. Consider the maximal
window size, a window spanning all 13 years of EJMR activity for which we have
data. The maximally active IP address has about 47k posts on EJMR\@. That IP
address would be ``explanatory'' by chance for about one in every 65k
topic-username pairs, or about 80 pairs over the
\niceTotalTopicUsernamePairs~observed in our data set. That would mean about 106
posts (based on the ratio of posts to unique topic/username pairs) out of the
47k should be expected to have been assigned in error or about 0.2\%. Of course,
most of our assignments happen for the 7-day window size. For less active IPs,
there is substantially less opportunity to ``steal'' assignments. As a result,
we expect this error rate to be low.

\subsection{Linguistic Analysis}
\label{sec:linguistic}

\textit{This section contains offensive speech that some readers may wish to avoid. Those readers should skip to Section~\ref{sec:reddit}.} We analyzed the linguistic character of EJMR posts and topics using a variety of machine learning techniques. We first removed any quoted blocks of text belonging to other posts to avoid misattribution of the content of posts. Due to the extensive use of obfuscation on the site, many posts required additional pre-processing. For example, consider the following posts:
\begin{itemize}
    \item ``Given women get free spots, blks and latins get free spots, it basically means you need to be far far right tail if u are a yt or azn homegrown American.'' (2022-12-27)
    \item ``Mold-fa//g//g//ot, I will split your a//s/s in two with my HUMONGOUS super HARD shalong. You will be squealing like the little beia/tch that you are.'' (2020-01-28)
    \item ``those d4mn j3ws had no morals either.'' (2022-08-13)
    \item ``Hey a\$\$h01e, I left you a message earlier too. I will be there in Boston to FIEK and RAEP you, so cover your \$hitty a\$\$ and your mouth now.'' (2014-12-26)
\end{itemize}
These posts are obfuscated to such an extent that we found most machine learning models failed to accurately classify them as toxic. To address this shortcoming, we developed software to deobfuscate such speech. First, we classified posts into commonly occurring natural languages on EJMR~\citep{stahl2023lingua}: English, German, Chinese, Korean, and a few others. Then we collected high-frequency non-English words in the English posts which we used to develop a dictionary mapping text like ``f**k,''  ``secks,'' and ``GTFO'' to  canonical forms. We used this dictionary to deobfuscate some of the most commonly obfuscated terms.

Then, we checked each word in each post for common symbol-based obfuscations like ``fa//g//g//ot,'' removing symbols where doing so resulted in an English word or well-known profanity. Finally, we transformed so-called leetspeak---such as ``d4mn j3ws''---to its canonical form. We did this by attempting common leetspeak substitutions and checking if those substitutions resulted in an English word or a well-known profanity. Our goal in this effort was not perfection, but rather \emph{some} improvement in the performance of machine learning models for this content. Numerous obfuscations remained after our triage. For example, the use of ``yt'' is context dependent: sometimes it means ``white'' and other times it means ``YouTube.'' So we chose not to transform certain out-of-corpus words like ``yt'' when we found them.  

Having deobfuscated all posts, we ran each post through a number of
transformer-based machine learning models. We selected models that are
state-of-the art and that, based on our informal inspection, also appeared to
perform well on EJMR content. For sentiment detection, we selected the default
Huggingface sentiment model. This is a checkpoint of
DistilBERT~\citep{Sanh2019DistilBERTAD} fine-tuned for sentiment
detection~\citep{huggingface2023sentiment} on the Stanford Sentiment
Treebank~\citep{socher-etal-2013-recursive}. We selected a similar fine-tuned
BERT model for detecting misogyny~\citep{attanasio-etal-2022-entropy}. For
toxicity we selected ToxiGen Roberta~\citep{hartvigsen2022toxigen}, a
checkpoint of the Roberta model~\citep{lui2019roberta} fined-tuned for toxicity
detection. A nearly identical ToxiGen-based model was used to measure
toxicity in Meta's recent Llama-2 large language model
\citep{touvron2023llama}. We corroborate the ToxiGen Roberta results using
Google's Perspective\citep{lees2022perspectiveapi}, a similar model used to identify toxic content on the New
York Times, Wall Street Journal, and Financial Times websites.\footnote{For a discussion of these results see 
Appendix~\ref{sec:perspective_toxicity} where we show that our conclusions regarding the toxicity of the content are not sensitive to our choices of models.} Finally, we used a Roberta-based
model for hate speech~\citep{barbieri2020tweet} classification.

\subsection{Comparison with Reddit}
\label{sec:reddit}

Reddit is a social media platform where users can post content, comment, and vote on posts. Much of that is done pseudonymously. As of 2023, it is the 10th most-visited website in the world. It is organized into ``subreddits,'' which are topic-specific communities. Subreddit moderation on Reddit is carried out by administrators who are either official Reddit employees or individuals selected by specific community members. Reddit bestows a degree of autonomy upon these subreddit moderators, enabling them to determine the permissible and unacceptable content within their respective subreddits, as long as they remain within the bounds of site-wide rules. This notable flexibility has paved the way for the emergence of a wide array of subreddits, some of which have stirred up controversy. The decentralized nature of Reddit's moderation, coupled with user anonymity and the absence of robust fact-checking mechanisms, has rendered the platform susceptible to the dissemination of misinformation and the reinforcement of echo chambers, ultimately fostering distorted worldviews among its user base \citep{cinelli2021echo}.

We retrieved a list of the 1,500 subreddits with the most subscribers from \url{https://www.reddit.com/best/communities}. From those we selected the top 1,000 subreddits that had at least 10,000 posts in the Pushshift Reddit Dataset~\cite{baumgartner2020pushshift}, which contains roughly all Reddit content from the site's inception through 2018. Clearly, some popular subreddits in 2023 did not exist in 2018 and some subreddits that were popular in 2018 were likely not popular in 2023. From each of these 1,000 most popular subreddits, we downloaded all ``utterances''---these are analogous to ``posts'' on EJMR\@. For each subreddit, we randomly, uniformly sampled 10,000 posts. We subjected these posts to the same linguistic analyses that we used for EJMR posts. They were deobfuscated and classified using the same models as EJMR posts.

\subsection{Geolocation of IP Addresses}

We use the commercial IP2Location database to obtain country, city, latitude-longitude, zip code, ISP, and domain information for all the IP addresses we identify.

Internet users have multiple IP addresses across the devices and the networks they use. Recent research suggests that the average consumer IP address in the United States is held for about 19 days and about 87\% of internet users will have, at any time, at least one IP address that is used for more than a month~\citep{mishra2020ip}. Because universities tend to have generous IP blocks---particularly elite universities---it seems likely that IP retention in these institutions will be longer. In addition, geolocation for university IPs is particularly accurate~\citep{saxon2022gps}.

\subsection{Time Stamps}
\label{sec:timestamps}

Unfortunately, EJMR does not display exact post times. Depending on the age of the post, it only displays whether it occurred ``m/h/d/m/y minutes/hours/days/months/years ago.'' This makes it, a priori, difficult to assign exact time stamps to posts, especially to older posts. However, EJMR provides two additional pieces of information. First, it has an RSS feed which displays the most recent 10 posts along with the exact time stamps (year, month, day, hour, minute, second) in every topic. Second, every post on EJMR has a unique, auto-incrementing integer post ID starting at 1 on \niceFirstPostDate. This post ID increases one-by-one for all the posts across all the topics on the site. 

We downloaded the RSS feed and the Wayback Machine for all the 642,247 topics. This gave us exact time stamps for a total of \niceNumEJMRPostsHavePubDate~posts. For the remaining \niceNumEJMRPostsLackPubDate~posts, we assigned the time stamp based on the auto-incrementing post ID by linearly interpolating between any two posts with known exact time stamps. Because the posts with exact time stamps are very evenly distributed we were able to accurately assign time stamps for all posts without exact time spots. The average time difference between posts with exact time stamps that have some posts without exact time stamps in between them is only about 3 minutes and even at the 95th and 99th percentile this difference is smaller than 10 minutes and 23 minutes, respectively.

\section{Results}
\label{sec:results}

\subsection{Descriptive Statistics}
\label{sec:descriptive}

We obtained EJMR content from both \url{http://www.econjobrumors.com} and \url{http://archive.org}. In total, our data included \niceTotalPosts~posts from \niceTotalTopics~topics on EJMR between \niceFirstPostDate~and \niceLastPostDate. There are \niceTotalUnassignablePosts~posts for which we do not have topic identifiers or usernames mostly because they originated from registered users for which the site does not display hexadecimal usernames. With our methods these posts are, by construction, unassignable to IP addresses. The remaining \niceTotalPostsWithTopicIDAndUsername~posts for which we have topic identifiers and usernames, are assignable posts which, in principle, can be assigned to IP addresses. From these assignable posts we recover \niceNumDistinctIPsAssigned~distinct IP addresses.

\subsection{Time Patterns}
\label{sec:time_patterns}

\begin{figure}[t!]
    \centering
    \begin{subfigure}[t]{0.49\textwidth}
        \centering
            \includegraphics[width=\linewidth]{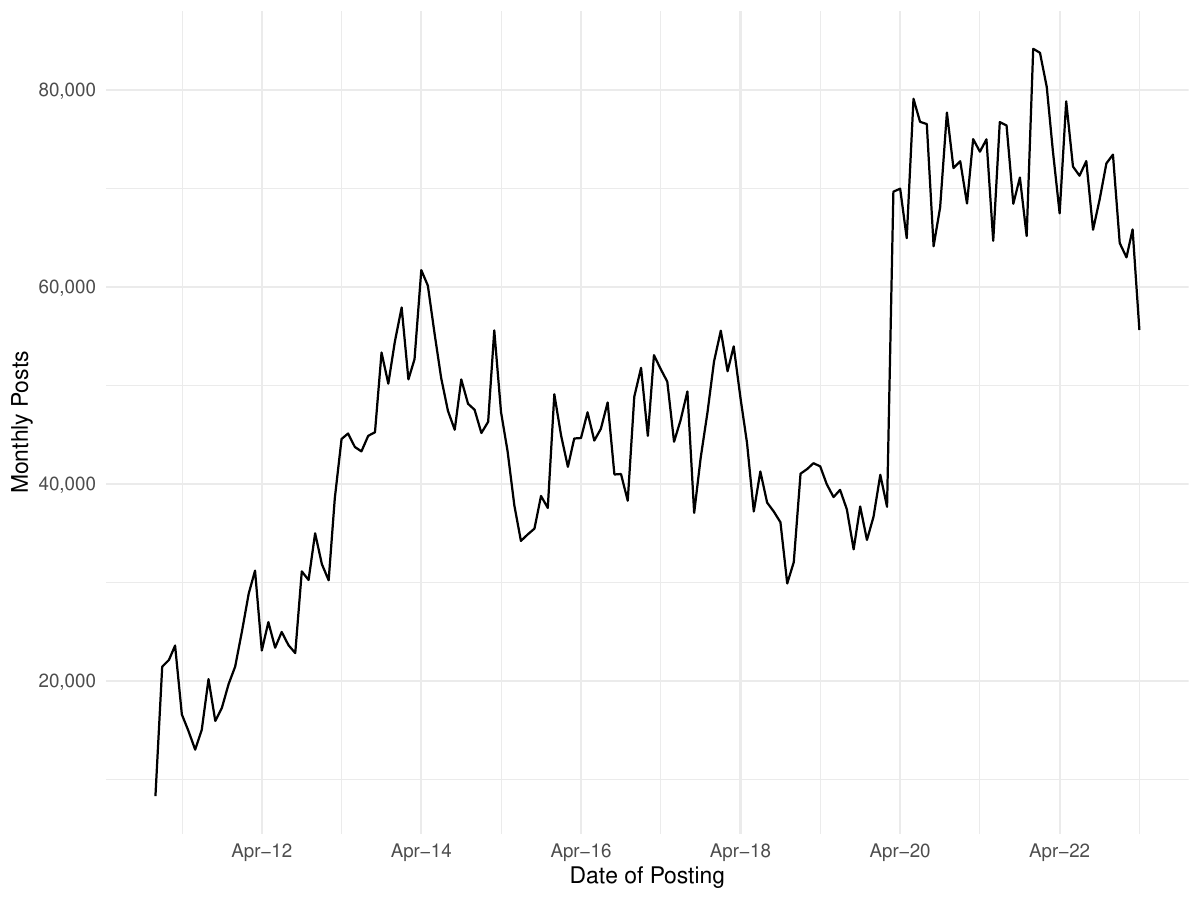}
        \caption{Monthly posts (all countries)}
    \end{subfigure}%
    \begin{subfigure}[t]{0.49\textwidth}
        \centering
            \includegraphics[width=\linewidth]{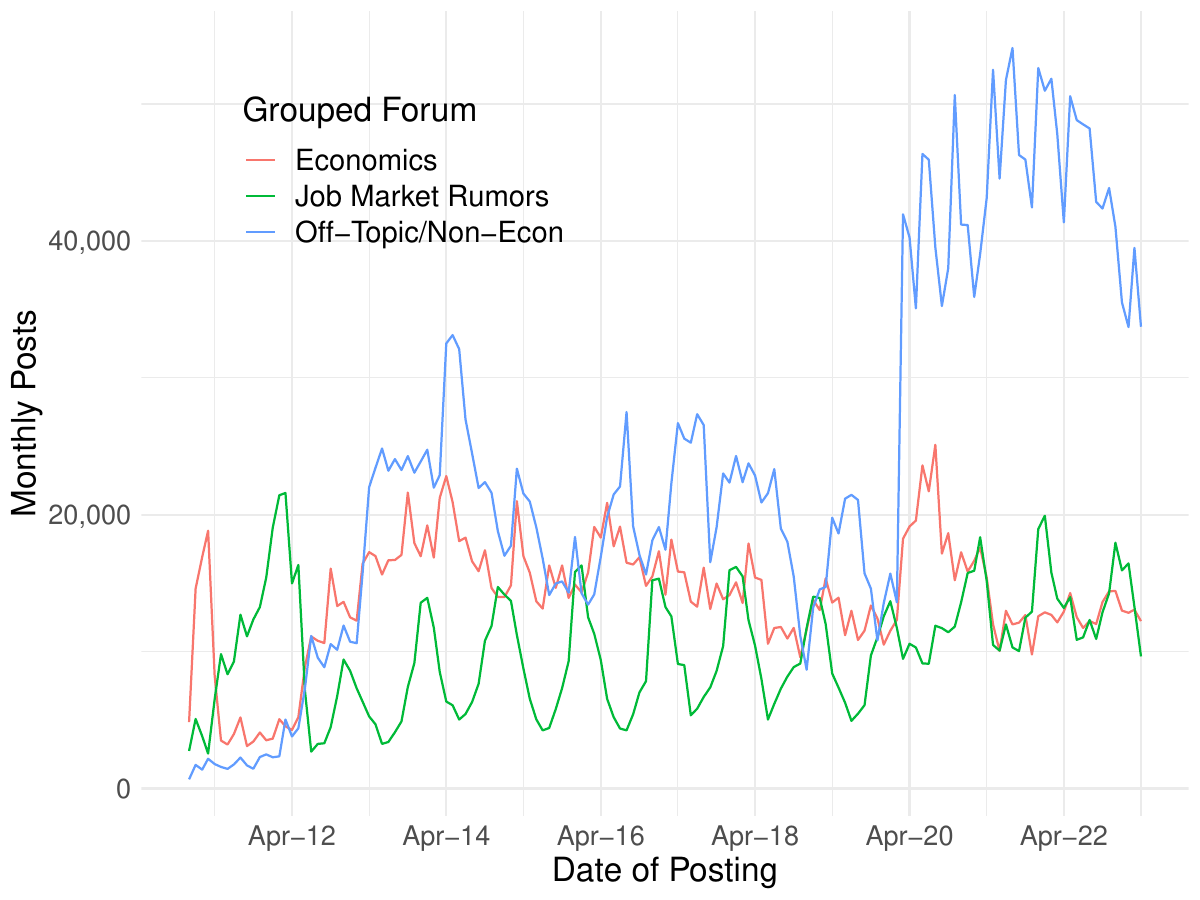}
        \caption{Monthly posts by forum (all countries)}
    \end{subfigure}
        \begin{subfigure}[t]{0.49\textwidth}
        \centering
                \includegraphics[width=\linewidth]{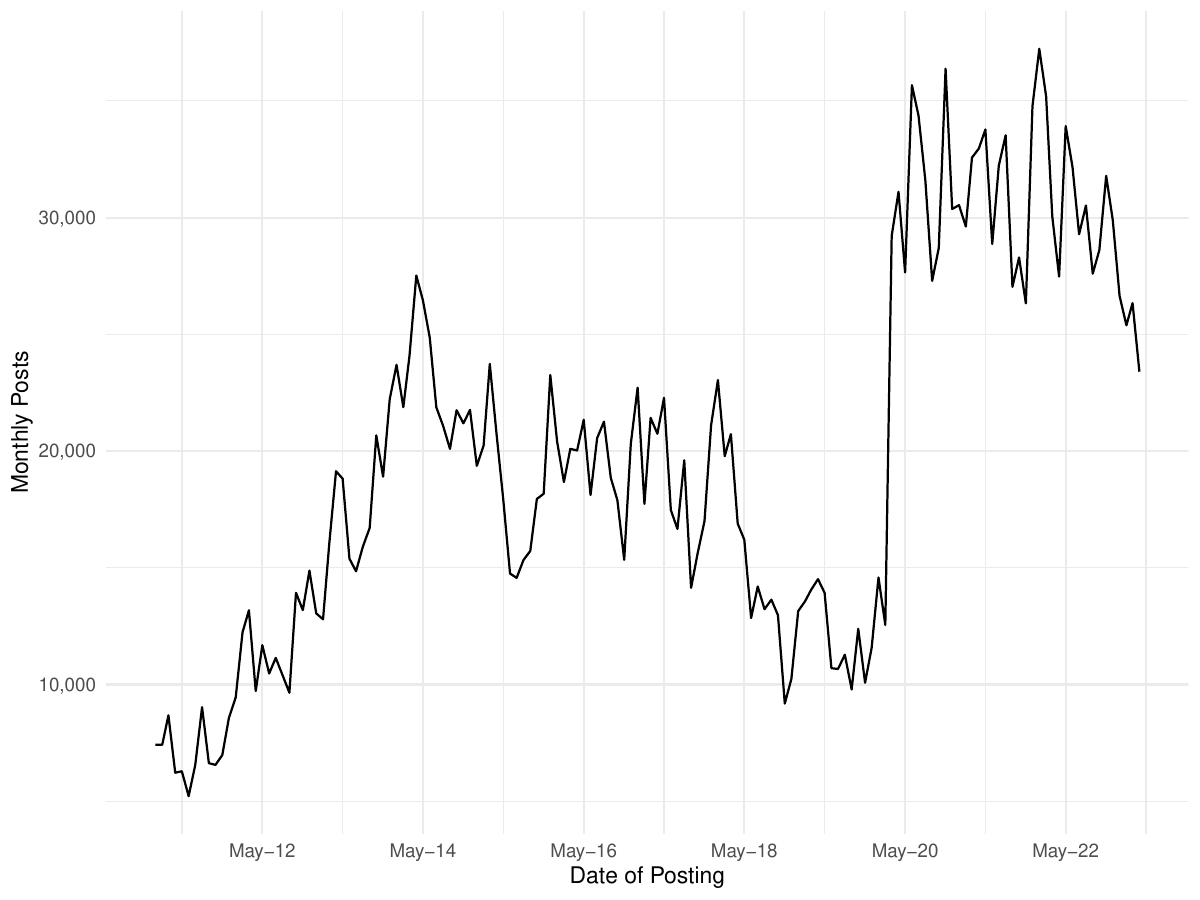}
        \caption{Monthly posts (United States)}
    \end{subfigure}
        \begin{subfigure}[t]{0.49\textwidth}
        \centering
    \includegraphics[width=\linewidth]{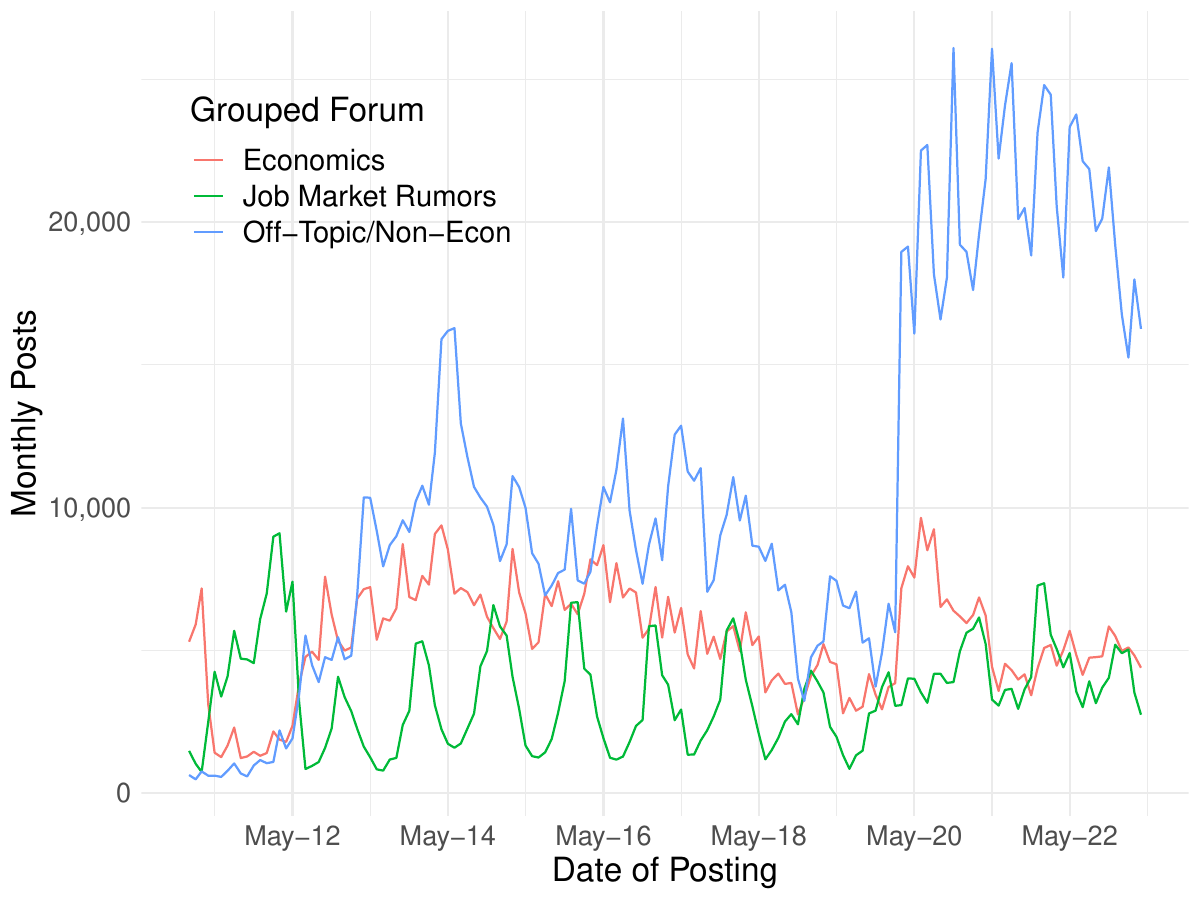}
        \caption{Monthly posts by forum (United States)}
    \end{subfigure}
    \caption{Number of monthly EJMR posts (left panels) and by grouped forums (right panels) over time for all countries (top panels) and the United States (lower panels). The left panel of the figure shows the number of monthly posts from December 2010 to April 2023. There is a marked increase in posting activity that coincides with the start of the COVID-19 pandemic in Europe and the United States in March 2020. The right panel shows the number of monthly posts in three groups of forums. Economics contains all forums with general economics discussions. Job Market Rumors contains all forums related to the academic job market including both junior and senior hiring. Off-Topic/Non-Econ contains all other forums.}
    \label{fig:posts_over_time}
\end{figure}

Panel A of \Cref{fig:posts_over_time} shows that the posting frequency on EJMR steadily rises between December 2010 and April 2014 and then remains relatively stable at around 40,000 monthly posts between 2013 and the beginning of 2020. However, the posting intensity jumps to around 70,000 posts per month in March 2020 with the beginning of the COVID-19 pandemic and, until recently, has remained at this elevated level. Panel B of \Cref{fig:posts_over_time} makes the cyclicality of the academic job market apparent. The green line which plots the monthly posts in the Job Market Rumors forums always peaks in December and January during the busiest part of the job market for academic economists.

The increase in EJMR posting frequency induced by the COVID-19 pandemic appears to be driven by two factors. First, as can be seen in \Cref{fig:posts_over_time}, the aggregate posting increase comes entirely from a sharp increase in the number of posts in the Off-Topic/Non-Econ forums which quadruples in size. There is also short transient increase in the number of posts in the Economics forums, but this increase subsides relatively quickly after a year. Second, as can be seen in panels C and D of \Cref{fig:posts_over_time}, the increase is primarily driven by IP addresses located in the United States whose monthly posting volume tripled from around 10,000 to over 30,000 posts per week and remained high. 

\begin{figure}[t]
   \centering
    \includegraphics[width=0.7\linewidth]{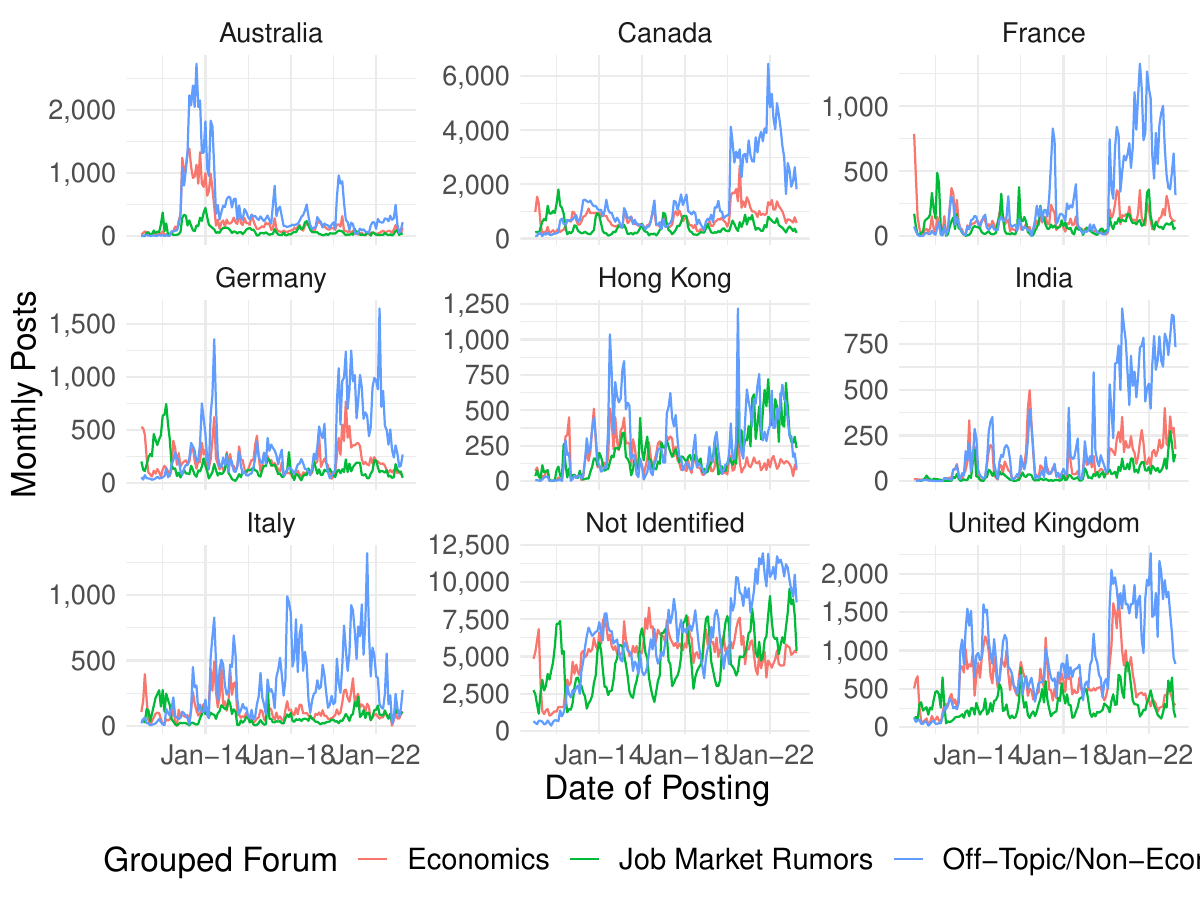}
    \caption{Monthly posts by grouped forums and by country (excluding US). The figure shows the distribution of monthly posts across the three large forum groups for the eight countries with the largest number of posts after the United States and for those posts for which we do not assign IP addresses.}
    \label{fig:monthly_posts_by_country}
\end{figure}

The COVID-19-induced increase in posting activity on EJMR however is not entirely confined to the United States. Other countries from which a large number of postings originate such as Canada, the United Kingdom, Italy, France, and Germany also experienced large increases in 2020, mostly in the Off-Topic category, as can be seen in \Cref{fig:monthly_posts_by_country}. However, unlike the United States the posting intensity in these countries mostly returned to pre-pandemic levels. Posts for which we cannot assign an IP address (bottom middle panel) display a relatively steady increase across all three forum groups over time and also have very strong posting cyclicality in the Job Market Rumors category.

\begin{figure}[t]
    \centering
    \includegraphics[width=0.9\linewidth]{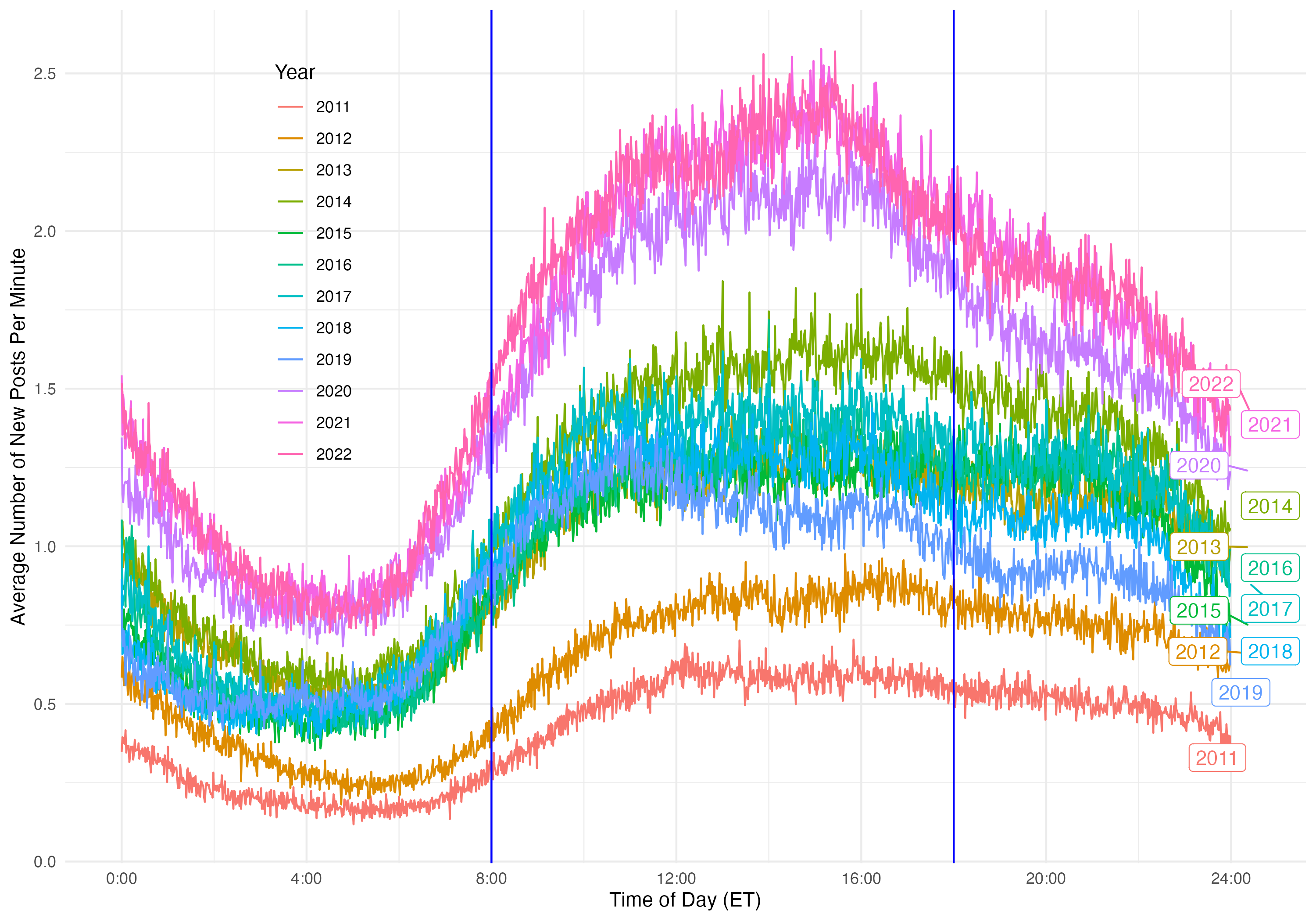}\\
    \caption{Distribution of posts across time of day (US Eastern Time). The figure shows the average number of EJMR posts per minute for a given year. The vertical blue lines show standard work hours for US Eastern time.}
    \label{fig:time_of_day}
\end{figure}

EJMR users tend to primarily post during US work hours and to a lesser extent in the evening. Figure~\ref{fig:time_of_day} shows total number of posts per minute by year from 2011 to 2022. The graph reveals that usage overall increased since the onset of the COVID-19 pandemic, but it did not change the overall pattern of EJMR users primarily posting during work hours.

\begin{figure}[tb]
    \centering
    \includegraphics[width=0.8\linewidth]{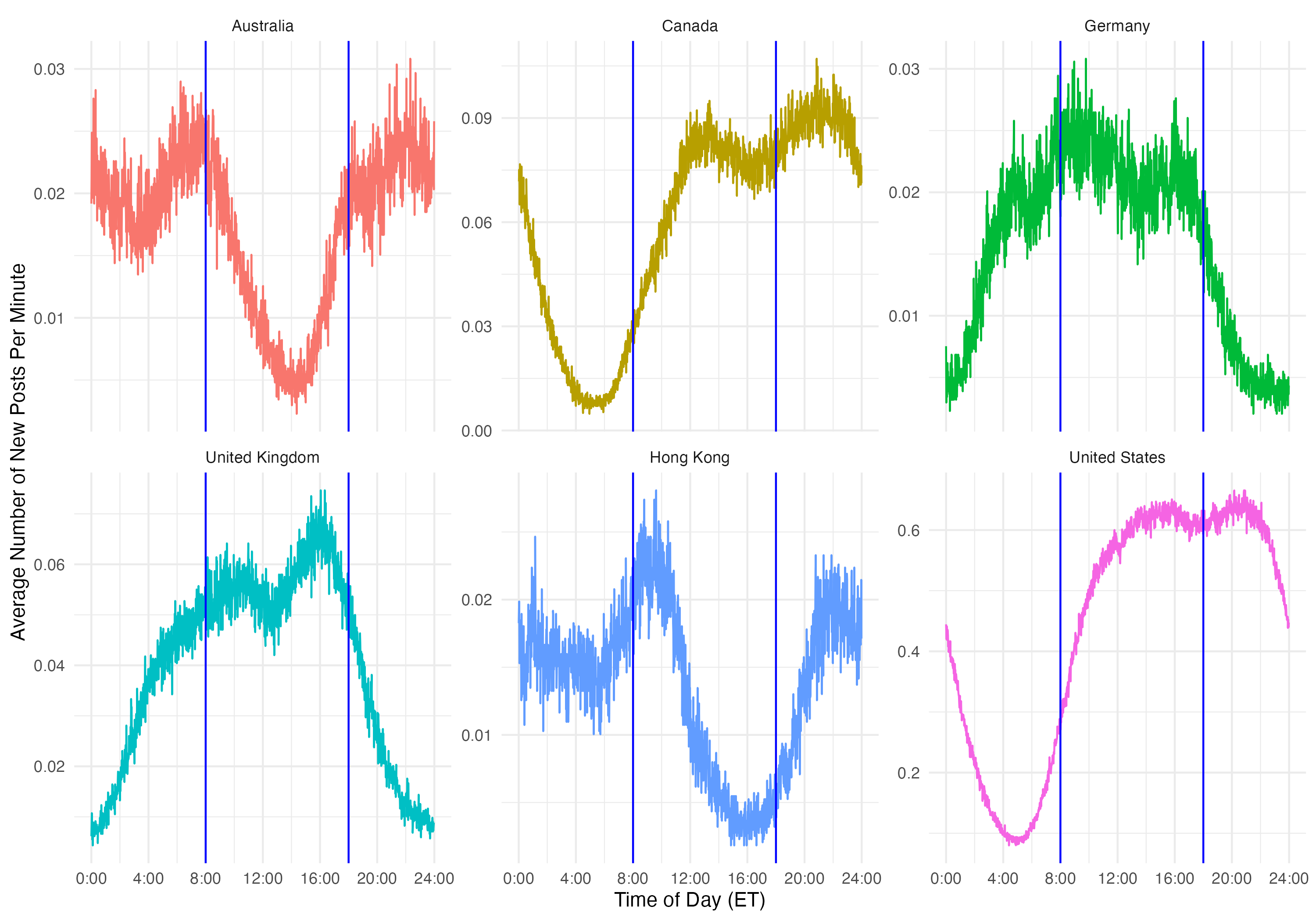}\\
    \caption{Distribution of posts across time of day (US Eastern Time) by country. The figure shows the average number of EJMR posts per minute for Australia, Canada, Germany, United Kingdom, Hong Kong, and the United States. The vertical blue lines show standard work hours for US Eastern time.}
    \label{fig:time_of_day_country}
\end{figure}

This pattern becomes even more apparent when using the country location of the posting IP addresses. Figure~\ref{fig:time_of_day_country} reports the distribution of posts across the time of day for the six countries with the largest number of posts: Australia, Canada, Germany, United Kingdom, Hong Kong, United States. Adjusted for their respective time zones EJMR users tend to post in the afternoon and in the evening, but less so during the morning or at night. However, as noted previously, the majority of the posts originate from IP addresses located in the United States.

\subsection{Geographical Distribution}
\label{sec:geography}

The majority of the assigned EJMR posts originate from IP addresses located in the United States. As noted in Section~\ref{sec:mapping}, with our methodology we assign \percentageOfEJMRPostsAssignedOfAssignable~of assignable posts to IP addresses.

\begin{figure}[t]
    \centering
    \includegraphics[width=\linewidth]{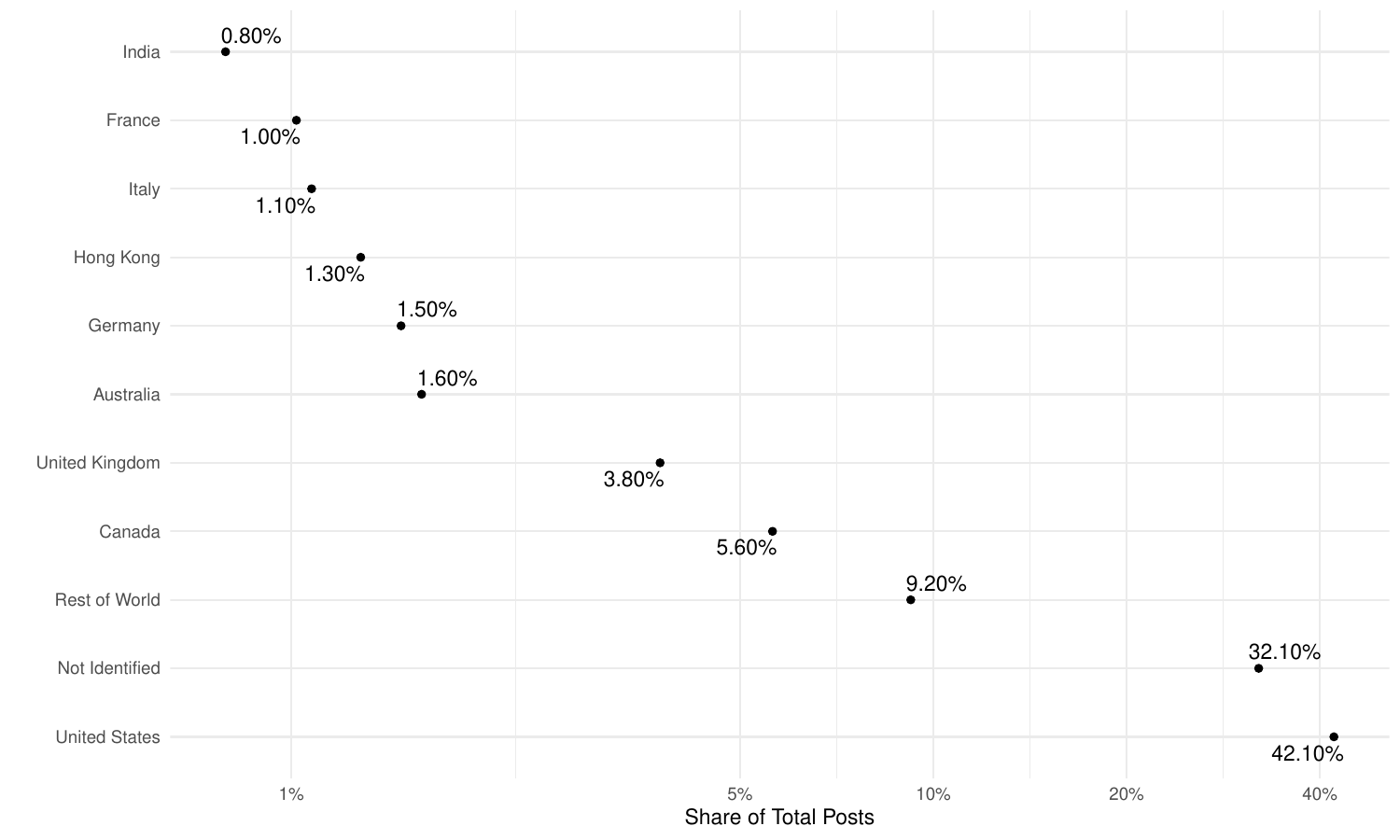}\\
    \caption{Distribution of posts across countries. The figure shows the share of all posts which can be assigned to a particular country. Posts for which we do not assign an IP address are in the Not Identified category.}
    \label{fig:posts_by_country_share}
\end{figure}

\Cref{fig:posts_by_country_share} shows that we are able to assign 67.8\% of assignable posts to particular countries using IP2Location. Among posts with geolocated IP addresses, 61.9\% originate from the United States with Canada (8.3\%) and the United Kingdom (5.5\%) a distant second and third. The rest of the posts with geolocated IP addresses come from other countries with significant research institutions in economics and finance such as Australia, Germany, Hong Kong, Italy, and France. There is also a substantial share of geolocated posts (13.6\%) from other countries in the rest of the world.

\begin{table}[tb]
    \centering
% \begin{tabular}{lcccccc}
% \toprule
% &\multicolumn{6}{c}{Language of Post} \\
% \cmidrule{2-7}
% Origin Country of Post & German & Chinese & Spanish & Portuguese & Russian & Korean\\
% \midrule
% Germany & \textbf{0.80} & 0.04 & 0.05 & 0.03 & 0.07 & 0.01\\
% China & 0.07 & \textbf{0.85} & 0.05 & 0.03 & 0.00 & 0.00\\
% Hong Kong & 0.09 & \textbf{0.86} & 0.03 & 0.02 & 0.00 & 0.00\\
% Spain & 0.29 & 0.03 & \textbf{0.35} & 0.33 & 0.01 & 0.00\\
% Portugal & 0.28 & 0.03 & 0.29 & \textbf{0.40} & 0.00 & 0.00\\
% Brazil & 0.09 & 0.00 & 0.12 & \textbf{0.77} & 0.01 & 0.00\\
% Russia & 0.29 & 0.06 & 0.05 & 0.03 & \textbf{0.57} & 0.00\\
% Korea & 0.20 & 0.04 & 0.06 & 0.07 & 0.01 & \textbf{0.62}\\
% Rest of World & 0.32 & 0.31 & 0.18 & 0.12 & 0.04 & 0.02\\
% \bottomrule
% \end{tabular}
\includegraphics[width=\linewidth]{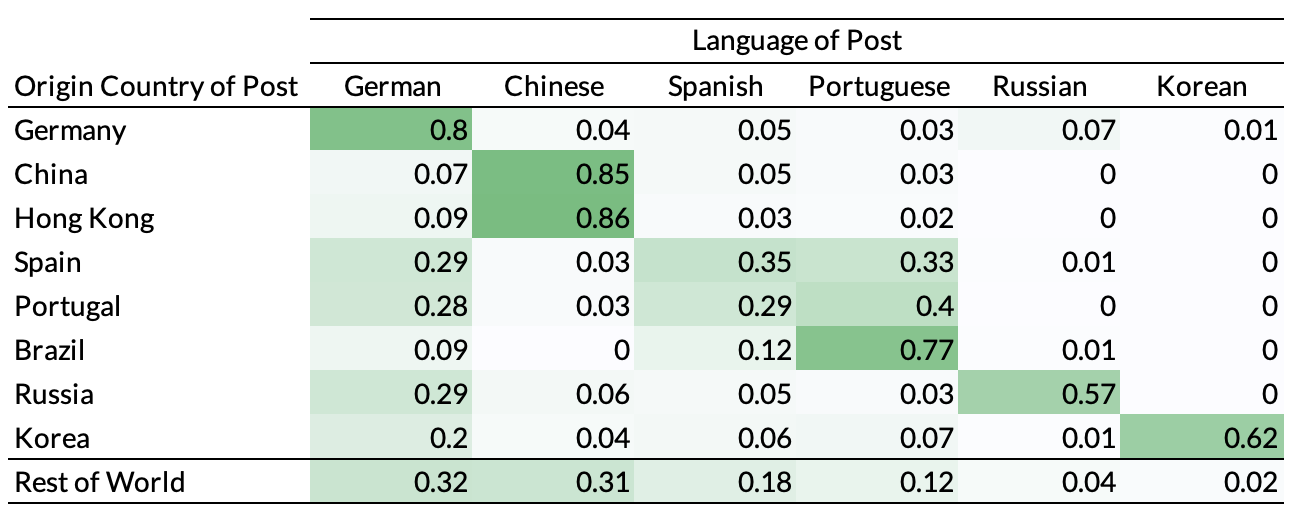}
    \caption{This figure shows the share of non-English posts for each country that are in the languages indicated in the six columns. These are the non-English languages with at least 1,000 posts on EJMR\@. Each country's primary language is in bold font.}
    \label{tab:lang_country}
\end{table}

An additional sanity check for the accuracy of our IP assignment and subsequent geolocation is whether the language that EJMR posters use corresponds to the country of origin of their IP address. Using the language classification of \cite{stahl2023lingua} we show in \Cref{tab:lang_country} that the dominant non-English language in all major non-English speaking countries in our data is indeed the country's native language. This pattern is particularly pronounced for Brazil, China, Germany, Hong Kong, and Korea and to a lesser extent for Spain, Portugal, and  Russia. 

\begin{figure}[tb]
    \centering
    \includegraphics[width=.7\linewidth]{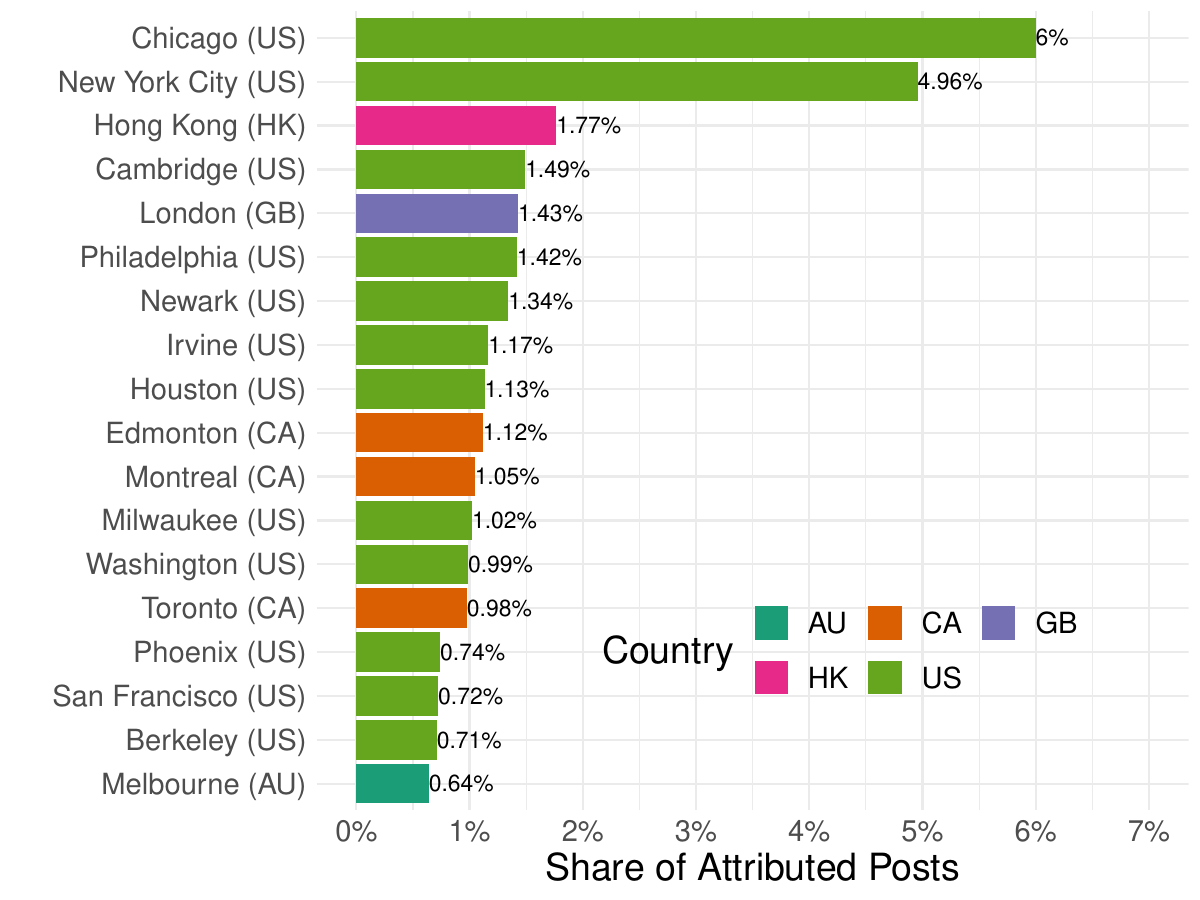}
    \caption{Share of posts with assigned IP address across cities. This figure shows the share of posts with an assigned IP address that originate from a given city. The share of posts from cities located in the United States, Hong Kong, the United Kingdom, Canada, and Australia are marked in light green, pink, purple, orange, and dark green, respectively.}
    \label{fig:posts_by_city}
\end{figure}

Beyond country of origin the IP addresses we recover also provide much more granular information about the exact location and internet service provider of EJMR posters. Figure~\ref{fig:posts_by_city} reports the cities with the largest number of posts. There is substantial heterogeneity driving the ranking of these cities. By far the largest number of posts originate from Chicago. However, the number of unique IP addresses from which posts attributed to Chicago originate is comparatively small (879 unique IP addresses).\footnote{A large number of posts originate from a few IP addresses located in Naperville, IL. Based on several cross-checks with other geolocation databases, these IP addresses are misclassified and are actually located in Chicago. We correct these locations in the IP2Location data.} In contrast, the next two cities on the list, Hong Kong and New York City, have much fewer posts but these posts originate from a larger set of IP addresses (952 and 1,381 unique IP addresses, respectively). Cambridge, the location of two of the leading economics departments in the world, is also among the top 5 cities and also has a smaller number of IP addresses (499 unique IP addresses) from which its posts originate. As expected from our country-level analysis, cities in the United States, particularly those with leading universities such as Cambridge or Berkeley are towards the top of the ranking despite their relatively small population size.

\begin{figure}[tb]
    \centering
    \includegraphics[width=0.8\linewidth]{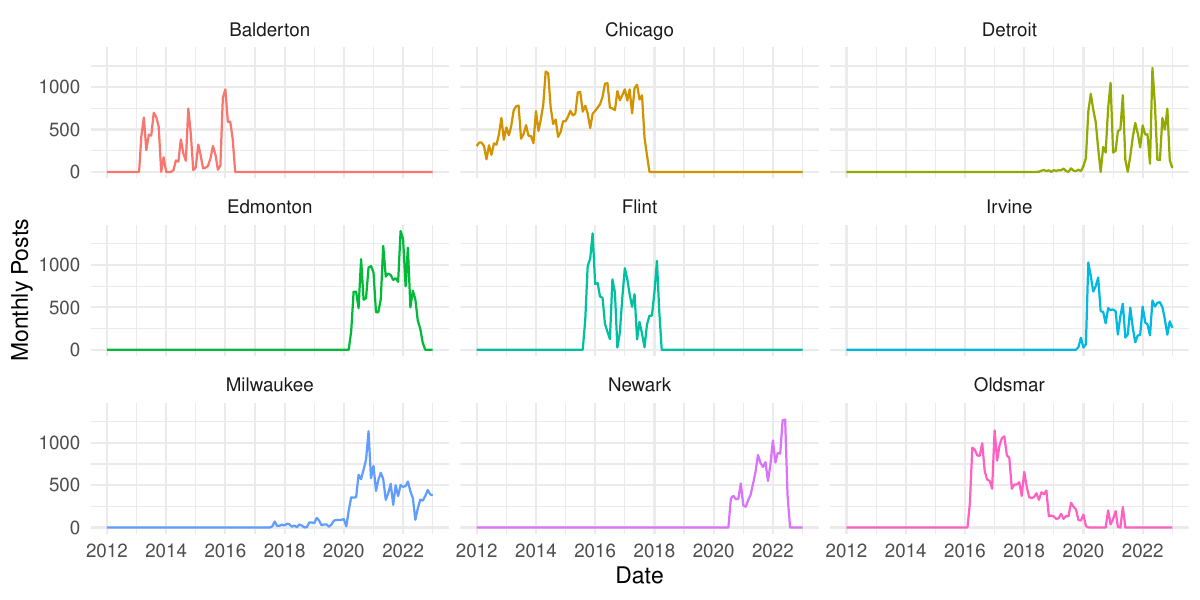}
    \caption{Monthly posts for selected EJMR power users. The figure shows the number of monthly posts over time for a select number of IP addresses from which a large number of posts originate.}
    \label{fig:monthly_posts_by_ip}
\end{figure}

The association of an IP address with a contributor is not necessarily persistent, both in the short and long run. Posters may change IP addresses because of new IP assignment by their internet service provider, the use of a different device, or various other reasons. Nonetheless, IP address do persist for a significant period of time for power users of the site. In Figure~\ref{fig:monthly_posts_by_ip}, we plot the posting frequency for a select number of IP addresses (and their respective locations) from which many EJMR posts originate. Posts from these IP addresses end abruptly when the users are assigned new IP addresses, but all of them make a large number of posts over an extended period of time (i.e., several years).

Posting on (and not just reading) EJMR appears to be pervasive and widespread, even from devices connected to university networks. 15.2\% of all posts to which we assign IP addresses originate directly from IP addresses associated with universities or research institutions.\footnote{We identify universities and research institutions based on the names associated with the ISP using a regular expression match. Any ISP whose name matches \texttt{University|College|School|Federal Reserve Board|The World Bank Group|Institute|Universitaet|Universitaris|International Monetary Fund|Universidad|NUS|Institute|NBER} is marked as university or research institution.} Although some universities also are the internet service provider for some of their faculty and students (e.g., through university-provided faculty or student housing), this means that a substantial number of posts on EJMR occur while users are connected at or through their workplace. Perhaps even more surprisingly, there are EJMR posts from identified IP addresses located at \emph{every} leading university in the United States.

\begin{figure}[htb]
   \centering
     \begin{subfigure}[t]{0.49\textwidth}
    \includegraphics[width=\linewidth]{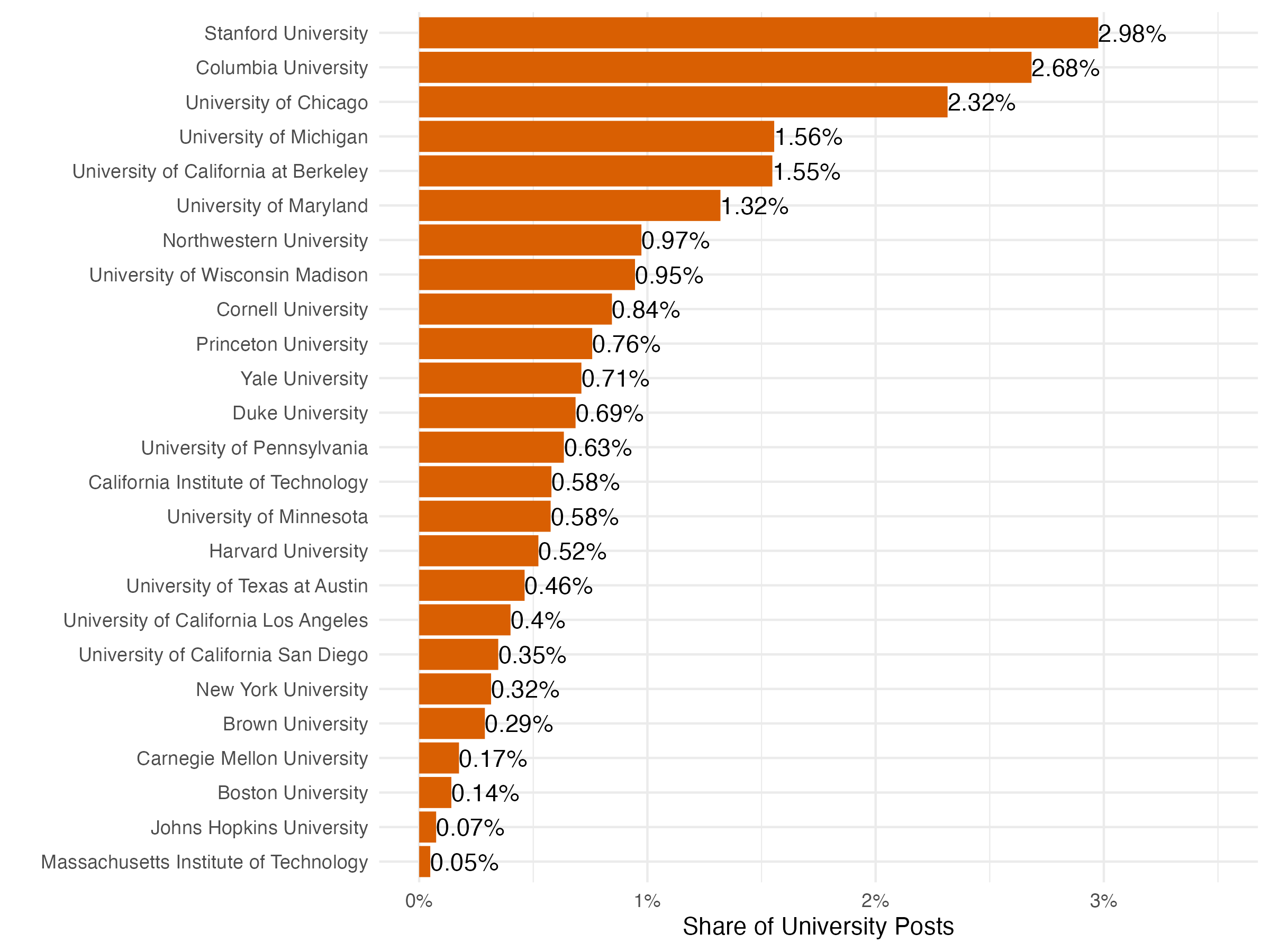}\\
    \caption{Top 25 economics departments}
    \label{fig:post_share_by_top25}
    \end{subfigure}
            \begin{subfigure}[t]{0.49\textwidth}
                \includegraphics[width=\linewidth]{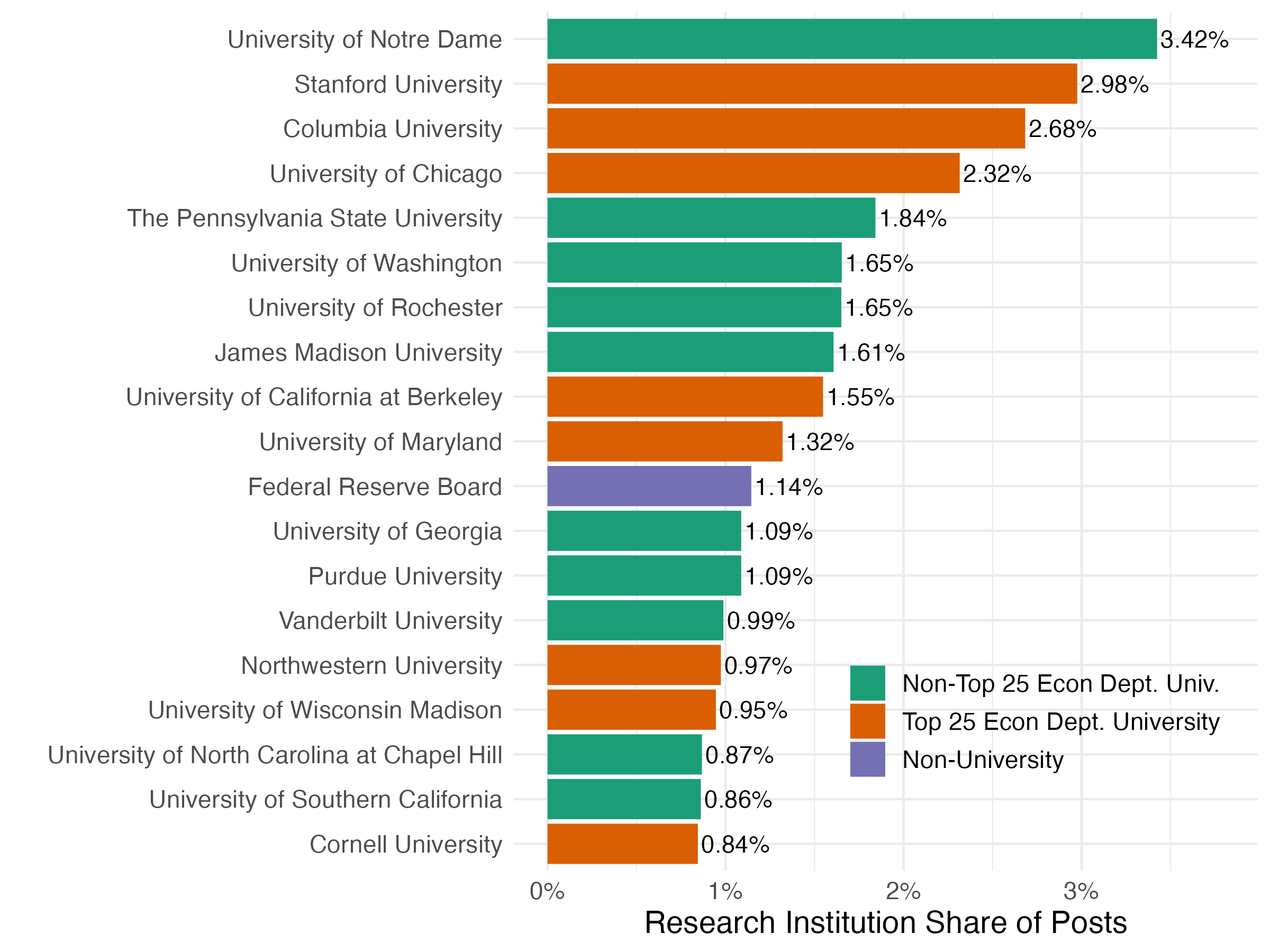}\\
    \caption{All research institutions}
    \label{fig:post_share_by_all_US}
    \end{subfigure}
    \label{fig:post_share_universities}
    \caption{Posting share among research institutions. Panel (i): Post share of all university or research institution posts by each of US universities with a top 25 economics department. The figure shows the share of posts accounted for by a given top 25 US university among all posts originating from IP addresses associated with universities or research institutions. Panel (ii):  Post share of US university or research institutions. The figure shows the share of posts accounted for by a given US university or research institution among all posts originating from IP addresses associated with US universities or research institutions.}
\end{figure}

Figure~\ref{fig:post_share_by_top25} reports how many posts come from each of the top 25 US universities (US universities with an economics department listed in the U.S.~News top 25) as a share of all the posts originating from IP addresses associated with universities or research institutions. The figure highlights the very large share that these top 25 US universities have across posts from all universities around the world as they account for more than 20\% of posts from over 500 universities around the world that show up in our data.

The representation of top US universities among EJMR posters is also apparent in Figure~\ref{fig:post_share_by_all_US} which reports the share of posts accounted for by a US university or research institution among all posts originating from IP addresses associated with universities or research institutions. Eight of the 19 institutions shown in the figure are universities ranked among the top 25 economics departments and among the top four universities contributing to EJMR, three (Stanford, Columbia, and University of Chicago) are ranked in the top 10. Among these institutions with the largest shares of EJMR posts there is also one organization that is not a university. The Federal Reserve Board employs over 400 PhD economists, many more than any single university. Given the sheer number of economists and that some of these economists contribute posts to EJMR while connected to their employer's network, it is perhaps not particularly surprising that the Federal Reserve Board appears in this list.

\begin{figure}
   \centering
    \includegraphics[width=\linewidth]{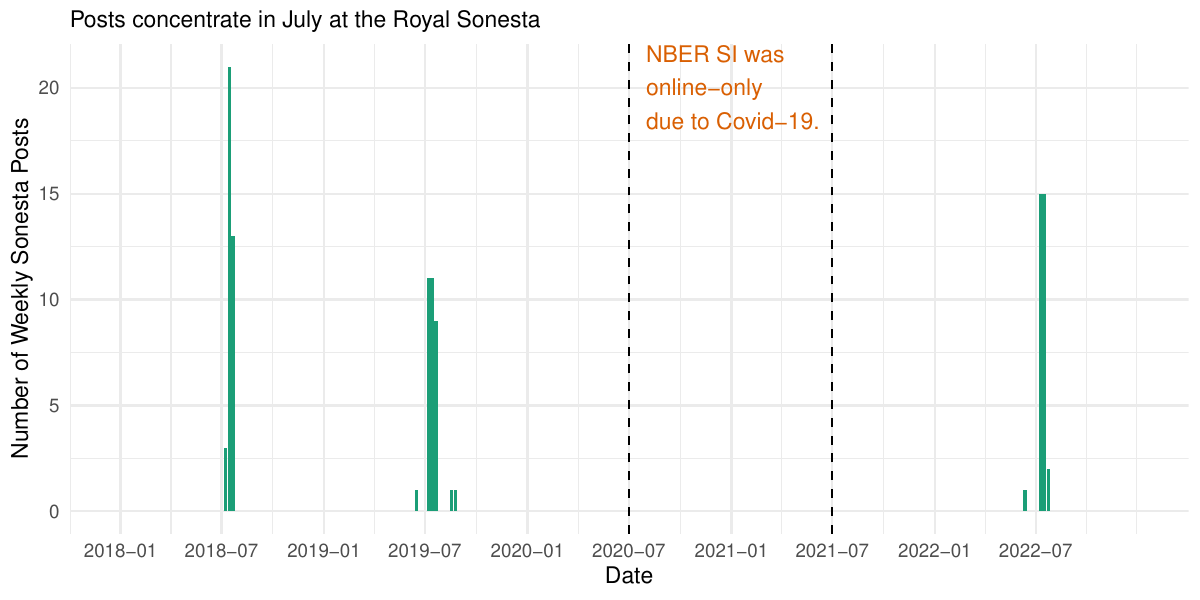}\\
    \caption{Number of weekly posts from the Royal Sonesta Boston, a hotel in Cambridge, Massachusetts. This hotel is the conference location of the annual NBER Summer Institute, a three-week conference held annually in July where attendance is by invitation only. EJMR posts from this location exclusively occur in July, except in 2020 and 2021 when the NBER Summer Institute was only held virtually rather than in person at the Royal Sonesta Boston.}
    \label{fig:nber_posts}
\end{figure}

Another indication that EJMR usage is pervasive throughout all echelons of the economics profession, including faculty at elite institutions, can be found in \Cref{fig:nber_posts}. The figure shows the number of weekly posts that originate from the Royal Sonesta Boston, a hotel in Cambridge, Massachusetts. This hotel serves as the location of the annual NBER Summer Institute, a three-week conference held annually in July. The NBER Summer Institute is the world's leading economics research conference and attendance is by invitation only. As is evident from the figure, EJMR posts from this hotel's IP address peak every year in July except in 2020 and 2021 when the NBER Summer Institute was only held virtually rather than in person at the Royal Sonesta Boston.

\subsection{Concentration of Posters and Posts}
\label{sec:concentration}

There are \niceTotalPostsWithTopicIDAndUsername~posts for which we have the topic and the username and from which we are able to recover \niceNumDistinctIPsAssigned~distinct IP addresses. However, these posts are far from evenly distributed across the many posters on the platform. Among the posts for which we assign IP addresses, a very large fraction of posts is generated by just a few IP addresses.

A mere 5\% of the \niceNumDistinctIPsAssigned~IP addresses generate over 50\% of all posts with assigned IP addresses and 20\% of IP addresses generate just over 80\% of these posts. While such a high concentration of contributions may appear extreme, it is quite common across many online platforms as documented by \cite{guo2008stretched,lei2009analyzing}. The degree of concentration is even higher than these numbers might suggest because they only take into account posts with assigned IP addresses. Recall that for \percentageOfEJMRPostsNotAssignedofAssignable~of assignable posts we do not assign an IP address because the likely IP addresses from which these posts originate do not generate a sufficient number of posts to meet the very conservative identification thresholds we employ. There are thus many more IP addresses with just a few posts each which are not contained in these figures.

\begin{figure}[htb]
   \centering
    \includegraphics[width=0.7\linewidth]{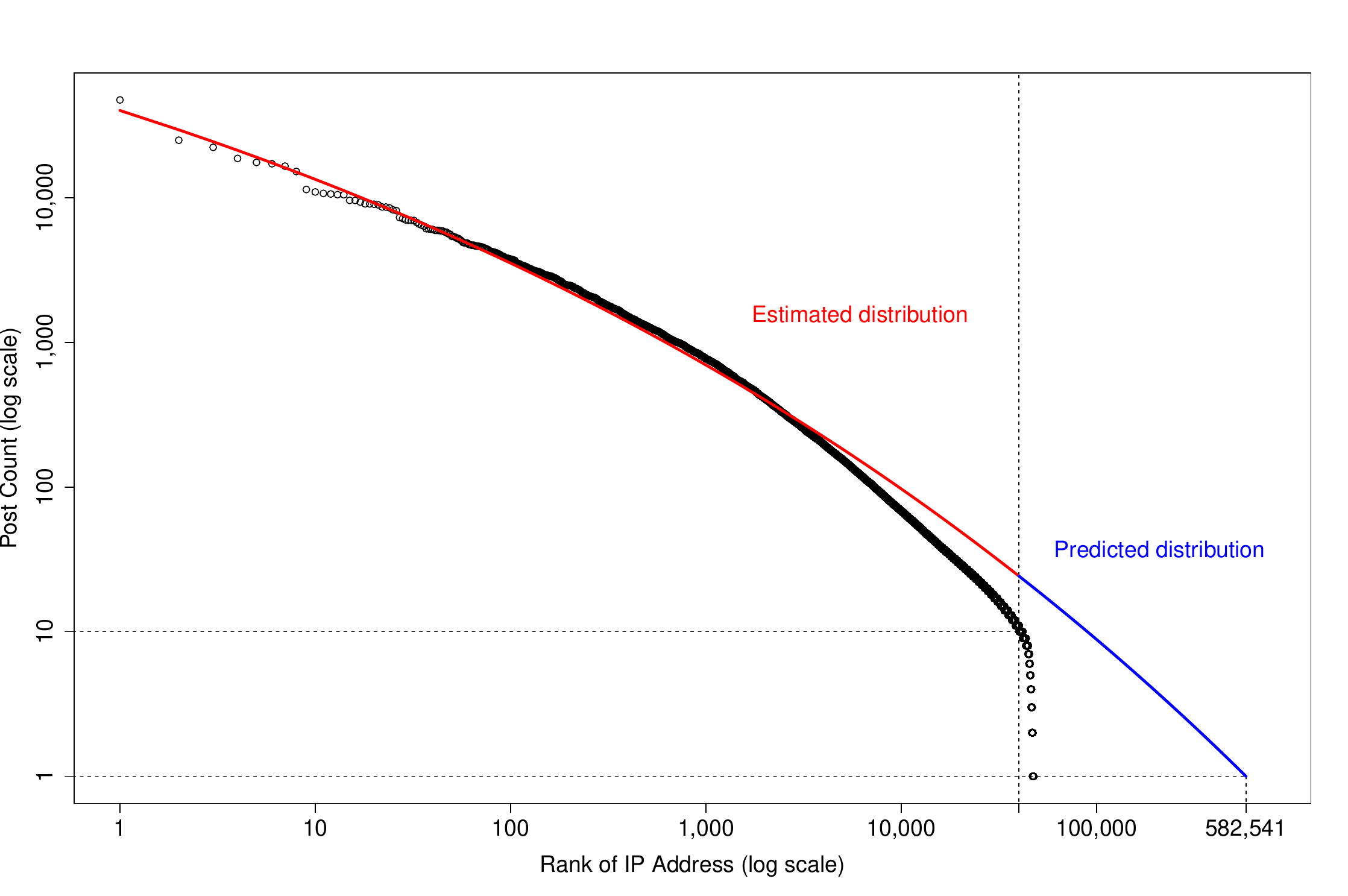}\\
    \caption{Distribution of posts by IP address rank. The figure shows the number of posts an IP addresses has contributed to EJMR. IP addresses are ordered on the x-axis by the number of posts assigned to them. We estimate a stretched exponential distribution (red line) of posts for the IP addresses ranked 1 to 40,000 and predict the  number of posts for all IP addresses with a lower rank (blue). There are \predictedIPaddresses~IP addresses which are predicted to have contributed at least one post to EJMR.}
    \label{fig:stretched_exponential}
\end{figure}

Prior research suggests that contributions on online platforms follow neither a power law nor an exponential function, but instead are best approximated by the stretched exponential function \citep{guo2008stretched,lei2009analyzing}. This is also the case for EJMR as can be seen in Figure~\ref{fig:stretched_exponential} which plots the relationship between IP rank of posters and number of corresponding posts in log-log space. The stretched exponential fits very well up to the point where our assignment procedure stops assigning IP addresses to posts. 
Loosely speaking, if an IP address posts in fewer than ten topics in the span of a week it will not be assigned to any posts. 

Fitting a stretched exponential distribution to the relationship between IP rank and post count further allows us to estimate how many IP addresses have ever posted on EJMR. We do so by estimating the stretched exponential up to IP rank 40,000 which has 10 posts assigned to it and then projecting out this fitted distribution until the estimated number of posts of an IP address is equal to 1. The estimated total number of posts is 7.4 million and matches the total number of observed posts (7.1 million) quite well. Under this projection there are \predictedIPaddresses~IP addresses which have contributed at least one post to EJMR. Thus, while the vast majority of posts come from just a few thousand IP addresses, our analysis suggests that a very large number of IP addresses has contributed to EJMR over the past decade.

\subsection{Content of EJMR Posts}
\label{sec:content}

We now turn to analyzing the content of EJMR posts. We focus, in particular, how this content varies across universities and IP addresses. 

\subsubsection{Mentions of Universities}

EJMR is intended to be a source of information about the academic job market and of professional news about the economics profession. Posts containing such information may be innocuous and and thus are more likely to come directly from within universities (i.e., from university IP addresses) than toxic posts. A natural way to analyze such information is to investigate how frequently posts from university IP addresses mention their own or other universities, especially for the US universities with the largest number of EJMR posts.

\begin{table}
\includegraphics[width=\linewidth]{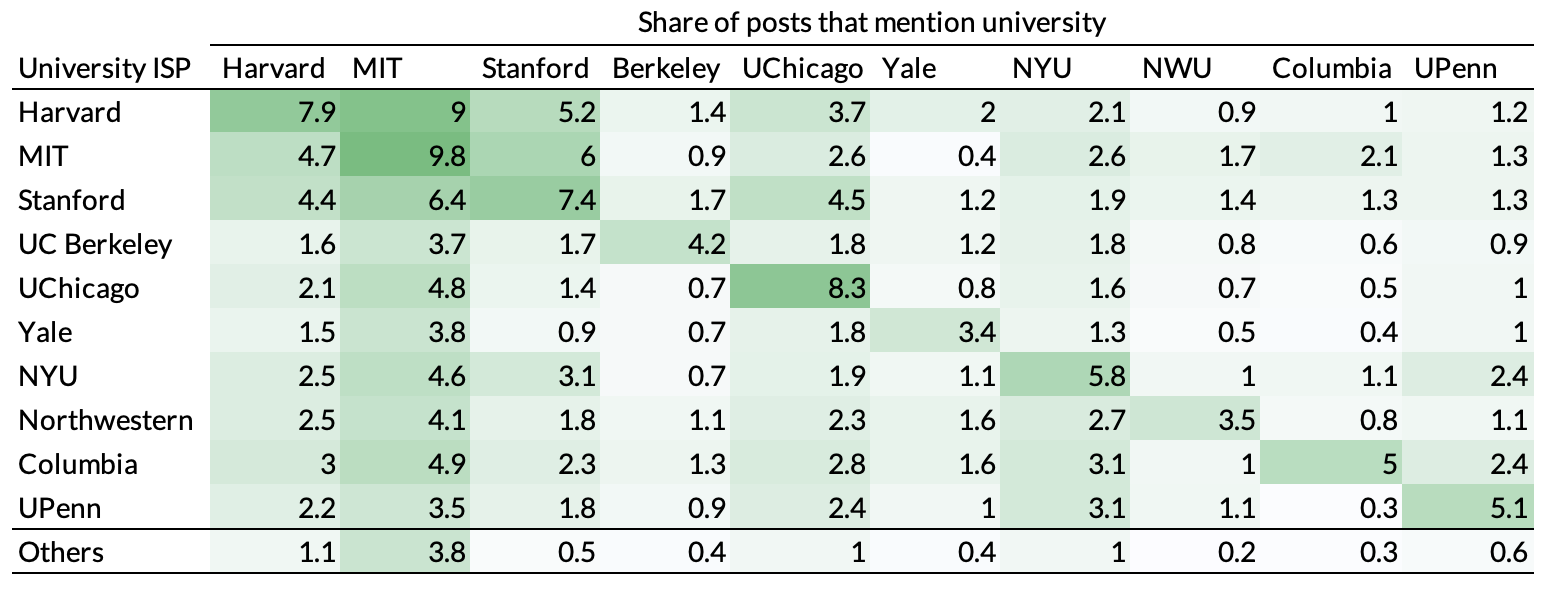}
\caption{Share of posts that mention a university from each university ISP. Keyword match for each variable (lower case): Harvard: ``harvard|hbs'', MIT: ``mit|sloan'', Stanford: ``stanford'', Berkeley: ``berkeley|haas'', UChicago: ``uchicago|university of chicago|chicago|booth'', Yale: ``yale'', NYU: ``nyu|stern'', Northwestern: ``northwestern|kellogg'', Columbia: ``columbia'', UPenn: ``upenn|penn|wharton''}
\label{tab:university_mentions}
\end{table}

For the ten largest US universities, as measured by the number of EJMR posts originating from IP addresses at theses universities, Table~\ref{tab:university_mentions} reports the share of posts that mention either the university itself or any other university. Several patterns stand out in this table. First, for all ten universities the largest share of posts mentioning a university is always the share of self-mentions. These self-mention shares are shown on the diagonal and in bold text. This large share of self-mentions may be the result of inside knowledge dissemination. Second, mentions of other universities tend to decline by university rank. For example, the share of posts mentioning Harvard is larger than the share of posts mentioning Columbia, Northwestern, and UPenn for every single university ISP except, of course, the own university ISP. Third, even among these institutions MIT stands out. Posts from other university ISPs shown in the last row of Table~\ref{tab:university_mentions} mention MIT almost four times more often than any of the other top 10 institutions.

\subsubsection{Misogyny, Toxicity, and Hate Speech}

As described in Section~\ref{sec:linguistic}, we deobfuscated the content and then used several transformer-based machine learning models to classify EJMR posts by sentiment, misogyny, and toxicity. Each of these transformer models is best-in-class yet nonetheless imperfect. We also re-created all the word-count measures used in the seminal study of \cite{wu2020gender}. 

EJMR employs automatic moderation that deletes profane speech but does not prohibit toxic speech more generally. For example, users are not permitted to write ``fuck'' but they may write ``these brahmin jeets are replacing the juifs in terms of financial engineering excellence. when do we start building the crematoriums?''\footnote{``Jews'' is variously obfuscated on EJMR, usually as merely ``Js'' while ``pajeet'' and ``jeet'' are racial epithets used in reference to persons of South Asian descent.} Clearly, such posts in our data were also not removed by EJMR's human moderators.

\begin{figure}[!htp]
    \centering
    \vspace{0.75cm}
    \includegraphics[width=0.49\linewidth]{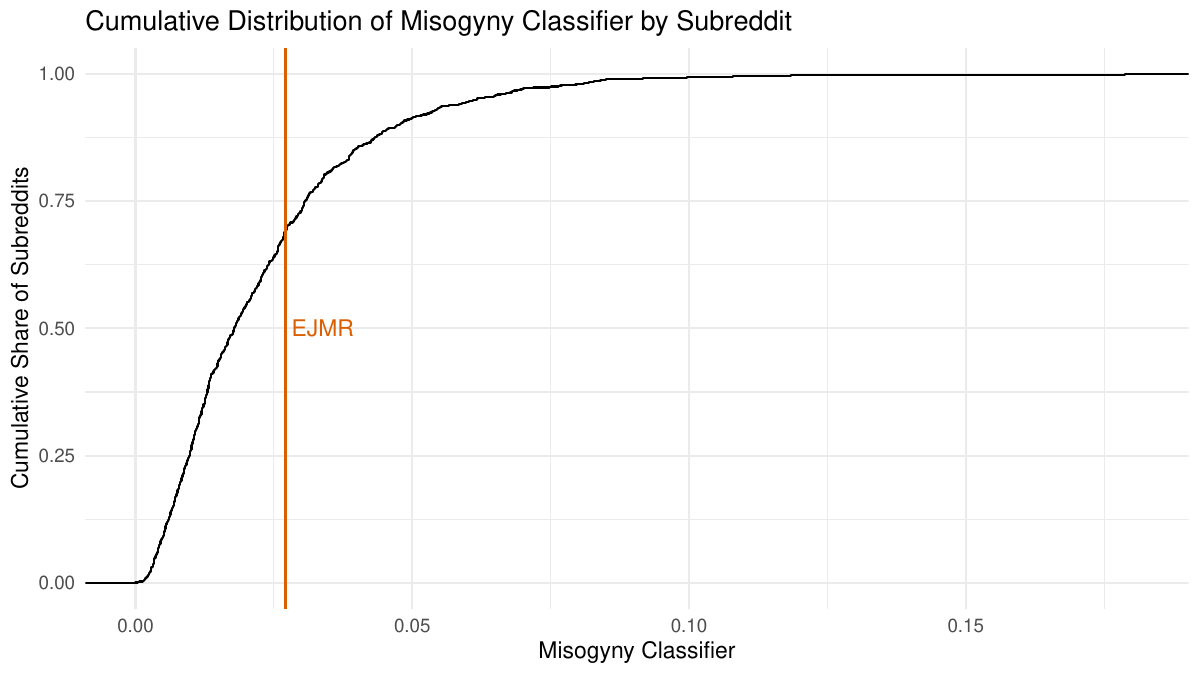} 
    \includegraphics[width=0.49\linewidth]{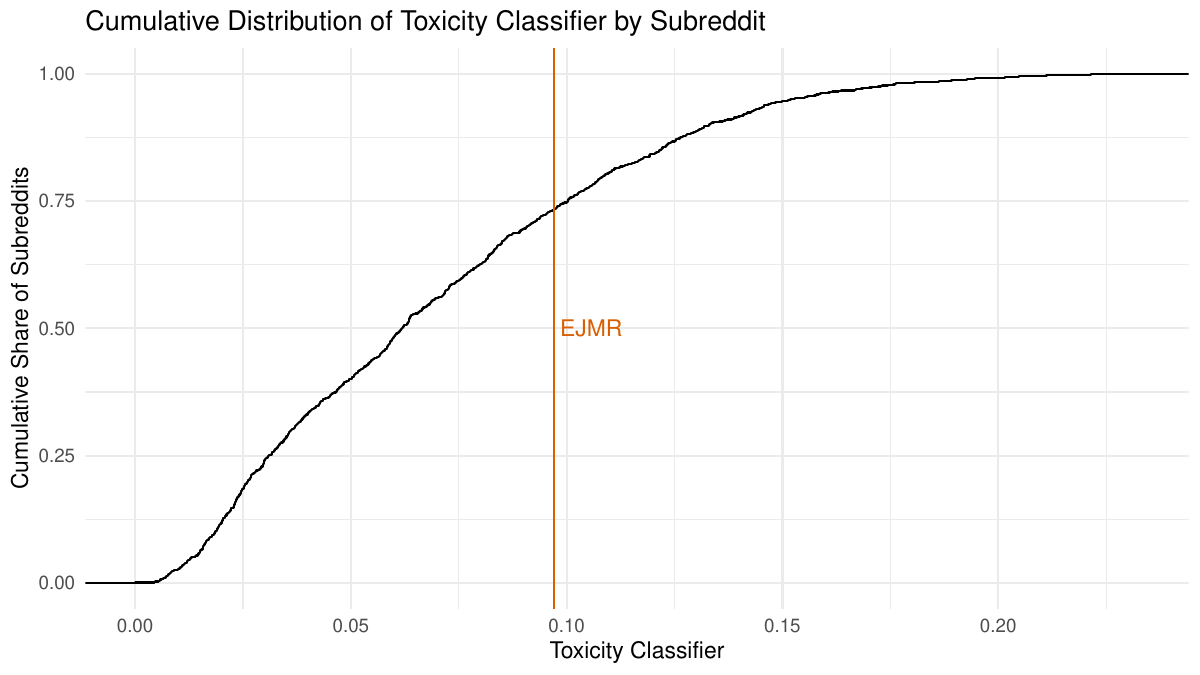} \\
    \includegraphics[width=0.49\linewidth]{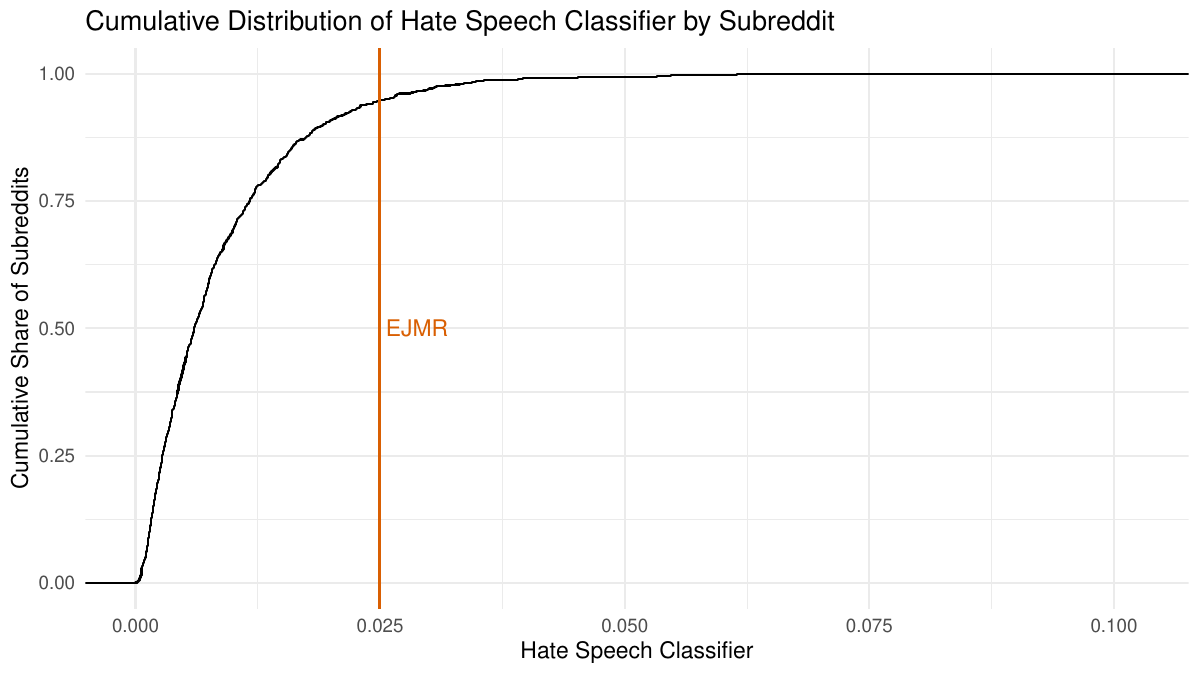}    
    \caption{Cumulative distributions of the share of posts containing misogyny, toxicity, and hate speech for the 1,000 most popular subreddits. The respective positions of EJMR in these distributions are indicated by the orange lines.}
    \label{fig:classifiers_cdf}
\end{figure}

Overall, the share of EJMR posts classified as misogynistic, toxic, or hate speech is 3.3\%, 11.8\%, and 3.1\% respectively.\footnote{We report these aggregated measures in Appendix Section \ref{sec:perspective_toxicity} along with other alternative classification measures.} These shares have remained relatively constant since the inception of EJMR with some mild seasonality over the course of each year with a higher share of misogynistic, toxic, or hate speech posts during the summer months. Across IP addresses the concentration of posts containing such offensive content is generally higher than that of other non-problematic posts. 20\% of identified IP addresses generate 81\% of all toxic posts with assigned IP addresses. Moreover, posts in the Off-Topic/Non-Econ forums are substantially more likely to be misogynistic (almost 4\%) or toxic (roughly 12\%) than those in the Economics (2\% and 10\%) or Job Market Rumors (2\% and 8\%) forums. However, even in the Job Market Rumors forums in which discussion focuses on the academic job market, approximately 2\% and 8\% of all posts are classified as misogynistic or toxic.

\begin{figure}[!tb]
    \centering
    \includegraphics[width=\linewidth]{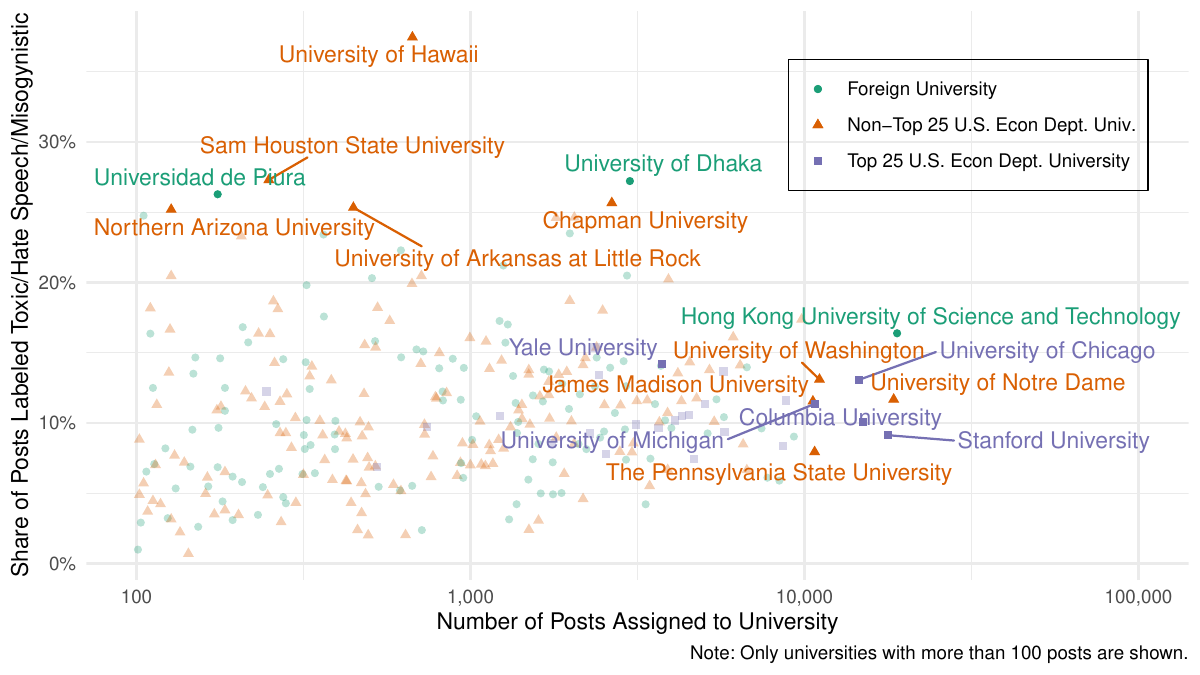}\\
    \caption{Total number of EJMR posts assigned to university and share of posts labeled toxic, misogynistic, or hate speech by university. The figure shows a scatterplot of the total number of EJMR posts and share of toxic, misogynistic, or hate speech posts by university for all university ISPs. US universities with an economics department ranked in the top 25 are marked in purple, US universities with an economics department ranked outside the top 25 are marked in orange, and non-US universities are marked in green.}
    \label{fig:post_and_number_by_univ}
\end{figure}

In addition, there is also considerable heterogeneity of the share of problematic posts across universities. Figure~\ref{fig:post_and_number_by_univ} reports a scatterplot of the total number of EJMR posts and combined share of toxic, misogynistic, or hate speech posts by university for all university ISPs. As is evident from the figure, usage (both problematic and not) of EJMR, is widespread throughout the economics profession and not limited to any particular subset of institutions.

Although our results so far show that many posts on EJMR contain problematic content, it is less clear whether the proportion of this content on EJMR is particularly egregious when compared to less professional settings. To evaluate whether the aforementioned shares of problematic content are relatively high or low, it is instructive to compare them to content on other internet platforms in which users can contribute posts while remaining relatively anonymous. As a validation, we therefore compare EJMR posts to Reddit posts. 

\begin{figure}[!htp]
    \centering
    \includegraphics[width=0.49\linewidth]{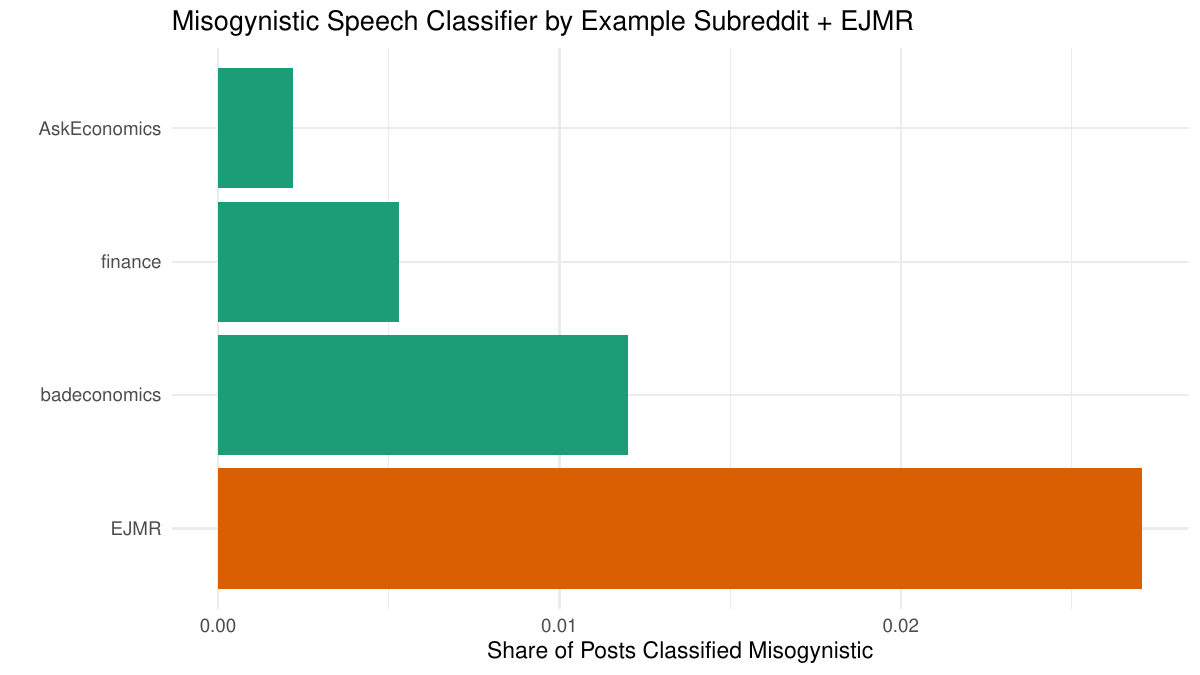} 
    \includegraphics[width=0.49\linewidth]{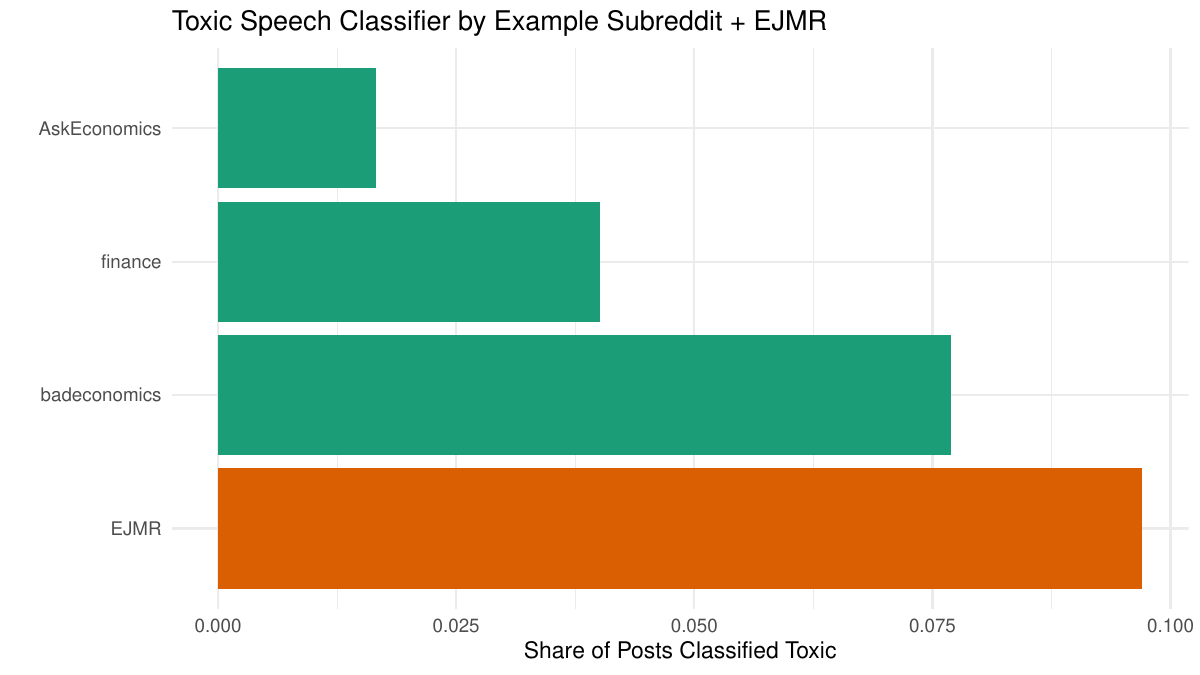} \\
    \includegraphics[width=0.5\linewidth]{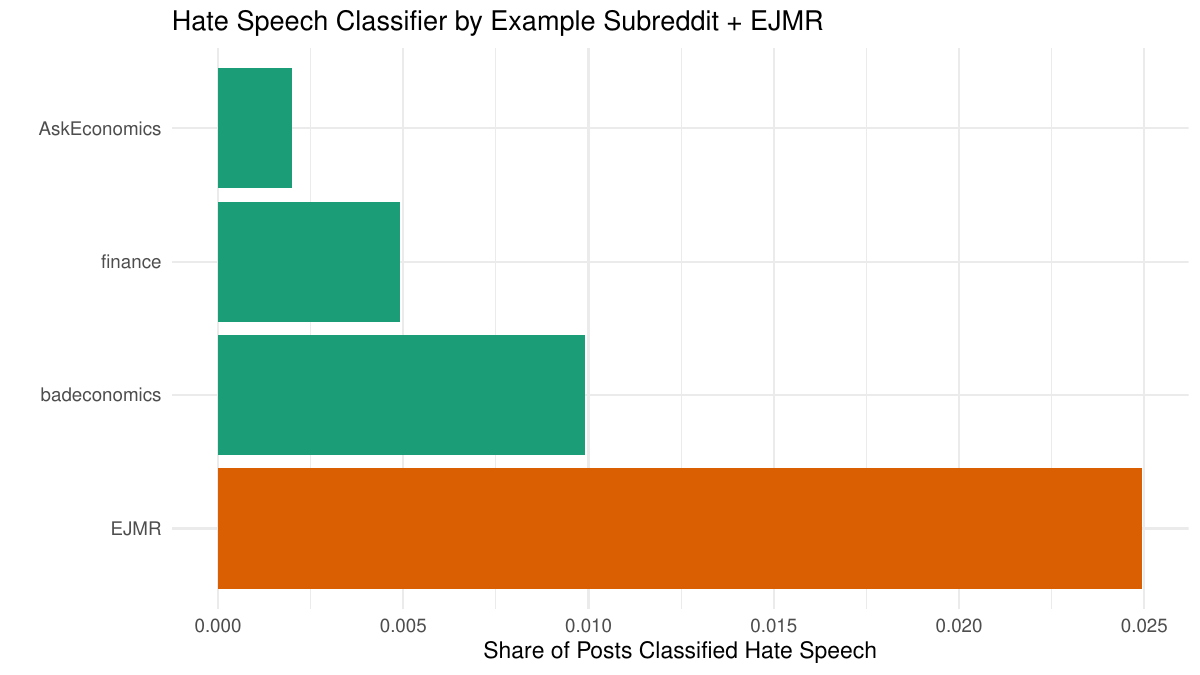}  
    \caption{Share of posts classified as containing misogyny, toxicity, and hate speech for r/AskEconomics, r/finance, r/badeconomics, and EJMR.}
    \label{fig:classifiers_example}
\end{figure}

\Cref{fig:classifiers_cdf} reports the cumulative distribution of the share of posts containing misogyny, toxicity, and hate speech classifiers for the 1,000 most popular subreddits. The respective positions of EJMR in these distributions are indicated by the orange lines. Compared to these subreddits, EJMR ranks at the 69th percentile of misogyny, the 73rd percentile of toxicity, and the 95th percentile of hate speech. Taken together, this suggests that even compared to other anonymous internet speech in a strictly non-professional setting such as Reddit, EJMR contains markedly more toxic material. This conclusion becomes even clearer when comparing EJMR to specific subreddits dedicated to economics or finance. Among the 1,000 most popular subreddits, the three most similar in terms of general topic are r/badeconomics, r/finance, and r/AskEconomics. \Cref{fig:classifiers_example} shows that EJMR has more posts labeled misogynistic, toxic, and hate speech than these related subreddits.\footnote{Perhaps the most concerning aspect of EJMR is that, in contrast to Reddit, it features specific commentary (and attacks) on economists who are, with a few notable exceptions, not public figures (e.g., singers, dancers, actors, and politicians who seek out public attention). This is important because private figures are more vulnerable and much less likely to have the resources to litigate defamatory content about them. They are also less likely to receive high damages awards for reputational injury assuming they could find an attorney to take the case on a contingency fee. Public figures, in contrast, 
% even though courts have made a normative decision that they should have a higher burden of proof, 
have the resources to hire lawyers and generally can use their access the media to rebut defamatory statements without assistance from the courts. 
% In short, attacking private figures is insidious precisely because they are so vulnerable.
}

\begin{figure}[!tp]
    \centering
        \begin{subfigure}[t]{0.9\textwidth}
            \includegraphics[width=\linewidth]{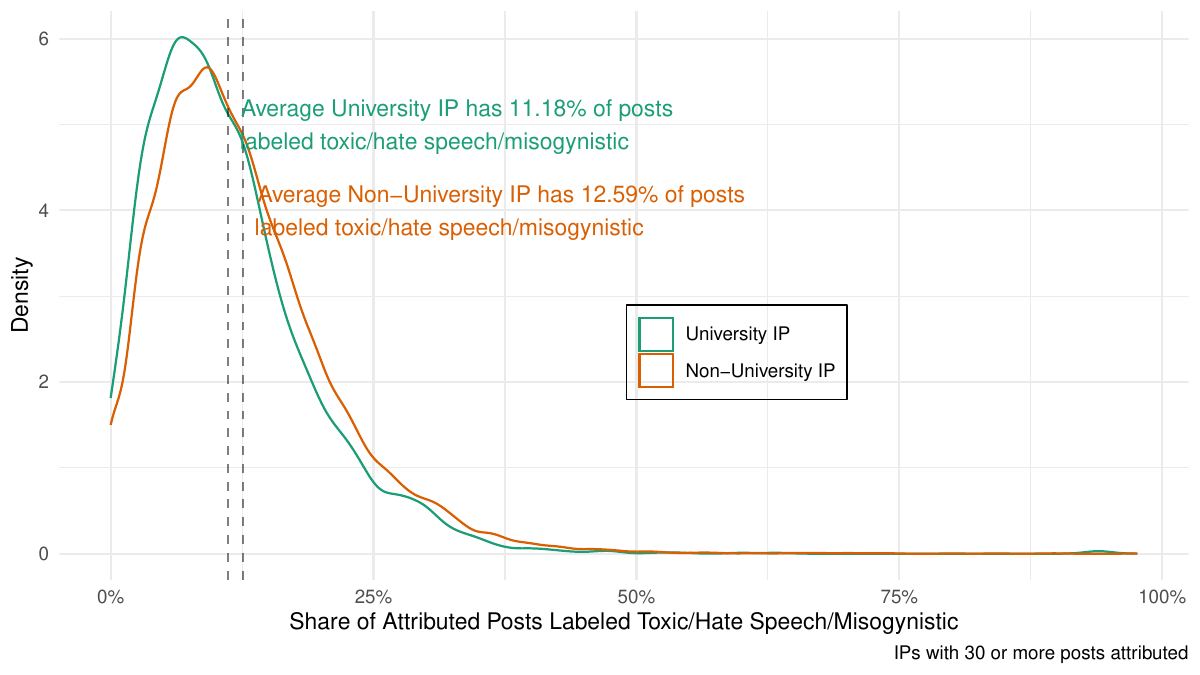}\\
    \end{subfigure}
        \begin{subfigure}[t]{0.9\textwidth}
            \includegraphics[width=\linewidth]{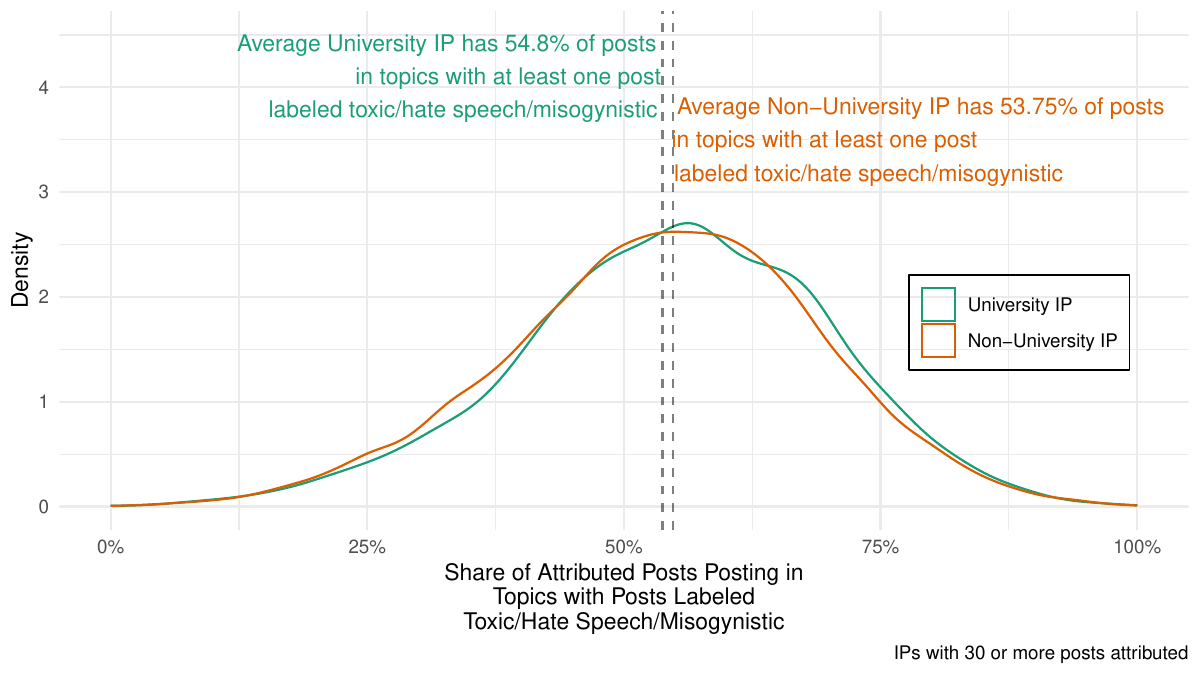}\\
    \end{subfigure}
    \caption{Distribution of misogynistic, toxic, and hate speech across IPs for university and non-university IP addresses. The figure plots density functions of the share of misogynistic, toxic, and hate speech posts across IP addresses (top panel) and density functions of the share of posts in topics with at least one post classified as misogynistic, toxic, and hate speech (bottom panel). The green lines shows the density functions for university IP addresses and the orange lines for non-university IP addresses. Only IP addresses with 30 or more posts attributed to them are shown in the figure.}
    \label{fig:ip_toxicity_density_by_univ}
\end{figure}

One might expect that EJMR users exhibit greater inhibition to contribute offensive content when connected through their work compared to when connected at home. If this were true, we should observe that posts from university IP addresses are less likely to be toxic on average than posts from non-university IP addresses. The top panel of \Cref{fig:ip_toxicity_density_by_univ} shows that this is indeed the case. The average university IP with 30 or more posts has 11.8\% of posts labeled as misogynistic, toxic, or hate speech whereas that share for the average non-university IP is equal to 12.59\%. This pattern also holds across all three groups of EJMR subforums. Overall, weighted across all IPs posts coming from university IP addresses are roughly one percentage points less likely to be problematic than posts originating from non-university IP addresses. This gap disappears for the share of posts that IP addresses contribute in topics with at least one post classified as misogynistic, toxic, and hate speech, as can be seen in the bottom panel of \Cref{fig:ip_toxicity_density_by_univ}.\footnote{We calculate this on a rolling basis; has this topic had at least one post labeled misogynistic, toxic or hate speech \emph{so far} when the post is created.} The orange university density is nearly identical to the green non-university density. EJMR contributors from university and non-university IP addresses thus appear to be equally willing to engage with problematic content. And they do so at very high rates given that the average IP address has over 50\% of its posts in topics with at least one post that is misogynistic, toxic, or hate speech.

Finally, among the top 10 IP addresses with the highest number of toxic posts, there is not a single one from a university IP address. However, among the top 10 university IP addresses with the highest number of toxic posts, there are several from leading universities and economics departments including the University of Chicago, the University of Rochester, the University of Washington, and University College London as well as from less prominent institutions such as Virginia Wesleyan University and Lingnan University. This fact pattern again underscores the diversity and pervasiveness of toxic speech on EJMR and in the economics profession.

Our analysis so far relies on simple comparisons of average shares of problematic posts on EJMR and Reddit. However, in addition to the vastly different target group comparing these two platforms might be complicated by the fact that the posts are different in their length and form. For example, it is possible that our toxicity classifiers may capture features that differ across length of post. We therefore examine how our classifiers vary by length of post (i.e., word count). We find that on both platforms short messages are more likely to be toxic than long messages. More importantly, for any length of a post, EJMR posts are always more likely to be labeled as misogynistic, toxic, and hate speech than Reddit on average.

\begin{figure}[!t]
    \centering
    \includegraphics[width=0.49\linewidth]{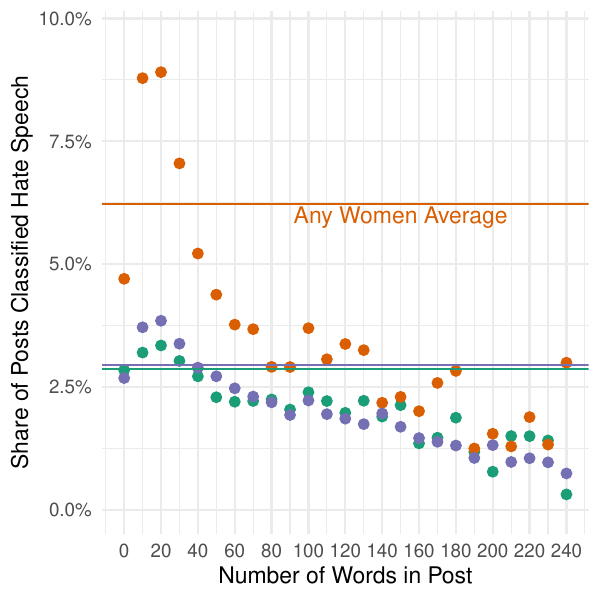}
    \includegraphics[width=0.49\linewidth]{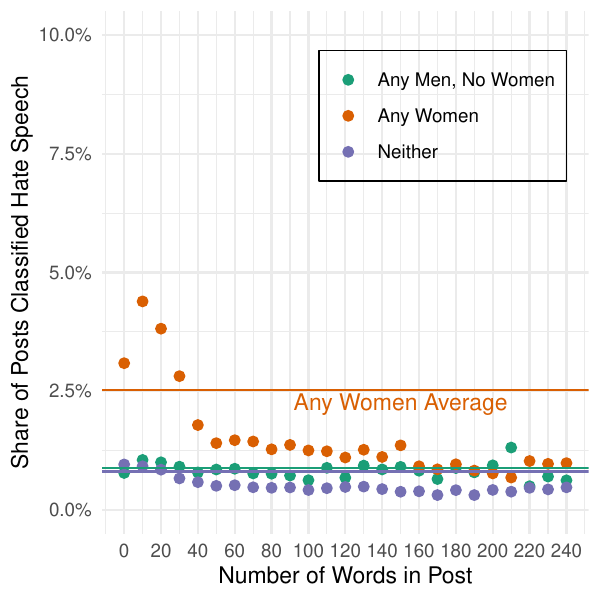}
    \caption{Share of EJMR and Reddit posts classified as hate speech by number of words. The figure shows the share of posts classified as hate speech by number of words on EJMR (left panel) and Reddit (right panel). Colors indicate the use of gendered language, such as "she" or "he." Posts mentioning women are in orange, men without women in green, and posts with no gendered language in purple.}
    \label{fig:hate_differences}
\end{figure}

One concern about the classifiers we use is that they may be labeling posts with noise. As a result, it is natural to ask how much of the difference between the platforms is due to the difference in toxicity, the difference in noise, or both? We use the response in toxicity to the mentions of women to identify the difference in toxicity across platforms more sharply. By comparing the toxicity of posts that mention women to those that mention neither men nor women, we capture the change in true toxicity rather than changes in errors. To do so, we classify posts containing female words (``she'', ``her'', ``hers'', ``woman'', ``girl'', ``gal'') as mentioning women and posts containing male words (``he'', ``him'', ``his'', ``man'', ``boy'', ``guy'') as mentioning men. \Cref{fig:hate_differences} shows the share of posts classified as hate speech by the number of words contained in the post on EJMR (left panel) and Reddit (right panel) for women (orange), men (green), and neither men nor women (purple). On both platforms, posts mentioning women are substantially more likely to contain hate speech and the same is true for misogyny and toxicity.

\begin{figure}[!t]
    \centering
    \includegraphics[width=\linewidth]{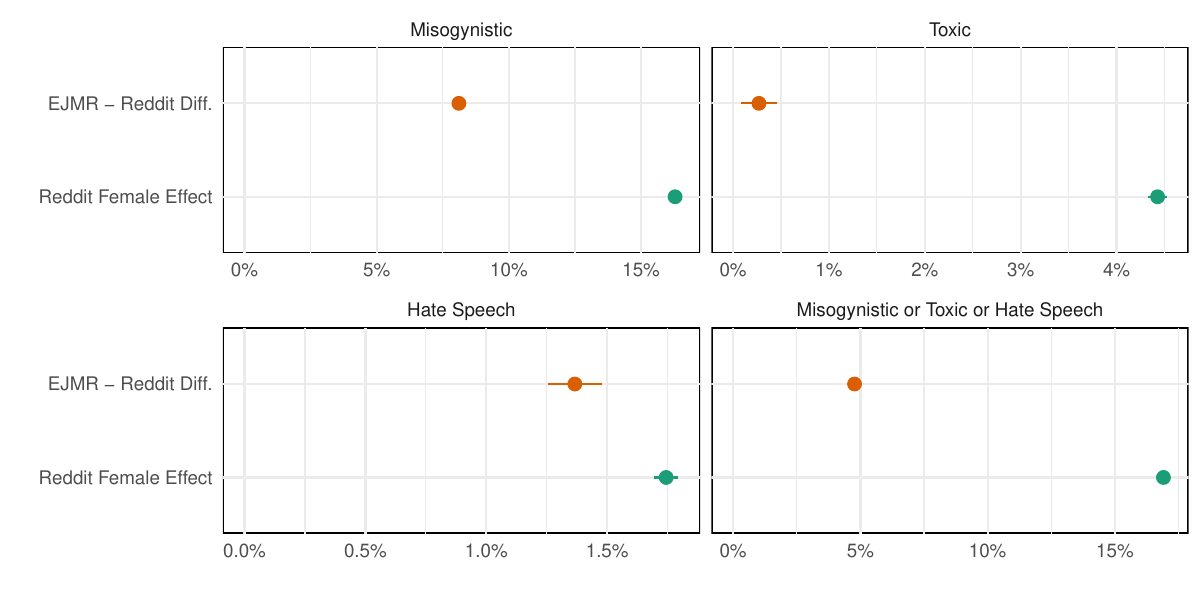}
    \caption{Point estimates and standard error bars for differences in misogyny, toxicity, hate speech, and combined offensive speech. The Reddit female effect estimates (green) show that compared to Reddit posts mentioning men, Reddit posts mentioning women are substantially more likely to be misogynistic, toxic, and hate speech. This pattern is even more pronounced on EJMR where the \emph{additional} effect relative to Reddit (orange) is sizable for misogyny and hate speech, but quite small for toxicity.}
    \label{fig:regression_comparison_slope}
\end{figure}

To be even more conservative, we compare the difference between posts mentioning women and those mentioning neither men nor women to the difference between posts mentioning men and those mentioning neither men nor women. We consider this gap to measure the ``true'' toxicity of the platform because it quantifies how much more toxic a platform is when women are mentioned compared to men. We then estimate and compare this difference in differences for both EJMR and Reddit and report the point estimates and standard error bars in \Cref{fig:regression_comparison_slope}. The Reddit female effect estimates (green) show that compared to Reddit posts mentioning men, Reddit posts mentioning women are substantially more likely to be misogynistic, toxic, and hate speech. However, this effect is even more pronounced on EJMR where posts mentioning women are \emph{even more likely} to include problematic content when compared to Reddit (orange). Across the board, we find that mentions of women attract even more misogyny (top-left panel) and hate speech (bottom-left panel) on EJMR than on Reddit, but a relatively similar level of toxicity (top-right panel).

\subsection{What explains posting behavior?}

What drives posting behavior? Due to EJMR's anonymous nature, we have limited explanatory variables at the individual level. However, among the set of posts associated with a university, we can examine several characteristics about those universities and see how they affect posting behavior.

First, we build a dataset from 2008 to 2022 using Integrated Postsecondary Education Data System (IPEDS) data to construct the number of economics degrees and number of economics Ph.D.s granted each year. This gives us a sense of the ``size'' of each institution, to determine whether it is simply larger schools that have more posts. A limitation is that these data only include American universities and colleges. We link these data by name, and are able to match 244 of 282 U.S.~ISPs.

Second, we use data on institution ranking as of March 2024 from Research Papers in Economics (RePEc) \citep{repec}.\footnote{We specifically use the ranking of institutions based on a maximum of 10 authors per institution, using the last 10 years publications, in order to avoid overweighting large institutions, and to focus on recent research activity.} This ranking is selected in two dimensions. It provides ranking activity for the top 10\% of institutions in RePEc, and the ranking is based on reported publication activity in RePEc (see \citet{covid_repec} for an example of usage of RePEc's instutition rankings). We link this data by name and match rank data for 77 university ISPs. 

Using these characteristics, we examine three outcomes at the university-by-year level: the average share of posts labeled toxic, misogynistic, or hate speech, the natural logarithm of the number of posts (which omits the extensive margin of posts), and total posts. Average share and log number of posts are estimated using ordinary least squares with year fixed effects, and total number of posts is estimated using Poisson regression with year fixed effects. We report the estimated effects in \Cref{fig:university_char_bivariate}. We report the corresponding regression table in Appendix \ref{sec:additional_tables}.

\begin{figure}
    \centering
    \includegraphics[width=\linewidth]{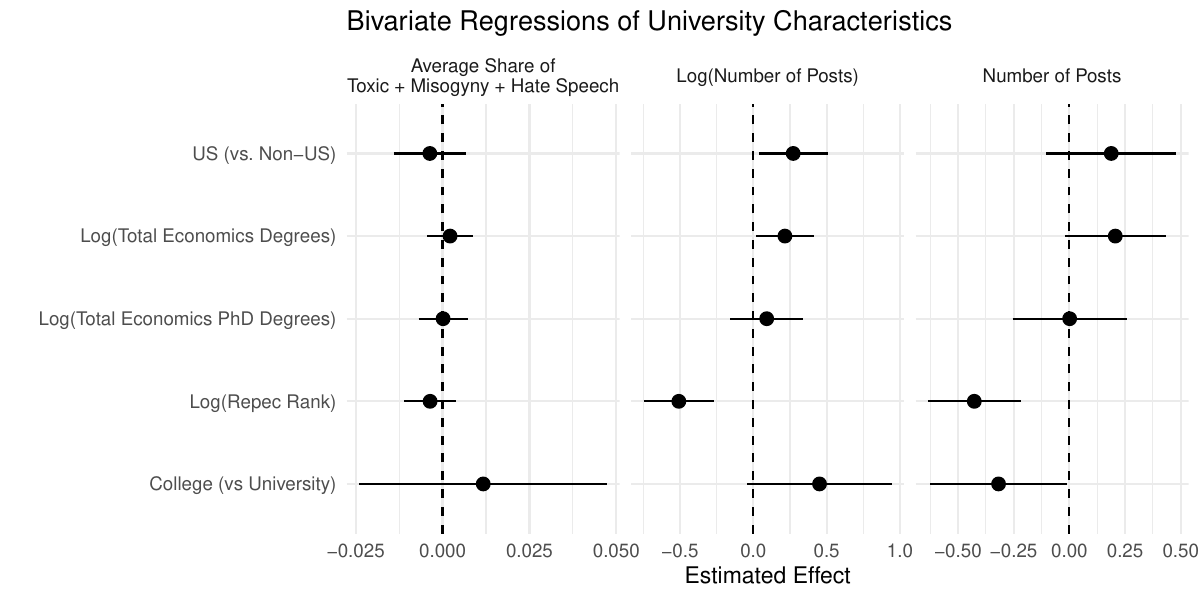}
    \caption{Bivariate estimates of outcomes regressed on university characteristics.}
    \label{fig:university_char_bivariate}
\end{figure}

None of the covariates are statistically significant in predicting the toxicity share of posts. For the number of posts, we see that U.S.~universities have roughly 25\% higher posts than non-US universities. Amongst these U.S.~universities, using the IPEDS data, we see that institutions with more matriculating economics students (both undergraduate and graduate) have more posts, but the number of matriculating PhD students is not statistically significant in predicting the number of posts. This could reflect both a larger undergraduate economics population and a larger economics faculty group that is interested in EJMR.\footnote{It is not possible to directly measure the share of faculty who are economists in the IPEDs data.} A higher rank measure (meaning a lower ranked school) predicts lower posting rate, implying that higher ranked universities tend to post more. Finally, colleges (as measured by schools without economics Ph.D.~students) tend to post 50\% less on average. This suggests that schools with economics Ph.D.~students are significantly more active, despite the size of the Ph.D.~program being less predictive. 

Our analysis of posting behavior at the university level suggests that more posts originate from larger and higher ranked institutions. But what drives individuals to post content on EJMR? Since the content posted on EJMR is anonymous, there is no extrinsic motive for any given poster to provide content, but their behaviour might be driven by intrinsic motivation. We now consider how attention may give posters reason to participate on EJMR. For this analysis, we use the full sample of posts, not just the subsample of IPs associated with a university ISP.

\begin{figure}
    \centering
    \includegraphics[width=\linewidth]{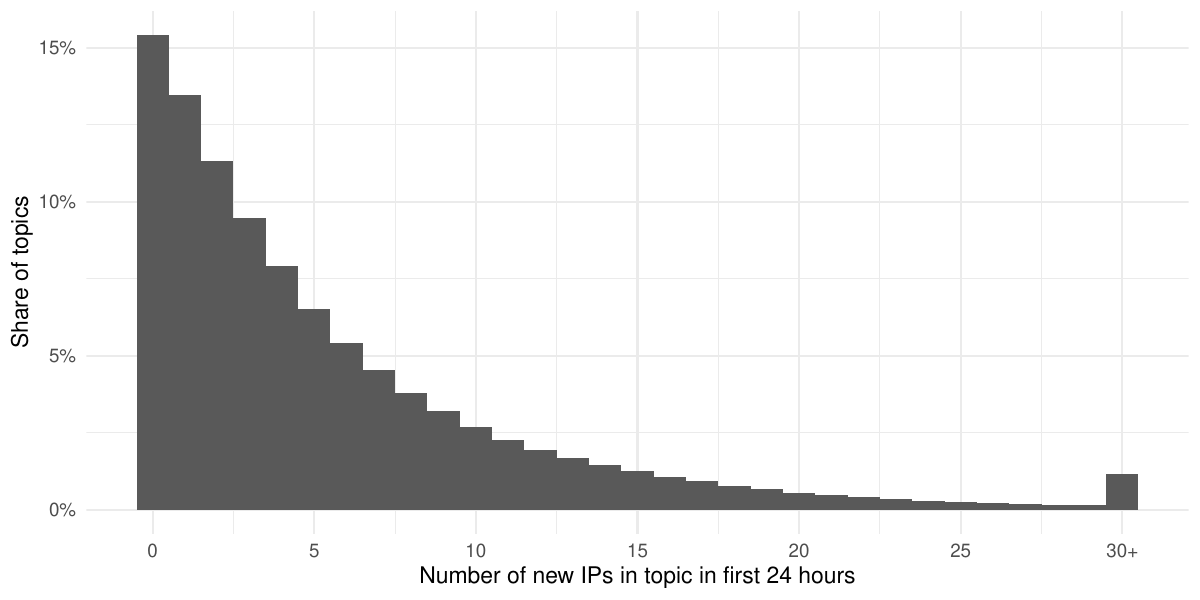}
    \caption{Distribution of the shares of topics by number of new IPs in the first 24 hours after the creation of a topic}
    \label{fig:first_post_within_day}
\end{figure}

We measure the first time that an IP creates a new topic.\footnote{A topic is also often called a thread. In what follows we use the two interchangeably.} On the first day that a topic is posted, there is a wide range of potential activity in response to it. Sometimes, there is no attention or activity on the topic, with few additional users posting in the thread. Other times, many users become active on the thread. We use the variation in how much attention a thread receives, as measured by the unique number of IPs that participate in a thread on the first day, and examine how this attention affects the original poster's activity on EJMR going forward. This attention variation on the first post is presented in \Cref{fig:first_post_within_day}.

Roughly fifty percent of new topics attract four or fewer new usernames in an hour.\footnote{The interpretation of this measure is not entirely straightforward because the same user could switch usernames by changing IP addresses. However, if we assume that users are unlikely to switch in this short time window, this measure can approximate individual posters.} However, there are some topics that receive far more attention, with many different users commenting on the thread within the first day---5\% of posts have 20 or more unique usernames posting on a topic. Importantly, note that there is no benefit that accrues to the original poster beyond this initial topic because in any other topic the original poster of the initial topic will have an entirely different username. Thus, posters cannot build a reputation as they remain anonymous to other users. However, in the initial topic, the original poster can be identified as ``OP'' (original poster) and may enjoy this status. In contrast to users on the platform, we can track IPs across topics following this initial post. This allows us to investigate how more attention for a poster's \emph{first} topic affects subsequent posting behavior by the same IP address. This approach is inspired and closely follows \cite{srinivasan2023paying}'s approach of measuring attention effects using Reddit and TikTok data. However, we differ from \cite{srinivasan2023paying} in our construction of a control group by only varying the relative level of attention \emph{among} first posters, rather than matching to a randomly sampled IP creating a topic.\footnote{We chose our approach because we found that a random match to an IP that creates a topic post is a poor control in our setting because our treatment is selected to be at the beginning (when we expect, by the nature of the selection, there to be more activity afterwards) whereas the random match can be in the middle or at the end of an IP's activity, leading to mechanically fewer posts afterwards. This alternative approach would lead us to find effects with enormous magnitudes.}

We construct a balanced panel 30 days before and 30 days after the initial topic creation by an IP, measuring the number of total posts and posts labeled as toxic, misogynistic, or hate speech created by the IP. We then estimate the following specification:
\begin{equation}
\label{eq:event-study}
    y_{it} =  \alpha_{i} + \alpha_{t} + \sum_{t, t\not=-7} \beta_{t} \text{Attention} + \epsilon_{it},
\end{equation}
where $\alpha_{i}$ corresponds to IP fixed effects, $\alpha_{t}$ are event-time fixed effects for days relative to the first post, and $\beta_{t}$ captures the effect of increasing attention in each period (excluding the day one week before the first topic post). Hence, a topic post with less attention is used as a control for the higher attention posts. We cluster these estimates at the IP level. Recall that each IP has only one event, the first topic post, but they may post in other threads prior to creating a topic. We estimate these regressions for both of our outcomes in the place of $y_{it}$. Note that in our measurement of these posts, \emph{we exclude the number of posts in the original topic}. Hence, the activity from an IP that receives much attention on their initial topic is not driven by engagement with that original topic, but rather in \emph{other} threads.

\begin{figure}
    \centering
    \includegraphics[width=\linewidth]{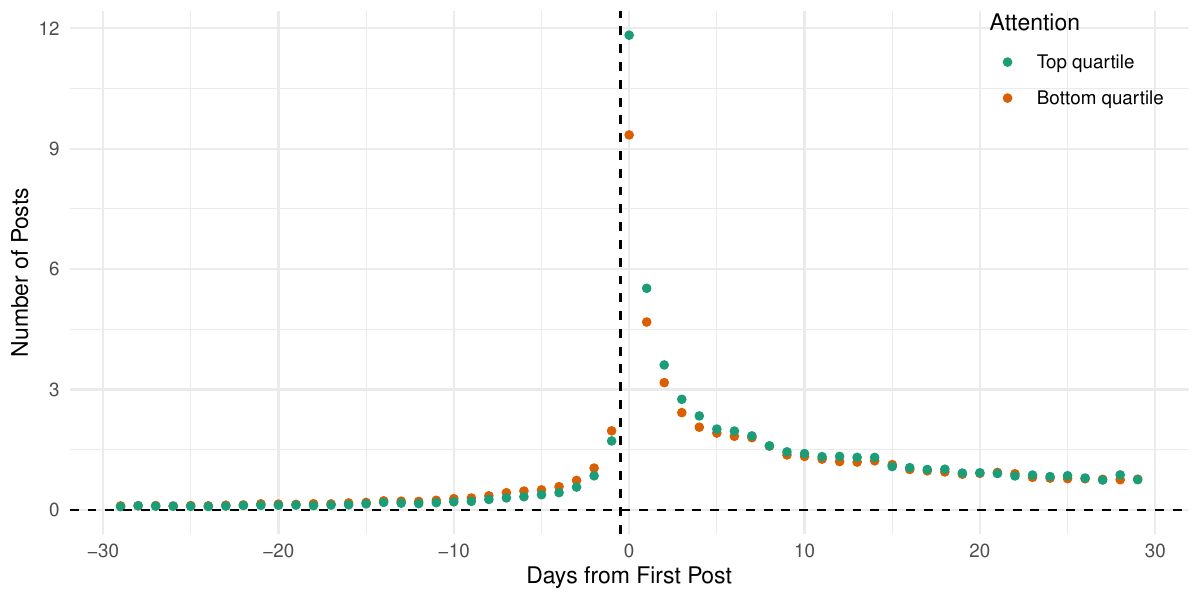}
    \caption{Average number of posts around initial topic post, split between top and bottom quartiles of attention paid to initial post. The top quartile is shown in green and the bottom quartile in orange.}
    \label{fig:binary-attention-on-posts2}
\end{figure}

To get a sense of the raw data, we split the sample into the top and bottom quartiles based on attention (ten or more usernames on the post in the first day vs fewer than four), and plot the average number of posts over time for each group in \Cref{fig:binary-attention-on-posts2}. On the day before the initial topic post, the average IP in each group averaged 1.5 posts a day, with slightly more posts created by the low-attention group. On the day of the event, the low-attention group's posts increased to nine per day, which suggests that activity generally increases with the initial posting of a topic. However, the high-attention group's posts increased to 12, roughly a third higher. Note that this comparison excludes posts within the original thread, implying that high attention in the initial topic leads to more posting activity in other topics. Subsequently, both groups' posting activity declines, and within a week the two are indistinguishable.

\begin{figure}
    \centering
     \begin{subfigure}[t]{0.49\textwidth}
    \includegraphics[width=\linewidth]{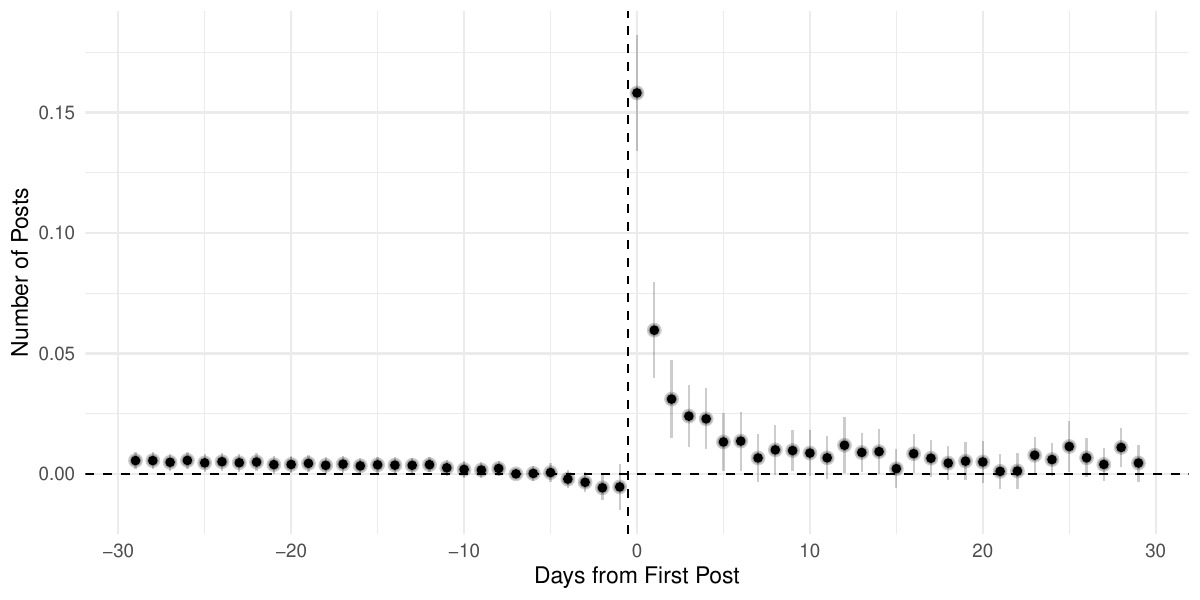}
    \caption{Effect on the total number of posts.}
    \label{fig:attention-on-posts}
    \end{subfigure}
     \begin{subfigure}[t]{0.49\textwidth}
       \includegraphics[width=\linewidth]{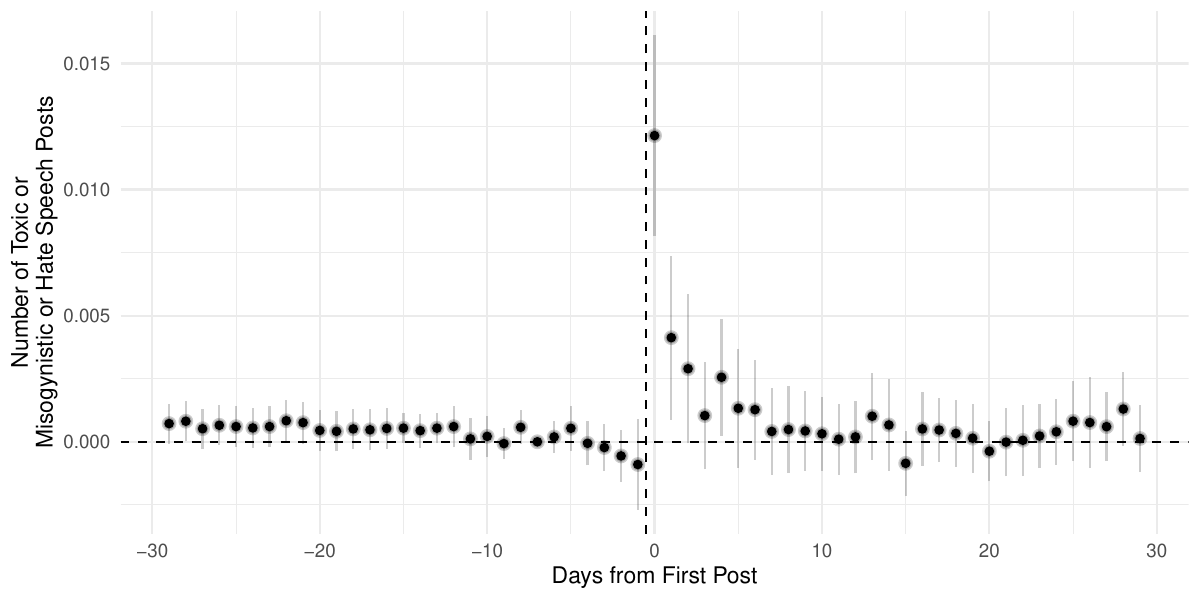}
    \caption{Effect on the total number of posts classified as toxic, misogynistic, or hate speech.}
    \label{fig:attention-on-toxic-posts}
    \end{subfigure}
    \caption{Coefficients and 95\% confidence intervals from \Cref{eq:event-study} for the effect of attention}
    \label{fig:attention-coef}
\end{figure}

In \Cref{fig:attention-on-posts}, we report the coefficients and 95\% confidence intervals from \Cref{eq:event-study} for the effect on total number of posts. There is a significant effect during the first few days following the initial posts with every additional piece of attention leading to 0.125 more posts in the first day, and 0.4 posts in on the second day, and 0.25 posts on the third day. This effect dies out over the course of a week. By day 5, the effect is statistically indistinguishable from zero relative to the week before the initial topic post. Hence, we are not able to detect any persistent effects on posting activity, although this may reflect our inability to perfectly track users over time due to changes in IP addresses.

Finally, we can look at the number of posts labeled toxic, misogynistic or hate speech as an outcome in \Cref{fig:attention-on-toxic-posts}. We observe a very similar pattern, with an increase in the number of posts during the first few days and a subsequent decline. If we consider the effect in the first period for toxic posts (0.0125) relative to the effect on total number of posts (0.15), this suggests that the average share of posts is around 8.3\%, slightly smaller than the overall average share of toxic posts at EJMR. Thus, attention does not disproportionally drive toxicity but does lead to more posting activity in general. 

The effect of attention adds detail to the motivations of individual posters on EJMR. \citet{wu2020gender} documents differences in discussion about gender, and argues that posters derive utility by boosting the prestige of their (male) in-group versus the out-group. Our results suggest that, similar to results in \cite{srinivasan2023paying} for Reddit and TikTok, posters respond to the attention paid to them on the site by using the site more afterwards. Strikingly though, in our setting the users are completely anonymous, with no persistent usernames. This is in stark contrast to the attention effects on Reddit or TikTok documented by \cite{srinivasan2023paying} where the attention that content creators receive is directly tied to their persistent usernames and can be leveraged into more followers, greater status on the platform, symbolic rewards, and even monetary payoffs. However, the attention effects we find on EJMR are solely driven by the intrinsic desire of users to see their posts receiving attention. Thus, while these effects are certainly not large enough to explain the overall posting behavior on the platform, they are useful for understanding why users would post any content in a fully anonymous setting.

\section{Conclusion}

In this paper we analyzed the behavior of posters on EJMR, a popular online platform for economists that allows users to read and post anonymously. Using only publicly available data we showed that the statistical properties of the scheme by which EJMR assigned usernames to posts until May 2023, identify the IP addresses from which most posts were made. To recover these IP addresses we employed a multi-step procedure. First, we developed GPU-based software to quickly compute the SHA-1 hashes used for the username allocation algorithm on EJMR\@. Second, we measured which IP addresses occur particularly often in a narrow time window and used the uniformity property of the SHA-1 hash to test whether these IP addresses appear more often than would likely occur by chance. 

We recovered \niceNumDistinctIPsAssigned~distinct IP addresses of EJMR posters and attributed them to \percentageOfEJMRPostsAssigned~of the roughly 7 million posts made over the past 12 years. Based on the geographic location of these IP addresses, we showed that the majority of posts come from large cities (and also smaller cities with elite universities) in the US and other developed countries with leading research institutions such as Canada, the United Kingdom, Germany, France, and Hong Kong. We further showed that posting on EJMR is pervasive throughout the economics profession including all top-ranked universities in the United States. A substantial number of posts also come from government agencies, companies, and non-profit organizations employing economists as well as universities around the world. Finally, we showed that EJMR contains much problematic content that violates the professional conduct code for economists, particularly in posts that target women.

Taken together, our paper provides further evidence of a toxic environment that is pervasive at all echelons of the economics profession, including, but not limited to, its most elite institutions.

\begin{singlespace}
\bibliographystyle{aer-nodash}
\bibliography{references.bib}
\end{singlespace}

\newpage

\appendix

\section{Alternative Toxicity Measures}\label{sec:perspective_toxicity}

In this section, we consider additional potential measures to classify posts' linguistic content. While each of the linguistic models used in the paper is best-in-class, we further corroborate the models using Perspective API, a Google-created moderation tool that is used by many large websites, including the New York Times, to detect problematic user-generated content~\citep{lees2022perspectiveapi}.  The comparisons with Perspective suggest that our results on toxicity are not sensitive to our choices of models. We also provide additional quantification of the overall linguistic content of the posts. 

\begin{figure}[h]
    \centering
    \includegraphics[width=0.8\linewidth]{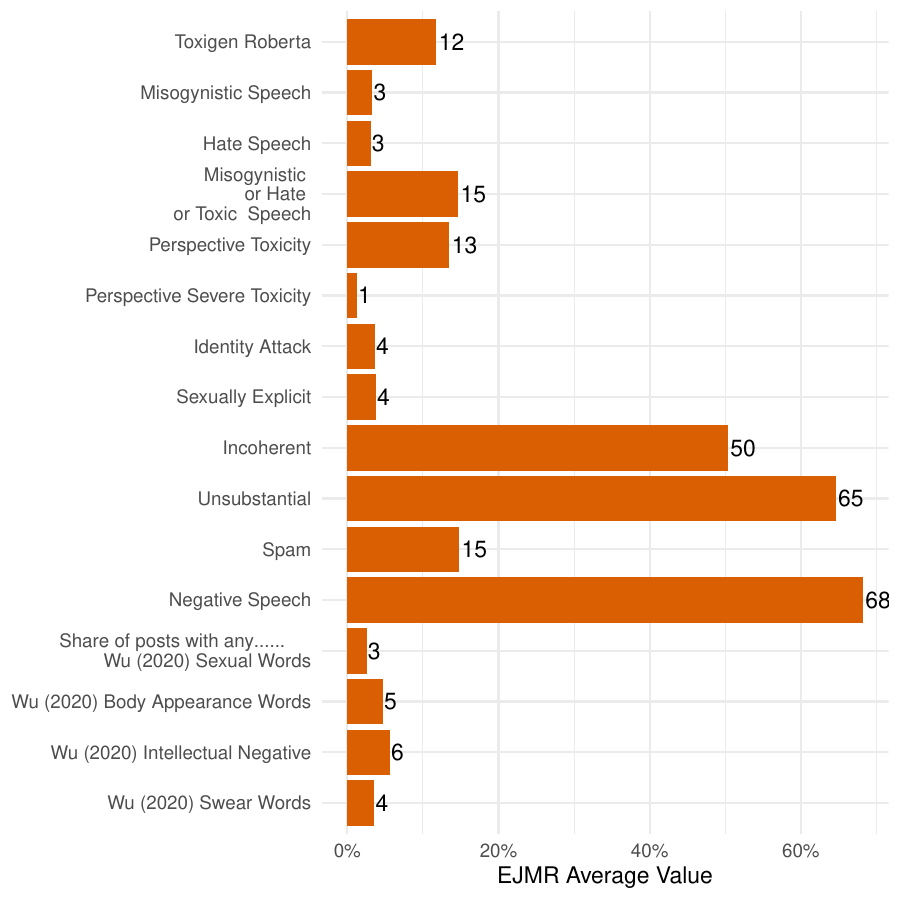}
    \caption{Average share of EJMR posts labeled by classifiers. Our respective classifiers are Toxigen Roberta~\citep{hartvigsen2022toxigen}; Misoygnistic Speech~\citep{attanasio-etal-2022-entropy}; Twitter Hate Speech~\citep{barbieri2020tweet}; either Toxigen Roberta, Misogynistic Speech or Twitter Hate Speech; Perspective Toxicity \citep{lees2022perspectiveapi}, Perspective Severe Toxicity, Perspective Identity Attack, Perspective Sexually Explicit, Perspective Incoherent, Perspective Unsubstantial, Perspective Spam; Huggingface Negative Sentiment~\citep{huggingface2023sentiment}; Indicator for presence of any word marked as (a) Sexual, (b) about physical appearance, (c) negative words about intellect, and (d) swear words as defined in \citet{wu2020gender}. }
    \label{fig:avg_linguistic}
\end{figure}

First, in \Cref{fig:avg_linguistic}, we consider the full set of EJMR posts, and calculate the average value of each classifier. The Perspective classifiers report percentage-based measures, rather than classification, and hence we report the average percentage value, rather than converting to a binary value first. Our respective classifiers are Toxigen Roberta~\citep{hartvigsen2022toxigen}; Misogynistic Speech~\citep{attanasio-etal-2022-entropy}; Twitter Hate Speech~\citep{barbieri2020tweet}; either Toxigen Roberta, Misogynistic Speech or Twitter Hate Speech; Perspective Toxicity (\url{https://perspectiveapi.com/research/}), Perspective Severe Toxicity, Perspective Identity Attack, Perspective Sexually Explicit, Perspective Incoherent, Perspective Unsubstantial, Perspective Spam;  Huggingface Negative Sentiment~\citep{huggingface2023sentiment}; and an indicator for presence of any word in the post marked as (a) sexual, (b) about physical appearance, (c) negative words about intellect, and (d) swear words as defined in \citet{wu2020gender}.

\begin{figure}[h]
    \centering
    \includegraphics[width=\linewidth]{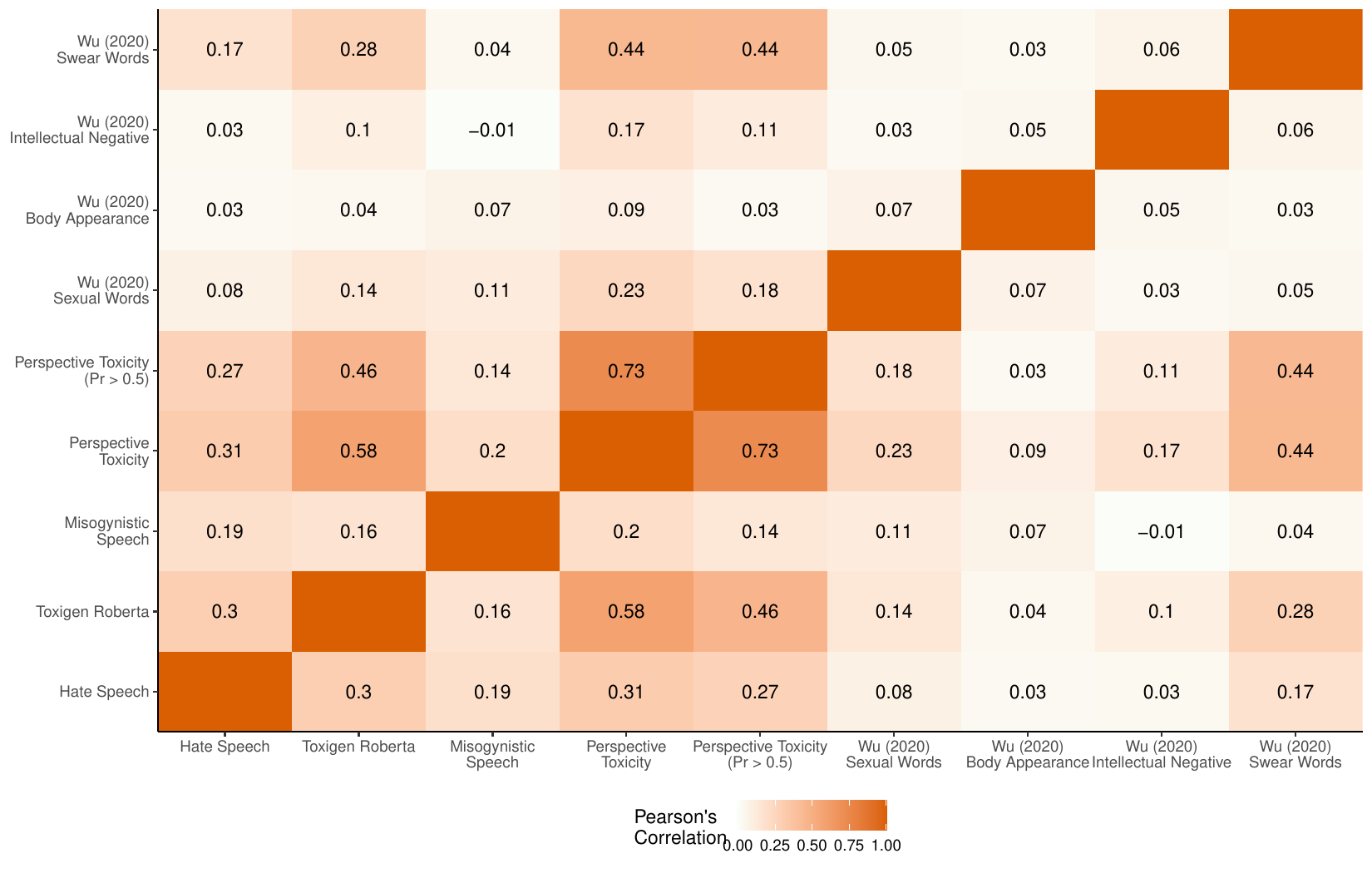}
    \caption{Correlation of Linguistic Measures}
    \label{fig:correlation_linguistic}
\end{figure}

Next, in \Cref{fig:correlation_linguistic}, we report the correlation between the three main measures used in the paper (Hate speech, Toxicity by Toxigen Roberta, and Misogynistic Speech) compared to Perspective's toxicity measure. We see a positive correlation of 0.58 between Toxigen Roberta's toxicity measure and Perspective Toxicity measure. When we convert the Perspective measure to a binary variable at the 0.5 cutoff, this correlation falls slightly. We see very similar correlations to the misogyny and hate speech measures for both Toxigen and Perspective as well. This suggests that while these measures are not identical, they imply very similar takeaways about the content of the site. 

We also see that the swear words and sexual words categorized by \citet{wu2020gender} are more correlated with our different linguistic measures. The other words are also positiviely correlated, but less so, and negative intellectual words are even negatively correlated with the categorized misogynistic speech by \citet{attanasio-etal-2022-entropy}. 

Third, we explore the underlying probability classifications from Perspective's toxicity score.\footnote{We do not use the same score measure for our other metrics because they are not well-calibrated to reflect probabilities.} Perspective defines the score in the following way (\url{https://developers.perspectiveapi.com/s/about-the-api-score?language=en_US}):
\begin{quote}
    The only score type currently offered is a probability score. It indicates how likely it is that a reader would perceive the comment provided in the request as containing the given attribute.  For each attribute, the scores provided represent a probability, with a value between 0 and 1. A higher score indicates a greater likelihood that a reader would perceive the comment as containing the given attribute. For example, a comment like “You are an idiot” may receive a probability score of 0.8 for attribute TOXICITY, indicating that 8 out of 10 people would perceive that comment as toxic. 
\end{quote}

\begin{figure}[h]
    \centering
    \includegraphics[width=\linewidth]{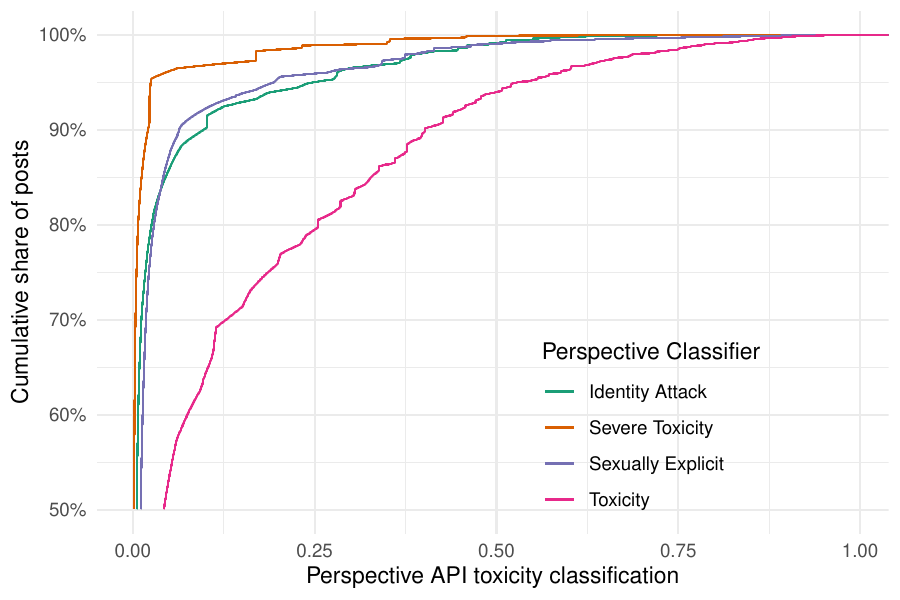}
    \caption{Cumulative density of Perspective's different classifiers across EJMR posts}
    \label{fig:persective_cdf2}
\end{figure}

In \Cref{fig:persective_cdf2}, we report the CDF of this metric across posts. Since the share of highly toxic material, as classified by Perspective, is very concentrated, we focus on the top portion of the graph. Under Perspective's definition, this would suggest that 20\% of the posts would be viewed as toxic by 25\% of individuals, and 6\% of posts would be viewed as toxic by 50\% of individuals.

\section{Frequently Asked Questions}\label{sec:faq}

There is a some misinformation about this manuscript online, particularly on EJMR\@. It is possible that you have encountered that misinformation and it seems prudent for us to address it in this FAQ.

\subsection{I read on EJMR that this is a ``hack.'' Is that true?}

No. Our study uses \emph{only} publicly available pages on EJMR, the same pages viewed by other EJMR users and indexed by search engines such as Google, Yandex, Baidu, Bing, and Archive.org. At no point did we access any non-public pages, hidden URLs, or APIs. Every page of EJMR we used in our study was in a chain of links starting from the EJMR homepage. Every page permitted access from EJMR's robots.txt, EULA (none), and terms of service (none). Furthermore, every single page from EJMR used in this study contained advertisements. All that is to say, everything in this study was visible to us as ordinary consumers of the ordinary EJMR content designed for public consumption. What we present in this paper is merely a statistical analysis of that ordinary, publicly available content, particularly the usernames shown on EJMR until May 2023.

\subsection{Did you abuse EJMR's servers in any manner?}

No. Our study used about 350,000 unique HTML pages from EJMR. According to Semrush Traffic Analytics, at the time of our scrape, EJMR received 824,000 monthly visits with an average of 12.5 pages per visit which amounted to 10.3 million monthly pageviews. Not only is our total number of pages a small fraction of EJMR's traffic, but we also used a commercial, third-party web indexer to obtain our data. The indexer we used is a major, well-known company that clearly labels its user agent, respects robots.txt directives, and meters its requests to avoid imposition on web administrators.

\subsection{Are you ``doxing'' people?}

No. Our study fits squarely in the rich tradition of scholarship on hate speech, harassment, and online communities. Our study does not dox any users. The paper reveals a single IP address, which is provided specifically in response to a public million-dollar prize offered by EJMR's owner for identifying the IP address of one particular post. We hope to collect and donate that prize to a suitable charity.

\newpage

\section{Additional Tables}\label{sec:additional_tables}

\begin{table}[h]
\input{tables/regression_results_posts}

\caption{This table reports the regression output from \Cref{fig:university_char_bivariate} for log(Number of Posts) as an outcome. }
\end{table}

\clearpage 

\begin{table}
\input{tables/regression_results_logposts}

\caption{This table reports the regression output from \Cref{fig:university_char_bivariate} for number of posts as an outcome. }
\end{table}
\clearpage 

\begin{table}
\input{tables/regression_results_badstuff}

\caption{This table reports the regression output from \Cref{fig:university_char_bivariate} for the average share of toxic, misogynist, or hate speech posts as an outcome. }
\end{table}

\clearpage 
\section{Posts by University ISP}
\begin{center}
\scriptsize
\begin{longtable}{llccccccr}
\caption{Total number of EJMR posts, share of posts labeled toxic, misogynistic, or hate speech, and share of posts labeled as containing positive or negative sentiment by university for all universities with more than 100 posts. 
} \label{tab:long_universities} \\

\toprule
\multicolumn{2}{l}{ISP and Country Code}   & \multicolumn{1}{c}{Combined} & \multicolumn{3}{c}{Separate Labels} & \multicolumn{2}{c}{Sentiment} & Count \\
    & & & Toxicity & Hate Speech & Misogyny & \multicolumn{1}{c}{+} & \multicolumn{1}{c}{-} & \\
\midrule 
\endfirsthead
\multicolumn{9}{c}%
{{\bfseries \tablename\ \thetable{} -- continued from previous page}} \\
\toprule
\multicolumn{2}{l}{ISP and Country Code}   & \multicolumn{1}{c}{Combined} & \multicolumn{3}{c}{Separate Labels} & \multicolumn{2}{c}{Sentiment} & Count \\
    & & & Toxicity & Hate Speech & Misogyny & \multicolumn{1}{c}{+} & \multicolumn{1}{c}{-} & \\
\midrule 
\endhead
HKUST & HK & 0.16 & 0.14 & 0.04 & 0.03 & 0.53 & 0.22 & 18934\\
University of Notre Dame & US & 0.12 & 0.09 & 0.02 & 0.03 & 0.64 & 0.25 & 18498\\
Stanford University & US & 0.09 & 0.08 & 0.01 & 0.02 & 0.53 & 0.27 & 17813\\
Columbia University & US & 0.10 & 0.08 & 0.02 & 0.02 & 0.57 & 0.29 & 14970\\
University of Chicago & US & 0.13 & 0.11 & 0.02 & 0.03 & 0.52 & 0.24 & 14597\\
\addlinespace
University of Washington & US & 0.13 & 0.12 & 0.02 & 0.02 & 0.50 & 0.21 & 11089\\
University of Michigan & US & 0.11 & 0.09 & 0.02 & 0.03 & 0.49 & 0.21 & 10734\\
Pennsylvania State University & US & 0.08 & 0.06 & 0.01 & 0.02 & 0.59 & 0.24 & 10723\\
James Madison University & US & 0.12 & 0.10 & 0.02 & 0.01 & 0.54 & 0.19 & 10598\\
University of Rochester & US & 0.17 & 0.16 & 0.04 & 0.02 & 0.57 & 0.23 & 9802\\
\addlinespace
University of Oxford & GB & 0.09 & 0.08 & 0.02 & 0.02 & 0.47 & 0.24 & 9296\\
UC Berkeley & US & 0.12 & 0.09 & 0.04 & 0.01 & 0.59 & 0.24 & 8784\\
University of Maryland & US & 0.08 & 0.07 & 0.01 & 0.02 & 0.50 & 0.23 & 8638\\
ROISNII & JP & 0.06 & 0.04 & 0.01 & 0.01 & 0.54 & 0.31 & 8408\\
Fraser Institute & CA & 0.06 & 0.05 & 0.01 & 0.02 & 0.50 & 0.20 & 7745\\
\addlinespace
Cardiff University & GB & 0.10 & 0.07 & 0.02 & 0.02 & 0.62 & 0.31 & 7435\\
Rouen Management School & FR & 0.14 & 0.10 & 0.03 & 0.04 & 0.52 & 0.45 & 6722\\
Federal Reserve Board & US & 0.07 & 0.05 & 0.01 & 0.01 & 0.56 & 0.25 & 6720\\
University of Georgia & US & 0.08 & 0.07 & 0.01 & 0.02 & 0.55 & 0.24 & 6551\\
Purdue University & US & 0.14 & 0.12 & 0.02 & 0.03 & 0.54 & 0.28 & 6418\\
\addlinespace
University of Southern California & US & 0.16 & 0.14 & 0.05 & 0.04 & 0.46 & 0.21 & 6118\\
Northwestern University & US & 0.09 & 0.07 & 0.02 & 0.02 & 0.53 & 0.27 & 5760\\
University of Cambridge & GB & 0.10 & 0.09 & 0.02 & 0.02 & 0.53 & 0.24 & 5738\\
University of Wisconsin Madison & US & 0.14 & 0.11 & 0.03 & 0.03 & 0.55 & 0.23 & 5722\\
University College London & GB & 0.12 & 0.10 & 0.02 & 0.02 & 0.49 & 0.16 & 5451\\
\addlinespace
Vanderbilt University & US & 0.10 & 0.08 & 0.02 & 0.03 & 0.59 & 0.29 & 5366\\
Bates College & US & 0.14 & 0.12 & 0.03 & 0.03 & 0.49 & 0.19 & 5194\\
University of Hong Kong & HK & 0.09 & 0.07 & 0.02 & 0.02 & 0.51 & 0.32 & 5127\\
Princeton University & US & 0.11 & 0.10 & 0.02 & 0.02 & 0.50 & 0.21 & 5029\\
UC Irvine & US & 0.12 & 0.09 & 0.03 & 0.03 & 0.55 & 0.27 & 4753\\
\addlinespace
University of Pennsylvania & US & 0.07 & 0.06 & 0.01 & 0.02 & 0.44 & 0.20 & 4662\\
Texas Tech University & US & 0.14 & 0.11 & 0.03 & 0.05 & 0.55 & 0.32 & 4534\\
Cornell University & US & 0.11 & 0.09 & 0.01 & 0.02 & 0.63 & 0.27 & 4510\\
Duke University & US & 0.10 & 0.09 & 0.01 & 0.02 & 0.52 & 0.24 & 4307\\
UNC Chapel Hill & US & 0.12 & 0.10 & 0.02 & 0.02 & 0.68 & 0.29 & 4281\\
\addlinespace
Michigan State University & US & 0.14 & 0.11 & 0.02 & 0.03 & 0.53 & 0.27 & 4177\\
California Institute of Technology & US & 0.10 & 0.09 & 0.01 & 0.02 & 0.47 & 0.21 & 4085\\
Goethe Universitaet Frankfurt & DE & 0.10 & 0.08 & 0.01 & 0.02 & 0.54 & 0.25 & 3994\\
University of Virginia & US & 0.20 & 0.18 & 0.03 & 0.03 & 0.64 & 0.23 & 3915\\
World Bank Group & US & 0.07 & 0.05 & 0.01 & 0.02 & 0.62 & 0.34 & 3894\\
\addlinespace
University of Wisconsin - Milwaukee & US & 0.11 & 0.09 & 0.03 & 0.02 & 0.72 & 0.27 & 3890\\
London School of Economics & GB & 0.10 & 0.08 & 0.02 & 0.03 & 0.44 & 0.24 & 3826\\
Yale University & US & 0.14 & 0.12 & 0.03 & 0.03 & 0.62 & 0.29 & 3749\\
University of British Columbia & CA & 0.14 & 0.12 & 0.04 & 0.02 & 0.55 & 0.24 & 3716\\
Georgetown University & US & 0.10 & 0.08 & 0.01 & 0.03 & 0.50 & 0.30 & 3677\\
\addlinespace
Harvard University & US & 0.10 & 0.09 & 0.01 & 0.02 & 0.49 & 0.20 & 3653\\
Australian National University & AU & 0.11 & 0.10 & 0.02 & 0.03 & 0.51 & 0.25 & 3569\\
Hong Kong Polytechnic University & HK & 0.07 & 0.04 & 0.03 & 0.02 & 0.62 & 0.30 & 3457\\
Wayne State University & US & 0.06 & 0.05 & 0.01 & 0.01 & 0.45 & 0.17 & 3438\\
University of Illinois & US & 0.12 & 0.09 & 0.02 & 0.03 & 0.48 & 0.20 & 3383\\
\addlinespace
LMU Muenchen & DE & 0.04 & 0.03 & 0.01 & 0.01 & 0.38 & 0.23 & 3345\\
Ohio State University & US & 0.12 & 0.09 & 0.03 & 0.02 & 0.51 & 0.22 & 3147\\
University of Minnesota & US & 0.10 & 0.08 & 0.02 & 0.03 & 0.61 & 0.27 & 3131\\
University of Tennessee & US & 0.09 & 0.07 & 0.02 & 0.02 & 0.41 & 0.23 & 3073\\
SUNY Buffalo & US & 0.08 & 0.06 & 0.01 & 0.02 & 0.47 & 0.18 & 3044\\
\addlinespace
University of Dhaka & BD & 0.27 & 0.25 & 0.09 & 0.08 & 0.65 & 0.25 & 3001\\
University of Toronto & CA & 0.20 & 0.17 & 0.06 & 0.05 & 0.56 & 0.29 & 2942\\
Imperial College  & GB & 0.13 & 0.11 & 0.03 & 0.02 & 0.57 & 0.24 & 2924\\
Chinese University of Hong Kong & HK & 0.07 & 0.04 & 0.02 & 0.02 & 0.61 & 0.25 & 2922\\
City University of Hong Kong & HK & 0.10 & 0.08 & 0.02 & 0.02 & 0.57 & 0.26 & 2901\\
\addlinespace
Universidade de Sao Paulo & BR & 0.14 & 0.12 & 0.03 & 0.03 & 0.47 & 0.19 & 2874\\
University of Miami & US & 0.11 & 0.09 & 0.03 & 0.02 & 0.46 & 0.19 & 2814\\
Southern Methodist University & US & 0.08 & 0.07 & 0.01 & 0.02 & 0.42 & 0.18 & 2800\\
Universitaet Mannheim & DE & 0.11 & 0.09 & 0.02 & 0.01 & 0.44 & 0.19 & 2706\\
Chapman University & US & 0.26 & 0.24 & 0.05 & 0.02 & 0.57 & 0.19 & 2649\\
\addlinespace
Universitaet Zuerich & CH & 0.14 & 0.12 & 0.03 & 0.02 & 0.62 & 0.26 & 2619\\
University of Texas at Austin & US & 0.08 & 0.07 & 0.01 & 0.02 & 0.54 & 0.26 & 2548\\
University of Arizona & US & 0.11 & 0.09 & 0.02 & 0.03 & 0.41 & 0.25 & 2517\\
ALU Freiburg & DE & 0.09 & 0.07 & 0.02 & 0.02 & 0.67 & 0.31 & 2511\\
University of Nebraska-Lincoln & US & 0.18 & 0.14 & 0.04 & 0.04 & 0.65 & 0.34 & 2487\\
\addlinespace
University of Alberta & CA & 0.09 & 0.07 & 0.01 & 0.02 & 0.53 & 0.29 & 2444\\
UC San Diego & US & 0.13 & 0.12 & 0.03 & 0.02 & 0.46 & 0.22 & 2431\\
University of Nottingham & GB & 0.15 & 0.13 & 0.04 & 0.02 & 0.68 & 0.29 & 2387\\
University of Illinois at Chicago & US & 0.15 & 0.13 & 0.03 & 0.03 & 0.49 & 0.20 & 2385\\
UC Los Angeles & US & 0.15 & 0.14 & 0.06 & 0.02 & 0.55 & 0.26 & 2368\\
\addlinespace
University of Calgary & CA & 0.13 & 0.10 & 0.02 & 0.03 & 0.46 & 0.19 & 2357\\
New York University & US & 0.09 & 0.08 & 0.02 & 0.02 & 0.45 & 0.21 & 2280\\
Queen's University & CA & 0.08 & 0.07 & 0.01 & 0.02 & 0.51 & 0.15 & 2223\\
University of Oregon & US & 0.15 & 0.11 & 0.03 & 0.04 & 0.65 & 0.27 & 2207\\
Georgia State University & US & 0.05 & 0.04 & 0.01 & 0.01 & 0.43 & 0.28 & 2173\\
\addlinespace
UC Davis & US & 0.14 & 0.11 & 0.04 & 0.05 & 0.50 & 0.21 & 2173\\
Indian School of Business & IN & 0.12 & 0.09 & 0.04 & 0.04 & 0.40 & 0.21 & 2125\\
Humboldt Universitaet Berlin & DE & 0.09 & 0.07 & 0.02 & 0.01 & 0.60 & 0.33 & 2105\\
Virginia Wesleyan University & US & 0.25 & 0.19 & 0.06 & 0.08 & 0.66 & 0.26 & 2046\\
University of Kentucky & US & 0.09 & 0.07 & 0.01 & 0.02 & 0.60 & 0.22 & 2042\\
\addlinespace
University of South Florida & US & 0.10 & 0.08 & 0.02 & 0.03 & 0.44 & 0.29 & 2005\\
Lingnan University & HK & 0.23 & 0.18 & 0.06 & 0.10 & 0.49 & 0.28 & 1983\\
University of Texas at Dallas & US & 0.19 & 0.15 & 0.04 & 0.05 & 0.61 & 0.30 & 1983\\
University of Western Ontario & CA & 0.11 & 0.09 & 0.02 & 0.03 & 0.52 & 0.24 & 1972\\
SUNY College at New Paltz & US & 0.14 & 0.12 & 0.03 & 0.02 & 0.36 & 0.24 & 1942\\
\addlinespace
Clemson University & US & 0.06 & 0.05 & 0.01 & 0.01 & 0.44 & 0.23 & 1908\\
Universitaet Heidelberg & DE & 0.13 & 0.10 & 0.03 & 0.03 & 0.51 & 0.25 & 1883\\
Universidad Pablo de Olavide & ES & 0.05 & 0.05 & 0.01 & 0.00 & 0.52 & 0.40 & 1873\\
Emory University & US & 0.13 & 0.11 & 0.03 & 0.02 & 0.60 & 0.26 & 1837\\
North Dakota State University & US & 0.09 & 0.08 & 0.01 & 0.01 & 0.68 & 0.27 & 1806\\
\addlinespace
Rice University & US & 0.25 & 0.22 & 0.07 & 0.04 & 0.69 & 0.29 & 1804\\
KAIST & KR & 0.05 & 0.04 & 0.01 & 0.01 & 0.45 & 0.27 & 1766\\
University of Lancaster & GB & 0.07 & 0.06 & 0.01 & 0.01 & 0.62 & 0.37 & 1763\\
Brown University & US & 0.09 & 0.06 & 0.02 & 0.03 & 0.54 & 0.24 & 1756\\
Washington University & US & 0.10 & 0.09 & 0.01 & 0.02 & 0.65 & 0.29 & 1725\\
\addlinespace
University of Edinburgh & GB & 0.14 & 0.12 & 0.03 & 0.02 & 0.62 & 0.26 & 1719\\
Kansas State University & US & 0.12 & 0.09 & 0.01 & 0.03 & 0.65 & 0.26 & 1716\\
Montclair State University & US & 0.13 & 0.10 & 0.02 & 0.04 & 0.47 & 0.30 & 1715\\
RFWU Bonn & DE & 0.14 & 0.11 & 0.04 & 0.04 & 0.57 & 0.27 & 1659\\
Monash University & AU & 0.12 & 0.09 & 0.03 & 0.02 & 0.53 & 0.23 & 1647\\
\addlinespace
Universitaet Kiel & DE & 0.05 & 0.04 & 0.01 & 0.01 & 0.48 & 0.25 & 1623\\
UC Santa Cruz & US & 0.12 & 0.10 & 0.04 & 0.02 & 0.44 & 0.20 & 1613\\
Loyola Marymount University & US & 0.03 & 0.03 & 0.00 & 0.00 & 0.79 & 0.21 & 1598\\
Seoul National University & KR & 0.09 & 0.07 & 0.01 & 0.02 & 0.67 & 0.21 & 1598\\
University of Waterloo & CA & 0.12 & 0.07 & 0.03 & 0.07 & 0.34 & 0.26 & 1538\\
\addlinespace
Charles University & CZ & 0.07 & 0.06 & 0.01 & 0.01 & 0.51 & 0.23 & 1536\\
Temple University & US & 0.10 & 0.09 & 0.01 & 0.02 & 0.46 & 0.20 & 1506\\
UC Merced & US & 0.02 & 0.02 & 0.00 & 0.01 & 0.41 & 0.23 & 1496\\
International Monetary Fund & US & 0.14 & 0.12 & 0.03 & 0.03 & 0.60 & 0.30 & 1495\\
William and Mary & US & 0.13 & 0.12 & 0.02 & 0.02 & 0.70 & 0.28 & 1494\\
\addlinespace
University of Montreal & CA & 0.06 & 0.04 & 0.01 & 0.02 & 0.43 & 0.20 & 1490\\
Indiana University & US & 0.10 & 0.07 & 0.02 & 0.03 & 0.60 & 0.31 & 1426\\
Cornell Weill Medical & US & 0.11 & 0.08 & 0.02 & 0.04 & 0.53 & 0.23 & 1425\\
UC (General) & US & 0.11 & 0.08 & 0.02 & 0.03 & 0.57 & 0.34 & 1393\\
Bogazici University & TR & 0.10 & 0.08 & 0.02 & 0.03 & 0.50 & 0.35 & 1389\\
\addlinespace
University of Otago & NZ & 0.09 & 0.06 & 0.01 & 0.04 & 0.45 & 0.27 & 1378\\
Kyungpook National University & KR & 0.04 & 0.03 & 0.00 & 0.01 & 0.53 & 0.14 & 1374\\
Aalto University & FI & 0.11 & 0.09 & 0.02 & 0.03 & 0.68 & 0.31 & 1365\\
Universitat de Catalunya & ES & 0.13 & 0.11 & 0.02 & 0.02 & 0.65 & 0.33 & 1341\\
UC Riverside & US & 0.10 & 0.08 & 0.01 & 0.03 & 0.50 & 0.21 & 1318\\
\addlinespace
Brock University & CA & 0.03 & 0.03 & 0.01 & 0.00 & 0.41 & 0.23 & 1305\\
NYU Abu Dhabi & AE & 0.17 & 0.12 & 0.02 & 0.07 & 0.70 & 0.30 & 1293\\
University of Exeter & GB & 0.16 & 0.13 & 0.04 & 0.03 & 0.65 & 0.29 & 1272\\
George Mason University & US & 0.08 & 0.07 & 0.01 & 0.02 & 0.50 & 0.21 & 1257\\
IIM Ahmedabad & IN & 0.21 & 0.17 & 0.05 & 0.03 & 0.75 & 0.25 & 1254\\
\addlinespace
SUNY Stony Brook & US & 0.14 & 0.12 & 0.03 & 0.02 & 0.69 & 0.27 & 1239\\
Carnegie Mellon University & US & 0.10 & 0.08 & 0.02 & 0.02 & 0.46 & 0.22 & 1222\\
Indian Institutes of Management & IN & 0.17 & 0.15 & 0.03 & 0.02 & 0.76 & 0.24 & 1222\\
Louisiana State University & US & 0.09 & 0.07 & 0.02 & 0.03 & 0.54 & 0.25 & 1214\\
Suny College at Fredonia & US & 0.07 & 0.05 & 0.01 & 0.02 & 0.69 & 0.29 & 1167\\
\addlinespace
Tulane University & US & 0.08 & 0.06 & 0.01 & 0.02 & 0.55 & 0.27 & 1148\\
Suffolk Community College & US & 0.14 & 0.11 & 0.03 & 0.03 & 0.69 & 0.30 & 1140\\
University of Missouri-Columbia & US & 0.08 & 0.08 & 0.02 & 0.01 & 0.57 & 0.26 & 1138\\
Mississippi State University & US & 0.16 & 0.09 & 0.04 & 0.09 & 0.57 & 0.31 & 1113\\
Baylor University & US & 0.07 & 0.06 & 0.00 & 0.01 & 0.57 & 0.19 & 1108\\
\addlinespace
Dartmouth College & US & 0.07 & 0.06 & 0.01 & 0.02 & 0.48 & 0.16 & 1077\\
Washington State University & US & 0.10 & 0.08 & 0.01 & 0.03 & 0.71 & 0.26 & 1068\\
University of Essex & GB & 0.10 & 0.08 & 0.04 & 0.02 & 0.39 & 0.20 & 1040\\
West Texas A\&M University & US & 0.08 & 0.07 & 0.01 & 0.01 & 0.67 & 0.30 & 1017\\
NUS & SG & 0.09 & 0.06 & 0.01 & 0.02 & 0.61 & 0.34 & 1011\\
\addlinespace
Florida International University & US & 0.07 & 0.05 & 0.02 & 0.02 & 0.66 & 0.29 & 999\\
Virginia Commonwealth University & US & 0.16 & 0.12 & 0.03 & 0.07 & 0.63 & 0.29 & 996\\
Vienna University Computer Center & AT & 0.14 & 0.12 & 0.02 & 0.02 & 0.60 & 0.31 & 955\\
University of Glasgow & GB & 0.06 & 0.05 & 0.01 & 0.01 & 0.52 & 0.22 & 951\\
Middle Tennessee State University & US & 0.09 & 0.07 & 0.01 & 0.02 & 0.58 & 0.25 & 946\\
\addlinespace
University of Warwick & GB & 0.07 & 0.06 & 0.01 & 0.01 & 0.58 & 0.25 & 937\\
Queen Mary University of London & GB & 0.12 & 0.10 & 0.01 & 0.03 & 0.52 & 0.21 & 935\\
University of Connecticut & US & 0.07 & 0.07 & 0.01 & 0.01 & 0.45 & 0.20 & 933\\
University of Texas at Arlington & US & 0.06 & 0.04 & 0.01 & 0.02 & 0.61 & 0.22 & 912\\
University of New South Wales & AU & 0.15 & 0.11 & 0.07 & 0.04 & 0.51 & 0.27 & 885\\
\addlinespace
University of Alabama & US & 0.12 & 0.11 & 0.02 & 0.02 & 0.56 & 0.31 & 848\\
Singapore Management University & SG & 0.12 & 0.10 & 0.02 & 0.01 & 0.65 & 0.31 & 827\\
London Business School & GB & 0.12 & 0.07 & 0.04 & 0.04 & 0.46 & 0.26 & 825\\
Rise at State College & US & 0.15 & 0.10 & 0.03 & 0.04 & 0.62 & 0.38 & 807\\
University of Bristol & GB & 0.14 & 0.12 & 0.04 & 0.03 & 0.55 & 0.28 & 806\\
\addlinespace
Auburn University & US & 0.12 & 0.10 & 0.01 & 0.02 & 0.73 & 0.20 & 791\\
Our Lady of the Lake University & US & 0.12 & 0.08 & 0.02 & 0.04 & 0.68 & 0.32 & 785\\
University of Colorado & US & 0.08 & 0.06 & 0.01 & 0.02 & 0.50 & 0.22 & 772\\
University of New Mexico & US & 0.06 & 0.05 & 0.01 & 0.01 & 0.60 & 0.31 & 762\\
Boston University & US & 0.10 & 0.07 & 0.01 & 0.04 & 0.59 & 0.32 & 740\\
\addlinespace
Binghamton University & US & 0.09 & 0.08 & 0.00 & 0.01 & 0.69 & 0.30 & 729\\
UNC Greensboro & US & 0.10 & 0.08 & 0.01 & 0.02 & 0.48 & 0.24 & 728\\
McMaster University & CA & 0.15 & 0.14 & 0.04 & 0.02 & 0.54 & 0.31 & 722\\
Sogang University & KR & 0.02 & 0.02 & 0.01 & 0.01 & 0.41 & 0.18 & 715\\
University of Arkansas & US & 0.20 & 0.14 & 0.06 & 0.06 & 0.75 & 0.21 & 713\\
\addlinespace
University of Houston & US & 0.14 & 0.12 & 0.04 & 0.03 & 0.47 & 0.18 & 710\\
Universidad de Guadalajara & MX & 0.15 & 0.13 & 0.02 & 0.03 & 0.49 & 0.26 & 689\\
University of Hawaii & US & 0.37 & 0.36 & 0.09 & 0.20 & 0.73 & 0.26 & 670\\
Universitaet St. Gallen & CH & 0.06 & 0.04 & 0.01 & 0.01 & 0.45 & 0.23 & 669\\
Northern Illinois University & US & 0.20 & 0.16 & 0.04 & 0.06 & 0.47 & 0.21 & 668\\
\addlinespace
Iowa State University & US & 0.02 & 0.02 & 0.00 & 0.00 & 0.33 & 0.28 & 639\\
Otto Von Guericke Universitaet & DE & 0.15 & 0.13 & 0.03 & 0.02 & 0.54 & 0.24 & 620\\
Indiana University Health Inc & US & 0.05 & 0.05 & 0.01 & 0.01 & 0.38 & 0.19 & 619\\
American University of Iraq & IQ & 0.22 & 0.19 & 0.04 & 0.03 & 0.70 & 0.30 & 619\\
University of Helsinki & FI & 0.05 & 0.05 & 0.00 & 0.01 & 0.42 & 0.23 & 615\\
\addlinespace
Oberlin College & US & 0.06 & 0.05 & 0.01 & 0.01 & 0.47 & 0.19 & 588\\
University of Central Missouri & US & 0.17 & 0.14 & 0.02 & 0.04 & 0.61 & 0.28 & 573\\
WWU & DE & 0.05 & 0.05 & 0.01 & 0.01 & 0.50 & 0.15 & 531\\
University of Florida & US & 0.18 & 0.17 & 0.04 & 0.03 & 0.49 & 0.17 & 527\\
Johns Hopkins University & US & 0.07 & 0.04 & 0.02 & 0.02 & 0.48 & 0.20 & 526\\
\addlinespace
University of Wisconsin - Stout & US & 0.15 & 0.14 & 0.03 & 0.04 & 0.50 & 0.19 & 520\\
The Urban Institute & US & 0.07 & 0.06 & 0.01 & 0.01 & 0.53 & 0.24 & 519\\
University of York & GB & 0.16 & 0.13 & 0.02 & 0.03 & 0.69 & 0.30 & 518\\
University of St. Andrews & GB & 0.20 & 0.17 & 0.06 & 0.04 & 0.58 & 0.32 & 507\\
Lehigh University & US & 0.07 & 0.06 & 0.01 & 0.01 & 0.67 & 0.31 & 497\\
\addlinespace
North Carolina State University & US & 0.10 & 0.08 & 0.01 & 0.01 & 0.63 & 0.32 & 495\\
Drexel University & US & 0.02 & 0.02 & 0.00 & 0.01 & 0.38 & 0.20 & 494\\
Georgia Institute of Technology & US & 0.08 & 0.06 & 0.01 & 0.01 & 0.48 & 0.25 & 492\\
Suffolk University & US & 0.05 & 0.03 & 0.00 & 0.01 & 0.65 & 0.29 & 485\\
Claremont University & US & 0.16 & 0.15 & 0.03 & 0.03 & 0.53 & 0.20 & 482\\
\addlinespace
St. John's College & US & 0.12 & 0.09 & 0.04 & 0.00 & 0.80 & 0.20 & 480\\
Tufts University & US & 0.09 & 0.07 & 0.01 & 0.03 & 0.61 & 0.37 & 474\\
University of South Carolina & US & 0.04 & 0.02 & 0.01 & 0.01 & 0.69 & 0.31 & 472\\
Rhode Island College & US & 0.06 & 0.05 & 0.01 & 0.00 & 0.56 & 0.28 & 471\\
University of Utah & US & 0.10 & 0.09 & 0.03 & 0.03 & 0.41 & 0.27 & 467\\
\addlinespace
University of Central Florida & US & 0.02 & 0.02 & 0.00 & 0.00 & 0.43 & 0.17 & 459\\
University of Arkansas Little Rock & US & 0.25 & 0.16 & 0.03 & 0.11 & 0.68 & 0.31 & 446\\
University of Maryland Baltimore & US & 0.07 & 0.07 & 0.01 & 0.02 & 0.50 & 0.23 & 445\\
Baruch College & US & 0.04 & 0.03 & 0.00 & 0.01 & 0.52 & 0.30 & 439\\
Arizona State University & US & 0.06 & 0.05 & 0.00 & 0.00 & 0.64 & 0.36 & 426\\
\addlinespace
University of Oklahoma & US & 0.09 & 0.08 & 0.01 & 0.02 & 0.69 & 0.31 & 426\\
Oklahoma State University & US & 0.06 & 0.05 & 0.01 & 0.01 & 0.45 & 0.29 & 424\\
Syracuse University & US & 0.09 & 0.07 & 0.02 & 0.02 & 0.41 & 0.23 & 422\\
UC Santa Barbara & US & 0.09 & 0.07 & 0.02 & 0.02 & 0.56 & 0.28 & 404\\
Wirtschaftsuniversitaet Wien & AT & 0.10 & 0.08 & 0.01 & 0.03 & 0.69 & 0.30 & 394\\
\addlinespace
ETH Zurich & CH & 0.08 & 0.06 & 0.03 & 0.01 & 0.66 & 0.34 & 393\\
Pontificia Universidad Javeriana & CO & 0.09 & 0.09 & 0.01 & 0.02 & 0.57 & 0.24 & 392\\
California State University & US & 0.06 & 0.06 & 0.01 & 0.01 & 0.46 & 0.25 & 386\\
Rutgers University & US & 0.13 & 0.11 & 0.03 & 0.05 & 0.46 & 0.19 & 383\\
Florida State University & US & 0.07 & 0.05 & 0.01 & 0.02 & 0.59 & 0.22 & 366\\
\addlinespace
Bilkent University & TR & 0.18 & 0.15 & 0.03 & 0.02 & 0.54 & 0.26 & 364\\
IIASA & AT & 0.23 & 0.18 & 0.04 & 0.04 & 0.68 & 0.32 & 363\\
University of Delaware & US & 0.09 & 0.04 & 0.01 & 0.06 & 0.63 & 0.33 & 361\\
Hunter College & US & 0.10 & 0.07 & 0.03 & 0.05 & 0.34 & 0.18 & 354\\
University of Economics Prague & CZ & 0.06 & 0.06 & 0.01 & 0.01 & 0.51 & 0.49 & 342\\
\addlinespace
University of Cincinnati & US & 0.14 & 0.11 & 0.03 & 0.03 & 0.66 & 0.31 & 335\\
University of Durham & GB & 0.08 & 0.06 & 0.01 & 0.03 & 0.42 & 0.21 & 332\\
Simon Fraser University & CA & 0.12 & 0.10 & 0.03 & 0.04 & 0.54 & 0.22 & 330\\
Texas A\&M University & US & 0.13 & 0.11 & 0.02 & 0.04 & 0.55 & 0.27 & 330\\
Universitaet Bremen & DE & 0.10 & 0.07 & 0.02 & 0.03 & 0.49 & 0.20 & 323\\
\addlinespace
University of Surrey & GB & 0.20 & 0.18 & 0.05 & 0.02 & 0.66 & 0.23 & 323\\
Queens University Belfast & GB & 0.14 & 0.13 & 0.01 & 0.02 & 0.64 & 0.24 & 321\\
Acadia University & CA & 0.08 & 0.08 & 0.01 & 0.01 & 0.65 & 0.23 & 319\\
University of Lausanne & CH & 0.09 & 0.07 & 0.02 & 0.03 & 0.61 & 0.28 & 317\\
University of Iowa & US & 0.06 & 0.05 & 0.01 & 0.02 & 0.50 & 0.19 & 316\\
\addlinespace
Thompson Rivers University & CA & 0.06 & 0.05 & 0.01 & 0.02 & 0.44 & 0.17 & 314\\
San Diego State University & US & 0.04 & 0.03 & 0.00 & 0.01 & 0.69 & 0.31 & 300\\
University of Massachusetts & US & 0.10 & 0.09 & 0.01 & 0.02 & 0.64 & 0.25 & 299\\
University at Albany & US & 0.08 & 0.07 & 0.00 & 0.02 & 0.71 & 0.29 & 291\\
Washington SIPC & US & 0.12 & 0.08 & 0.04 & 0.04 & 0.38 & 0.28 & 282\\
\addlinespace
East China University & CN & 0.04 & 0.02 & 0.03 & 0.00 & 0.41 & 0.17 & 280\\
West Virginia University & US & 0.09 & 0.09 & 0.01 & 0.00 & 0.64 & 0.35 & 280\\
Universitaet Bern & CH & 0.05 & 0.05 & 0.00 & 0.00 & 0.50 & 0.26 & 275\\
University of Leeds & GB & 0.15 & 0.12 & 0.03 & 0.01 & 0.67 & 0.32 & 275\\
University of Pittsburgh & US & 0.03 & 0.03 & 0.01 & 0.00 & 0.62 & 0.26 & 271\\
\addlinespace
Brandeis University & US & 0.09 & 0.09 & 0.02 & 0.01 & 0.45 & 0.20 & 269\\
Villanova University & US & 0.12 & 0.10 & 0.03 & 0.02 & 0.62 & 0.33 & 269\\
Philipps-Universitaet Marburg & DE & 0.07 & 0.05 & 0.01 & 0.02 & 0.51 & 0.22 & 267\\
Wright State University & US & 0.18 & 0.15 & 0.05 & 0.07 & 0.48 & 0.24 & 265\\
IFO Institut Muenchen & DE & 0.10 & 0.10 & 0.01 & 0.00 & 0.66 & 0.28 & 262\\
\addlinespace
Kennesaw State University & US & 0.14 & 0.10 & 0.05 & 0.01 & 0.55 & 0.25 & 259\\
Whitworth University & US & 0.19 & 0.12 & 0.04 & 0.06 & 0.69 & 0.31 & 257\\
University of Ghent & BE & 0.06 & 0.03 & 0.00 & 0.03 & 0.49 & 0.38 & 251\\
Virginia Polytechnic Institute & US & 0.16 & 0.14 & 0.02 & 0.04 & 0.62 & 0.38 & 251\\
Sam Houston State University & US & 0.27 & 0.25 & 0.05 & 0.04 & 0.77 & 0.23 & 249\\
\addlinespace
Miami University & US & 0.05 & 0.05 & 0.00 & 0.00 & 0.59 & 0.33 & 247\\
MIT & US & 0.12 & 0.10 & 0.03 & 0.02 & 0.58 & 0.37 & 245\\
Texas A\&M San Antonio & US & 0.11 & 0.09 & 0.04 & 0.03 & 0.69 & 0.30 & 242\\
University of East Anglia & GB & 0.05 & 0.05 & 0.00 & 0.02 & 0.42 & 0.24 & 239\\
Nova University & US & 0.16 & 0.12 & 0.06 & 0.03 & 0.49 & 0.14 & 232\\
\addlinespace
Flinders University & AU & 0.03 & 0.03 & 0.01 & 0.01 & 0.42 & 0.24 & 231\\
Lake Forest College & US & 0.12 & 0.09 & 0.01 & 0.04 & 0.48 & 0.25 & 221\\
University of Innsbruck & AT & 0.16 & 0.13 & 0.04 & 0.03 & 0.62 & 0.35 & 216\\
Valdosta State University & US & 0.12 & 0.10 & 0.04 & 0.07 & 0.40 & 0.19 & 212\\
LaTrobe University & AU & 0.17 & 0.14 & 0.06 & 0.08 & 0.48 & 0.38 & 208\\
\addlinespace
Stockholm University & SE & 0.06 & 0.05 & 0.00 & 0.01 & 0.42 & 0.19 & 207\\
Ball State University & US & 0.23 & 0.23 & 0.07 & 0.01 & 0.47 & 0.14 & 206\\
University of North Florida & US & 0.03 & 0.02 & 0.01 & 0.01 & 0.47 & 0.24 & 202\\
Kenya School of Monetary Studies & KE & 0.03 & 0.03 & 0.00 & 0.01 & 0.33 & 0.26 & 194\\
University of Melbourne & AU & 0.06 & 0.05 & 0.01 & 0.01 & 0.37 & 0.22 & 194\\
\addlinespace
American University & US & 0.07 & 0.04 & 0.01 & 0.03 & 0.54 & 0.29 & 184\\
Grand Valley State University & US & 0.04 & 0.03 & 0.00 & 0.01 & 0.29 & 0.26 & 184\\
Lund University & SE & 0.12 & 0.09 & 0.01 & 0.04 & 0.70 & 0.30 & 184\\
University de Los Andes & CO & 0.11 & 0.07 & 0.02 & 0.03 & 0.40 & 0.21 & 184\\
University of Kent & GB & 0.04 & 0.03 & 0.01 & 0.01 & 0.50 & 0.43 & 181\\
\addlinespace
Central Michigan University & US & 0.11 & 0.10 & 0.02 & 0.03 & 0.65 & 0.25 & 179\\
Norwegian School of Management & NO & 0.15 & 0.11 & 0.03 & 0.01 & 0.70 & 0.27 & 178\\
Universidad Carlos III de Madrid & ES & 0.10 & 0.06 & 0.05 & 0.00 & 0.51 & 0.24 & 176\\
McGill University & CA & 0.07 & 0.06 & 0.01 & 0.01 & 0.49 & 0.31 & 175\\
Universidad de Piura & PE & 0.26 & 0.24 & 0.06 & 0.05 & 0.75 & 0.25 & 175\\
\addlinespace
Sacred Heart University & US & 0.11 & 0.10 & 0.01 & 0.01 & 0.40 & 0.15 & 174\\
Morgan State University & US & 0.04 & 0.03 & 0.00 & 0.01 & 0.64 & 0.36 & 171\\
University of Manchester & GB & 0.05 & 0.04 & 0.02 & 0.02 & 0.44 & 0.27 & 164\\
University of Wyoming & US & 0.06 & 0.06 & 0.00 & 0.01 & 0.39 & 0.21 & 163\\
Institute for Advanced Study & US & 0.05 & 0.04 & 0.00 & 0.01 & 0.38 & 0.11 & 161\\
\addlinespace
Hong Kong Baptist University & HK & 0.03 & 0.00 & 0.02 & 0.01 & 0.48 & 0.45 & 153\\
Universidad de Alicante & ES & 0.15 & 0.11 & 0.01 & 0.04 & 0.35 & 0.22 & 150\\
University of Northumbria & GB & 0.14 & 0.13 & 0.02 & 0.02 & 0.51 & 0.18 & 148\\
Carleton University & CA & 0.10 & 0.07 & 0.01 & 0.03 & 0.65 & 0.33 & 147\\
University of Prince Edward Island & CA & 0.07 & 0.06 & 0.02 & 0.03 & 0.63 & 0.32 & 145\\
\addlinespace
George Washington University & US & 0.01 & 0.00 & 0.00 & 0.01 & 0.45 & 0.16 & 143\\
Swarthmore College & US & 0.07 & 0.05 & 0.01 & 0.02 & 0.59 & 0.23 & 139\\
Clark University & US & 0.02 & 0.02 & 0.00 & 0.00 & 0.49 & 0.30 & 135\\
IIM Kashipur & IN & 0.05 & 0.04 & 0.02 & 0.00 & 0.53 & 0.44 & 131\\
University of Texas at San Antonio & US & 0.08 & 0.08 & 0.01 & 0.00 & 0.48 & 0.15 & 130\\
\addlinespace
Berry College & US & 0.03 & 0.03 & 0.00 & 0.00 & 0.57 & 0.24 & 127\\
Northern Arizona University & US & 0.25 & 0.16 & 0.02 & 0.11 & 0.67 & 0.31 & 127\\
Purdue University Fort Wayne & US & 0.20 & 0.15 & 0.04 & 0.09 & 0.53 & 0.46 & 127\\
Hawaii Pacific University & US & 0.17 & 0.17 & 0.01 & 0.02 & 0.70 & 0.30 & 126\\
Boston College & US & 0.14 & 0.10 & 0.02 & 0.03 & 0.65 & 0.34 & 125\\
\addlinespace
Universitaet Hamburg & DE & 0.03 & 0.02 & 0.01 & 0.01 & 0.48 & 0.17 & 124\\
International Christian University & JP & 0.08 & 0.07 & 0.01 & 0.01 & 0.46 & 0.11 & 122\\
Austin College & US & 0.04 & 0.03 & 0.01 & 0.01 & 0.49 & 0.08 & 118\\
Ithaca College & US & 0.11 & 0.07 & 0.02 & 0.04 & 0.40 & 0.17 & 115\\
Southern Utah University & US & 0.07 & 0.07 & 0.00 & 0.00 & 0.62 & 0.27 & 114\\
\addlinespace
University of the Witwatersrand & ZA & 0.07 & 0.07 & 0.02 & 0.00 & 0.44 & 0.14 & 113\\
Augustana College & US & 0.04 & 0.04 & 0.00 & 0.01 & 0.52 & 0.25 & 112\\
University of Auckland & NZ & 0.12 & 0.11 & 0.02 & 0.03 & 0.34 & 0.27 & 112\\
BTU Cottbus-Senftenberg & DE & 0.16 & 0.11 & 0.04 & 0.05 & 0.45 & 0.21 & 110\\
Western Illinois University & US & 0.18 & 0.15 & 0.02 & 0.05 & 0.42 & 0.28 & 110\\
\addlinespace
Fordham University & US & 0.04 & 0.04 & 0.00 & 0.00 & 0.44 & 0.23 & 108\\
Dalhousie University & CA & 0.07 & 0.05 & 0.01 & 0.03 & 0.59 & 0.15 & 107\\
University of Kansas & US & 0.06 & 0.05 & 0.01 & 0.02 & 0.35 & 0.22 & 105\\
University of Newcastle upon Tyne & GB & 0.25 & 0.20 & 0.06 & 0.03 & 0.73 & 0.19 & 105\\
Guangzhou Education School & CN & 0.03 & 0.02 & 0.01 & 0.01 & 0.43 & 0.11 & 103\\
\addlinespace
Amherst College & US & 0.09 & 0.09 & 0.00 & 0.00 & 0.46 & 0.19 & 102\\
Colby College & US & 0.05 & 0.04 & 0.00 & 0.02 & 0.39 & 0.13 & 102\\
Melbourne Institute of Technology & AU & 0.01 & 0.00 & 0.01 & 0.00 & 0.64 & 0.36 & 101\\
\bottomrule
\end{longtable}
\end{center}

\end{document}

%% file: constants.tex
\newcommand{\niceTotalTopicUsernamePairs}{5,184,896}
\newcommand{\percentPostsAssigned}{65}

\newcommand{\niceTotalPosts}{7,098,111}
\newcommand{\niceTotalPostsWithTopicIDAndUsername}{6,912,773}
\newcommand{\niceTotalUnassignablePosts}{185,338}

\newcommand{\niceTotalTopics}{695,364}

\newcommand{\niceFirstPostDate}{December 17, 2010}
\newcommand{\niceLastPostDate}{May 10, 2023}

\newcommand{\niceFirstPostDateHex}{December 21, 2010}

\newcommand{\niceNumEJMRPostsAssigned}{4,692,946}
\newcommand{\percentageOfEJMRPostsAssigned}{66.1\%}

\newcommand{\percentageOfEJMRPostsAssignedOfAssignable}{67.9\%}
\newcommand{\percentageOfEJMRPostsNotAssignedofAssignable}{32.1\%}
\newcommand{\niceNumDistinctIPsAssigned}{47,630}

\newcommand{\niceNumEJMRPostsHavePubDate}{3,689,727}
\newcommand{\niceNumEJMRPostsLackPubDate}{3,408,384}

\newcommand{\niceUniqueTopicIDUsernamePairs}{5,184,896}

\newcommand{\niceMinIPsPerTopicUsername}{64,195}
\newcommand{\niceMeanIPsPerTopicUsername}{65,537}
\newcommand{\niceMaxIPsPerTopicUsername}{66,774}

\newcommand{\englishNumHashes}{9 quadrillion}

\newcommand{\pSevenDays}{1.37\times10^{-10}}
\newcommand{\pThirtyOneDays}{2.51\times10^{-11}}
\newcommand{\pNinetyOneDays}{1.39\times10^{-11}}

\newcommand{\predictedIPaddresses}{582,541}

%% file: tables/regression_results_posts.tex
\begingroup
\centering
\begin{tabular}{lccccc}
   \tabularnewline \midrule \midrule
   Dependent Variable: & \multicolumn{5}{c}{log(Number of Posts)}\\
   Model:                            & (1)           & (2)      & (3)           & (4)             & (5)\\  
   \midrule
   \emph{Variables}\\
   US                                & 0.2721$^{**}$ &          &               &                 &   \\   
                                     & (0.1197)      &          &               &                 &   \\   
   log(Total Economics Students)     &               & 0.0914   &               &                 &   \\   
                                     &               & (0.1268) &               &                 &   \\   
   log(Total Economics PhD Students) &               &          & 0.2157$^{**}$ &                 &   \\   
                                     &               &          & (0.1001)      &                 &   \\   
   log(RePEc Ranking)                &               &          &               & -0.5079$^{***}$ &   \\   
                                     &               &          &               & (0.1227)        &   \\   
   College                           &               &          &               &                 & 0.4517$^{*}$\\   
                                     &               &          &               &                 & (0.2512)\\   
   \midrule
   \emph{Fixed-effects}\\
   Year                              & Yes           & Yes      & Yes           & Yes             & Yes\\  
   \midrule
   \emph{Fit statistics}\\
   Observations                      & 2,157         & 870      & 859           & 603             & 2,157\\  
   R$^2$                             & 0.01684       & 0.01722  & 0.03207       & 0.11879         & 0.01035\\  
   \midrule \midrule
   \multicolumn{6}{l}{\emph{Clustered (ISP) standard-errors in parentheses}}\\
   \multicolumn{6}{l}{\emph{Signif. Codes: ***: 0.01, **: 0.05, *: 0.1}}\\
\end{tabular}
\par\endgroup

%% file: tables/regression_results_logposts.tex
\begingroup
\centering
\begin{tabular}{lccccc}
   \tabularnewline \midrule \midrule
   Dependent Variable: & \multicolumn{5}{c}{log(Number of Posts)}\\
   Model:                            & (1)           & (2)      & (3)           & (4)             & (5)\\  
   \midrule
   \emph{Variables}\\
   US                                & 0.2721$^{**}$ &          &               &                 &   \\   
                                     & (0.1197)      &          &               &                 &   \\   
   log(Total Economics Students)     &               & 0.0914   &               &                 &   \\   
                                     &               & (0.1268) &               &                 &   \\   
   log(Total Economics PhD Students) &               &          & 0.2157$^{**}$ &                 &   \\   
                                     &               &          & (0.1001)      &                 &   \\   
   log(RePEc Ranking)                &               &          &               & -0.5079$^{***}$ &   \\   
                                     &               &          &               & (0.1227)        &   \\   
   College                           &               &          &               &                 & 0.4517$^{*}$\\   
                                     &               &          &               &                 & (0.2512)\\   
   \midrule
   \emph{Fixed-effects}\\
   Year                              & Yes           & Yes      & Yes           & Yes             & Yes\\  
   \midrule
   \emph{Fit statistics}\\
   Observations                      & 2,157         & 870      & 859           & 603             & 2,157\\  
   R$^2$                             & 0.01684       & 0.01722  & 0.03207       & 0.11879         & 0.01035\\  
   \midrule \midrule
   \multicolumn{6}{l}{\emph{Clustered (ISP) standard-errors in parentheses}}\\
   \multicolumn{6}{l}{\emph{Signif. Codes: ***: 0.01, **: 0.05, *: 0.1}}\\
\end{tabular}
\par\endgroup

%% file: tables/regression_results_badstuff.tex
\begingroup
\centering
\begin{tabular}{lccccc}
   \tabularnewline \midrule \midrule
   Dependent Variable: & \multicolumn{5}{c}{Average Share of Toxic + Misogyny + Hate Speech}\\
   Model:                            & (1)      & (2)      & (3)      & (4)      & (5)\\  
   \midrule
   \emph{Variables}\\
   US                                & -0.0037  &          &          &          &   \\   
                                     & (0.0053) &          &          &          &   \\   
   log(Total Economics Students)     &          & 0.0001   &          &          &   \\   
                                     &          & (0.0036) &          &          &   \\   
   log(Total Economics PhD Students) &          &          & 0.0021   &          &   \\   
                                     &          &          & (0.0033) &          &   \\   
   log(RePEc Ranking)                &          &          &          & -0.0036  &   \\   
                                     &          &          &          & (0.0038) &   \\   
   College                           &          &          &          &          & 0.0117\\   
                                     &          &          &          &          & (0.0182)\\   
   \midrule
   \emph{Fixed-effects}\\
   Year                              & Yes      & Yes      & Yes      & Yes      & Yes\\  
   \midrule
   \emph{Fit statistics}\\
   Observations                      & 2,157    & 870      & 859      & 603      & 2,157\\  
   R$^2$                             & 0.03858  & 0.05355  & 0.05377  & 0.04485  & 0.03825\\  
   \midrule \midrule
   \multicolumn{6}{l}{\emph{Clustered (ISP) standard-errors in parentheses}}\\
   \multicolumn{6}{l}{\emph{Signif. Codes: ***: 0.01, **: 0.05, *: 0.1}}\\
\end{tabular}
\par\endgroup